\documentclass[tightenlines,nofootinbib,floatfix,notitlepage,11pt,superscriptaddress]{revtex4-1}

\usepackage{tcolorbox}
\tcbuselibrary{theorems}
\usepackage{epsfig}
\usepackage{graphicx}
\usepackage{multirow}
\usepackage[inline]{enumitem}
\usepackage{color}
\usepackage[dvipsnames]{xcolor}
\usepackage{amsmath,amssymb}
\usepackage{tabularx}
\usepackage[pdfencoding=auto, psdextra]{hyperref}
\usepackage{units}
\usepackage{soul}
\usepackage[T1]{fontenc}
\usepackage{ulem} %For crossing out words.
\usepackage{aas_macros}
\usepackage{upgreek}
\usepackage{mathrsfs}
\usepackage{bbm}
\usepackage{bbold} 
\usepackage{enumitem}
\newcommand{\be}{\begin{equation}}
\newcommand{\ee}{\end{equation}}
\newcommand{\bea}{\begin{eqnarray}}
\newcommand{\eea}{\end{eqnarray}}

\newcommand{\sigmam}{\sigma/m}
\newcommand{\cmg}{\; \textrm{cm}^2/\textrm{g}}
\newcommand{\Msun}{M_\odot}

\newcommand{\cdm}{\rm cdm}
\newcommand{\kpc}{\, \mathrm{kpc}}
\newcommand{\Mvir}{M_{\rm 200}}
\newcommand{\cvir}{c_{\rm 200}}
\newcommand{\rvir}{r_{\rm 200}}
\newcommand{\rsph}{r_{\rm eff}}

\newcommand{\qcdm}{q_\mathrm{cdm}}
\newcommand{\N}{\mathcal{N}}
\newcommand{\SIDM}{$\mathtt{SIDM1}$}
\newcommand{\CDM}{$\mathtt{CDM}$}
\newcommand{\subsubsubsection}[1]{\textbf{#1}}
\newcommand\york{Department of Physics and Astronomy, York University,\\Toronto, Ontario, M3J 1P3, Canada}
\newcommand\gu{Institute for Theoretical Physics, Goethe University, 60438 Frankfurt am Main, Germany}

\begin{document}

%%%%%%%%%%%%%%%%%%%%%%%% Title page %%%%%
\title{Jeans Model for the Shapes of Self-interacting Dark Matter Halos}

\author{Yilber Fabian Bautista}
\email[]{yfabian.bautista@ed.ac.uk}	
\affiliation{Higgs Centre for Theoretical Physics,
School of Physics and Astronomy,\\
The University of Edinburgh, Edinburgh EH9 3JZ, Scotland, UK}
\affiliation{Institut de Physique Théorique, CEA, Université Paris–Saclay,
F–91191 Gif-sur-Yvette cedex, France}
\author{Andrew Robertson}
\email{arobertson@carnegiescience.edu}
\affiliation{The Observatories of the Carnegie Institution for Science, 813 Santa Barbara Street, Pasadena, CA 91101, USA}
\author{Laura Sagunski}
\email{sagunski@itp.uni-frankfurt.de}
\affiliation{\gu}
\author{Adam Smith-Orlik}
\email{asorlik@yorku.ca}
\affiliation{\york}
\author{Sean Tulin}
\email{stulin@yorku.ca}
\affiliation{\york}

\date{\today}

\begin{abstract}
The Jeans model is a semi-analytical approach to modeling self-interacting dark matter (SIDM) that works remarkably well to reproduce the spherically-averaged halo profiles from observations and simulations of relaxed galaxies and galaxy clusters.
However, SIDM halos are not spherically symmetric in general since they respond to nonspherical baryon distributions and retain nonsphericity from their initial collapse.
In this work, we generalize the Jeans model to describe SIDM density profiles and halo shapes beyond spherical symmetry.
Observational tests via halo shapes are especially important for testing SIDM in massive galaxies, $M_{\rm 200} \sim 10^{12} - 10^{13} \; \Msun$, where SIDM and collisionless dark matter halos can have indistinguishable spherically-averaged profiles but distinct halo shapes.
We validate our model by comparing to cosmological simulations with baryons for both SIDM with $\sigmam = 1 \cmg$ and collisionless cold dark matter.
Our approach differs from previous work in this direction, taking into account the fact that multiple scatterings are required to impact the shape of the halo, as well as being computationally inexpensive to implement.
The nonspherical Jeans model can be used in conjunction with halo shape observations (e.g., from gravitational lensing or X-ray data) to directly constrain dark matter self-interactions.

\end{abstract}

%\pacs{}
\maketitle

%%%%%%%%%%%%%%%%%%%%%%%%%%%%%%%%%%%%%%%%
\section{Introduction \label{sec:intro}} 
%%%%%%%%%%%%%%%%%%%%%%%%%%%%%%%%%%%%%%%%

Microphysical collisions between dark matter particles can impact structure formation in the universe on small scales, a framework known as self-interacting dark matter (SIDM)~\cite{Spergel:1999mh}.
This paradigm has long been motivated to explain the characteristics of dark matter-dominated dwarf and low surface brightness galaxies.
These systems are observed to have halos with flatter inner profiles and/or reduced central densities~\cite{Moore:1994yx,Flores:1994gz,McGaugh:1998tq,deBlok:2001hbg,Kleyna:2003zt,Oh:2010ea,Walker:2011zu,BoylanKolchin:2011de} compared to the predictions from collisionless cold dark matter (CDM) simulations~\cite{Carlson:1992fn,Navarro:1995iw,Navarro:1996gj}. To reconcile this discrepancy, elastic DM self-interactions are introduced, leading to heat transfer across the halo and forming a core instead of a cusp in the center of the halo (in dark matter-dominated systems)~\cite{Dave:2000ar,Colin:2002nk}.
The required scattering cross section per unit mass is in the range $\sigma/m \sim 0.5 - 50 \cmg$~\cite{Elbert:2014bma}.
At the same time, SIDM preserves the successful predictions of CDM for large-scale structure~(see~\cite{Tulin:2017ara,Adhikari:2022sbh} for reviews).

Yet, the core-cusp problem remains unresolved, largely due to the question of baryonic feedback (e.g., from supernovae) for CDM halos~\cite{Navarro:1996bv,Governato:2009bg,Brooks:2012vi,Zolotov:2012xd}. 
The more pressing challenge for CDM, however, is whether simulations with baryons can account for the diverse range of spiral galaxy rotation curves observed in nature, spanning both cored and cuspy halos~\cite{Oman:2015xda,2018MNRAS.473.4392S,Kaplinghat:2019dhn,Zentner:2022xux}.
In the  SIDM framework, the halo behaves as a hydrostatic fluid in the total gravitational potential, forming a core if the central baryon density is low and a cusp if it is high~\cite{Kaplinghat:2013xca}, which is consistent with observed rotation curves~\cite{Kamada:2016euw,Ren:2018jpt}.

In any case, other astronomical systems with different systematics are needed to provide complementary tests of SIDM.
For example, halos of larger galaxies (and beyond) are less impacted by supernova feedback.
Several studies of galaxy groups and clusters have constrained the SIDM cross section to be around or below~$\sim 0.1 \; {\rm cm^2/g}$~\cite{Kaplinghat:2015aga,Harvey:2018uwf,Sagunski:2020spe,Andrade:2020lqq,K:2023huw}.
This is accommodated in many dark sector models that predict velocity-dependent scattering cross sections~\cite{Ackerman:2008gi,Feng:2009hw,Loeb:2010gj,Buckley:2009in,Lin:2011gj,Cline:2013pca,Cline:2013zca,Boddy:2014yra}.
Since the typical scattering velocity scales with halo mass, combining dwarf and cluster measurements is a powerful way to constrain SIDM models.

The predictions of SIDM with velocity-dependent cross section can be tested on intermediate scales, namely, massive spiral and elliptical galaxies.
On the observational side, there are no significant discrepancies with CDM. 
Mass-modeling studies of spiral galaxy lenses~\cite{Trott:2008yd,Barnabe:2012mi} and elliptical galaxies~\cite{Humphrey:2006rv,Wasserman:2017vnt,Shajib:2020ywe} prefer cuspy dark matter profiles, as predicted from CDM simulations.
Halo shapes of elliptical galaxies, inferred from X-ray observations, are also nonspherical and show no evidence of becoming rounder toward their centers~\cite{Buote:2002wd,McDaniel:2021kpq}, also in line with CDM predictions~\cite{Peter:2012jh}.
At face value, the absence of round, cored halos disfavors SIDM on these scales, leading to the constraint $\sigmam \lesssim 1 \cmg$~\cite{Peter:2012jh,McDaniel:2021kpq}.
However, these considerations are based on SIDM-only simulations, whereas realistic halos have baryons.
Due to the response of SIDM to the baryon potential, SIDM halos need not be cored nor round in their interiors~\cite{Kaplinghat:2013xca,Vogelsberger:2014pda,Elbert:2016dbb,Sameie:2018chj}.
Recent simulations including baryons found that halo shapes inferred from X-ray and gravitational lensing are very similar for both CDM and SIDM with $\sigmam = 1 \cmg$~\cite{Despali:2022vgq}.

In this paper, we propose a new semi-analytic method for modeling the nonspherical density profiles and shapes of SIDM (and also CDM) halos.
Our setup is based on the Jeans model of Refs.~\cite{Kaplinghat:2013xca,Kaplinghat:2015aga}.
The Jeans model treats the collisional inner halo as an isothermal ideal gas in hydrostatic equilibrium (with baryons included as a fixed external potential) which is smoothly matched onto a collisionless CDM outer halo.
Despite some of the assumptions of the model being questioned~\cite{Sokolenko:2018noz}, comparisons with simulated systems have demonstrated that the spherically-symmetric version of the Jeans model can accurately reproduce the spherically averaged density profiles of simulated SIDM halos~\cite{Creasey:2016jaq, Robertson:2020pxj}. In addition, the spherically-symmetric version has been widely used to model real systems, yielding excellent agreement with the dark matter density profiles inferred from many observations~\cite{Kaplinghat:2015aga,Ren:2018jpt,Sagunski:2020spe,Jiang:2022aqw,Andrade:2020lqq}. Here we generalize the Jeans model beyond spherical symmetry to model the nonspherical structure of SIDM halos including baryons.

Previous work has studied an axisymmetric version of the isothermal Jeans model~\cite{Kaplinghat:2013xca}, in particular, to assess the effect of baryonic  thin disk potentials on dark matter halos and rotation curves~\cite{Amorisco:2010sb,Kamada:2016euw,Ren:2018jpt}.
However, as we show, this approach has a boundary condition issue when one tries to match the inner isothermal profile to an outer CDM profile.
The issue is that hydrostatic equilibrium is imposed for a given number of scatters per particle $\N$, say, at $\N = 1$, which leads to a halo shape discontinuity at the matching surface.
In contrast, $N$-body simulations for SIDM show that halo shapes have a smooth transition between collisionless and collisional regions of the halo as a function of $\N$~\cite{Peter:2012jh}.

To avoid this issue, we propose a new setup dubbed the ``squashed'' Jeans model, which goes as follows.
First, we solve for the density profile according to the spherical Jeans model.
Next, iso-density spheres are mapped (``squashed'') to iso-density spheroids (in the axisymmetric case) or ellipsoids (in the general case).
Using kinetic theory arguments, we derive this mapping as a function of $\N$, which in turn describes the shape of the halo.
This approach yields a smooth transition at the matching surface where the isothermal and hydrostatic limits are reached only in the limit $\N \gg 1$.
In the outer halo, where $\N \ll 1$, the halo retains its nonsphericity from its initial collapse, i.e., in the collisionless limit.
We validate our model by comparing to cosmological $N$-body simulations with baryons for both SIDM with $\sigma/m = 1\; \cmg$, as well as CDM~\cite{Robertson:2020pxj}.

Let us mention that Jeans modeling for SIDM is complementary to $N$-body simulations, which are the traditional ``from-first-principles'' approach for testing the SIDM paradigm.
The current state-of-the-art are cosmological simulations including both dark matter self-interactions and baryons to model the interplay between dark matter and galaxy evolution~\cite{Vogelsberger:2014pda,2017MNRAS.472.2945R,Robertson:2017mgj,Robertson:2018anx,Robertson:2020pxj,Despali:2022vgq}.
However, there are some limitations to this approach.
First, they are computationally expensive, which limits the number of particle physics models that can be simulated.
Second, the fact that they are fully dynamical means that it is not possible to choose the final state, say, to fit a particular observed system.
This limitation is alleviated for large simulation volumes containing many systems spanning the halo mass range of interest.
However, there is no guarantee that the baryon profiles---which depend on how star formation, feedback, and other sub-grid astrophysics effects are implemented---match observed systems.
This is particularly important when testing SIDM in systems that are largely baryon-dominated in their interiors.
Precise modeling of the baryon density is possible for idealized simulations with a fixed baryon potential~\cite{Elbert:2016dbb,Sameie:2018chj}, but these remain intensive as well.

The Jeans model has its advantages and disadvantages compared with N-body simulations. 
On the plus side, the Jeans model is much less computationally intensive than simulations.
Also, the non-dynamical nature of the method means that we can completely fix the final state of the system.
This allows us to model individual observed systems on a one-by-one basis, matching all observables in as much detail as desired.
However, the drawback of being non-dynamical is that the Jeans model can only assess whether a given system is {\it consistent} with SIDM, not whether such a system will actually arise dynamically.
Nevertheless, the goal is that once the Jeans model has been tested against simulations, it can be used to confront SIDM predictions against observations across a wider range of cross sections and baryon profiles compared to simulations.
(We defer this to future work.)

The remainder of this work is organized as follows. In Sec.~\ref{sec:spherical_jeans} we review the standard spherically symmetric isothermal Jeans model. Then in  Sec.~\ref{sec:JMBSS}, we introduce  the characteristics of nonspherical CDM halos, which will serve as guidance for modeling nonspherical SIDM halos, 
and describe two 
versions of the Jeans model beyond spherical symmetry:
{\it (i)} The nonspherical isothermal Jeans model obtained by solving the isothermal Jeans-Poisson equations without spherical symmetry.
This setup is nontrivial to solve numerically, and it turns out to be unsatisfactory anyway when compared to $N$-body simulations due to the discontinuity of the density contours  at the matching surface mentioned above. 
{\it (ii)}  We introduce the squashed Jeans model, which agrees better with simulations producing smooth shapes, and is much easier to compute numerically. Along the way, we compare the 2D halo profiles predicted from each model, as well as the prediction for the spherically-averaged density profiles and halo shapes in the isothermal and  squashed model, to those obtained directly via the spherically symmetric Jeans model. 
Next, in Sec.~\ref{sec:sims}, we introduce the  \textsc{Eagle-50} simulations from Ref.~\cite{Robertson:2020pxj}, which are cosmological simulations with baryons for CDM and SIDM with $\sigmam = 1 \cmg$, with halos spanning the mass range $3.5 \times 10^{11} - 1.6 \times 10^{14} \; \Msun$.
In Sec.~\ref{sec:shapes}, we compare the halo shapes inferred from the squashed Jeans model to the simulations.
Our conclusions are given in Sec.~\ref{sec:conclude}.
Further information about our numerical methods is given in Appendices~\ref{app:relaxation}, \ref{app:spherical}, and \ref{app:nonspherical}.\footnote{Code is available from the authors upon request.}  
In Appendix \ref{app:squashed}, we provide a  kinetic-theory derivation of the squashed halo-shape functions for SIDM. 
Finally, we included four ancillary files (\texttt{Ancillary1,2,3,4.pdf}) with this submission showing the full sample of simulated halos analyzed in this work.

%\footnote{Our numerical package is publicly available under the following link: {\bf [url]}} 

\section{Review of the isothermal Jeans model with spherical symmetry}

\label{sec:spherical_jeans}

The spherical Jeans model for SIDM delineates the halo into two regions: {\it (i)} the inner halo where the self-interaction rate is large and collisions enforce hydrostatic equilibrium for dark matter particles, and {\it (ii)} the outer halo where the self-interaction rate is small and dark matter is effectively collisionless.
The matching radius $r_m$ between the two regions is  implicitly defined via  the rate equation 
\be \label{eq:rate}
{\rho}_{\rm dm}(r_m) \frac{\sigma}{m} v_{\rm rel} t_{\rm age} = \N_m \, ,
\ee
where ${\rho}_{\rm dm}(r)$ is the spherically-symmetric dark matter density, $\sigma/m$ is the self-interaction cross section per unit mass, $v_{\rm rel}$ is the typical relative velocity between dark matter particles, $t_{\rm age} \approx 5-10$ Gyr is the characteristic age of a given system, and $\N_m$ is the mean number of scatters at $r_m$.\footnote{For velocity-dependent cross sections, one should average $\sigma v_{\rm rel}$ over the distribution of particles in the halo. There are some subtleties associated with this averaging, discussed (for example) in Ref.~\cite{2023ApJ...946...47Y}.}
Henceforth, we follow the typical assumption that $\N_m = 1$, i.e., that dark matter particles within $r<r_m$ have scattered more than once per particle, while those at $r>r_m$ have scattered (on average) less than once each~\cite{Rocha:2012jg,Kaplinghat:2015aga}.

To model the collisional inner halo ($r < r_m$), the starting point is the (steady-state) Jeans equation~\cite{Kaplinghat:2013xca,Kaplinghat:2015aga},
\be \label{eq:Jeans_sph}
\sigma_0^2 \frac{ \partial {\rho}_{\rm dm}}{\partial 
 r} = - {\rho}_{\rm dm} \,  \frac{\partial \Phi}{\partial r} \, ,
\ee
which describes an isothermal fluid in spherical hydrostatic equilibrium.
Here, $\sigma_0$ is the 1D velocity dispersion, assumed to be constant, and the total gravitational potential $\Phi$ (dark matter plus baryons) solves Poisson's equation,
\be \label{eq:Poisson}
\nabla^2 \Phi = 4 \pi G \big(\rho_{\rm dm} + \rho_b \big) \, .
\ee
The baryon density $\rho_b$ is also taken to be spherically symmetric.

In the outer halo ($r > r_m$), the self-interaction rate is presumed to be small and the profile is approximated as what the halo would be for collisionless CDM.
In CDM-only simulations, halos are well-described by the Navarro-Frenk-White (NFW) profile~\cite{Navarro:1995iw,Navarro:1996gj}
\be\label{eq:nfw}
\rho_{\rm NFW}(r) = \frac{\rho_s}{(r/r_s)(1+r/r_s)^2} \, ,
\ee
parametrized in terms of scale radius $r_s$ and density $\rho_s$, or equivalently in terms of virial mass $\Mvir$ and concentration parameter $\cvir = \rvir/r_s$, where $\rvir$ is the virial radius.
In cosmological settings, the parameters are approximately correlated through the mass-concentration relation (MCR)~\cite{Ludlow:2013vxa}. 
With the presence of  baryons, CDM halos can be affected by galaxy formation, e.g., due to adiabatic contraction (AC) from baryonic in-fall~\cite{Blumenthal:1984bp,Gnedin:2004cx,Cautun:2019eaf}. 
Here we consider an adiabatically-contracted NFW profile following Ref.~\cite{Cautun:2019eaf}
\be
M_{\rm AC}(r) = M_{\rm NFW}(r) \left[
0.45 + 0.38 \left(\frac{(1-f_b) M_b(r)}{f_b M_{\rm NFW}(r)} + 1.16\right)^{0.53} \right] \, ,
\ee
where $M_{b}, M_{\rm NFW}, M_{\rm AC}$ are the enclosed mass profiles for baryons, and dark matter without and with AC, respectively. $f_b\approx 0.16$ is the cosmological baryon fraction \cite{Planck:2018vyg}.

The full dark matter density profile spanning both regions is a piecewise function
\be \label{eq:matching}
\rho_{\rm dm}(r) = \left\{ \begin{array}{cc}
\rho_{\rm iso}(r) & r < r_m \\
\rho_{\cdm}(r) & r > r_m
\end{array}
\right. \, ,
\ee
where $\rho_{\rm iso}$ is the (isothermal) solution to Eqs.~\eqref{eq:Jeans_sph} and~\eqref{eq:Poisson}, and $\rho_{\rm cdm}$ is taken to be an NFW profile with or without AC.\footnote{Whether or not outer halos experience AC can impact the inner halo profile~\cite{Kaplinghat:2013xca,Sagunski:2020spe}. We address this in Sec.~\ref{sec:sims} by analyzing CDM simulations.}
The two functions are matched at $r=r_m$ by imposing that the DM density and enclosed mass are continuous. This leads to the matching conditions 
\be 
\label{eq:matching_rm}
\rho_{\rm iso}(r_m) = \rho_{\cdm}(r_m) \, ,  \qquad
M_{\rm iso}(r_m) = M_{\cdm}(r_m) \, . 
\ee
There are two prescriptions to impose this matching at a given $r_m$, dubbed ``inside-out'' and ``outside-in'' matching~\cite{Sagunski:2020spe}.
For inside-out matching, the core density $\rho_0$ and dispersion $\sigma_0$ are taken as inputs and one solves for $\rho_{\rm iso}(r)$ as an initial value problem from $r=0$ to $r_m$, with initial conditions
\be \label{eq:BC_0}
\rho_{\rm iso}(0) = \rho_0 \, , \quad M_{\rm iso}(0) = 0 \, .
\ee
Then, the parameters of the outer halo are determined to satisfy the matching conditions~\eqref{eq:matching_rm}.
On the other hand, for outside-in matching, the parameters of the outer halo are taken as inputs and one solves for $\rho_{\rm iso}(r)$ as a boundary value problem subject to all four  conditions in Eq.~\eqref{eq:matching_rm} and \eqref{eq:BC_0}.
With two extra boundary conditions, the unknown parameters $\rho_0,\sigma_0$ are determined along with $\rho_{\rm iso}(r)$ itself. In practice, this is accomplished, for instance, using  the relaxation method described in
Appendix~\ref{app:relaxation},    applied to the spherical Jeans model  in Appendix~\ref{app:spherical}. 
In this work, we use the outside-in method due to its more natural implementation beyond spherical symmetry, discussed in Appendix~\ref{app:nonspherical}.

\section{Jeans model beyond spherical symmetry}
\label{sec:JMBSS}
\subsection{Nonspherical CDM halos}
\label{sec:CDM}

To model SIDM halo shapes, we must first consider the structure of collisionless CDM halos, which are known to have sizable triaxiality, e.g.~\cite{Chua:2018sbi,Prada:2019tim}.
Not only are CDM density profiles a key input to the Jeans model, but the nonspherical Jeans model must be able to describe the nonsphericity of CDM halos in the limit $\sigmam=0$. 

In this work, we model axisymmetric departures from spherical symmetry for CDM as follows:
Beginning with a spherically-symmetric CDM profile (e.g., an NFW profile or its AC variant), we map
\be
\rho_{\cdm}(r) \; \to \; \rho_{\cdm}(\rsph)
\ee
with the effective spheroidal radius
\be \label{eq:r_sph_cdm}
\rsph(r,\theta) = r \sqrt{ \qcdm^{2/3} \sin^2 \theta  + \qcdm^{-4/3} \cos^2\theta } \, ,
\ee
which maps an iso-density sphere to an iso-density spheroid of the same volume.
Here, $\qcdm$ is the axial ratio of the halo, which can be oblate ($\qcdm < 1$) or prolate ($\qcdm > 1$). 
Furthermore, $N$-body simulations find CDM halo shapes that vary with radius~\cite{Allgood:2005eu}.
Here we adopt a power-law ansatz for the axial ratio~\cite{Gonzalez:2023oyv}, given by
\be \label{eq:q_cdm}
\qcdm(\rsph) = q_0 \left( \frac{\rsph}{0.1 r_{200}} \right)^\alpha \, ,
\ee
where $q_0, \alpha$ are constant parameters, and $r_{200}$ is the virial radius of the halo.
We note that for $\alpha \ne 0$, there is a nonuniform shape profile, and Eq.~\eqref{eq:r_sph_cdm} must be solved implicitly to determine $\rsph$ as a function of $(r,\theta)$.
In Sec.~\ref{sec:sims}, we demonstrate that our assumptions provide excellent fits to collisionless halo shapes from CDM simulations with baryons.

\subsection{Nonspherical isothermal Jeans model}
\label{sec:nonspherical_jeans}

Next, we attempt to generalize the isothermal Jeans model to nonspherical halos by relaxing the assumption of spherical symmetry but otherwise following the same arguments as above.
The collisional inner halo is treated as an isothermal fluid in hydrostatic equilibrium in a nonspherical potential, according to
\be \label{eq:Jeans_nonsph}
\sigma_0^2 \boldsymbol{\nabla} \rho_{\rm dm} = - \rho_{\rm dm}\,  \boldsymbol{\nabla} \Phi\, ,
\ee
generalizing Eq.~\eqref{eq:Jeans_sph}.
In general, we have two sources of nonsphericity. First, baryon distributions can exhibit strong departures from spherical symmetry, e.g., in the case of galactic disks. 
Second, the collisionless outer halo may be nonspherical as well, as expected for CDM halos as discussed in Sec.~\ref{sec:CDM}.
The distribution of mass in the outer halo does impact the gravitational potential in the inner halo since Newton's shell theorem does not apply to a nonspherical system.

To determine the inner profile, we proceed as follows.
First, we formally write the solution to Eq.~\eqref{eq:Jeans_nonsph} as
\be
\rho_{\rm dm}(\mathbf r) = \rho_0 \, \exp\left(- \frac{\Phi(\mathbf r) - \Phi(\mathbf 0)}{\sigma_0^2}\right) \, 
\label{eq:hydro_equilib}\,,
\ee
where $\rho_0 = \rho_{\rm dm}(0)$.
We separate the total potential into the sum of dark matter plus baryons, $\Phi = \Phi_{\rm dm} + \Phi_b$, each of which satisfies Poisson's equation by the superposition principle.
Next, we define rescaled dimensionless potentials
\be \label{eq:dimless_pot}
\phi_{\rm dm}(\mathbf r) = \frac{ \Phi_{\rm dm}(\mathbf r) - \Phi_{\rm dm}(0)}{\sigma_0^2} \, , \quad
\phi_b(\mathbf r) = \frac{ \Phi_b(\mathbf r) - \Phi_b(0)}{\sigma_0^2} \, .
\ee
Then Poisson's equation for the dark matter potential can be written as
\be
\nabla^2 \phi_{\rm dm}(\mathbf r) = \left( \frac{ 4\pi G \rho_0}{\sigma_0^2 } \right)  e^{-\phi_{\rm dm}(\mathbf r) -\phi_b(\mathbf r)} \,  \, . \label{eq:main}
\ee
To proceed further, we expand the dark matter potential in spherical harmonics
\be \label{eq:phi_dm_ell_m}
\phi_{\rm dm}(r,\theta,\varphi) = \sum_{\ell,m} \phi_{\ell m}(r) \, Z_{\ell m}(\theta,\varphi) \, .
\ee
Since $\phi_{\rm dm}$ is a real function, we use a basis of real-valued spherical harmonics $Z_{\ell m}$ (see Appendix~\ref{app:nonspherical}). 
Finally, we project Eq.~\eqref{eq:main} onto this basis in the usual way, which, after using the orthogonality conditions for $Z_{\ell m}$, gives 
\be
\frac{\partial^2 \phi_{\ell m}}{\partial r^2} + \frac{2}{r} \frac{ \partial \phi_{\ell m}}{\partial r} - \frac{\ell (\ell + 1)}{r^2} \phi_{\ell m} 
=  \left( \frac{4\pi G \rho_0}{\sigma_0^2} \right) \int d\Omega \, Z_{\ell m}(\theta,\varphi) \,  e^{-\phi_b(\mathbf r)} \, e^{-\phi_{\rm dm}(\mathbf r)} \, . \label{eq:main2}
\ee
Unfortunately, this expansion does not provide a full separation of variables since the right-hand side is nonlinear in $\phi_{\rm dm}$. In practice, we truncate the expansion at a given order $\ell_{\rm max}$. 
In the case with axial symmetry, this produces  a system of $\ell_{\rm max} + 1$ coupled second-order equations, since $m \ne 0$ modes vanish, while in the general case there are $(\ell_{\rm max} + 1)^2$ equations.
If the departure from spherical symmetry in the dark matter potential is gentle, it may be that only a few $\ell$ modes are sufficient.

The full dark matter profile spanning both regions is taken to be a piecewise function 
\be \label{eq:matching_2}
\rho_{\rm dm}(\mathbf r) = \left\{ \begin{array}{cc}
\rho_{\rm iso}(\mathbf r) & r < r_m \\
\rho_{\cdm}(\mathbf r) & r > r_m
\end{array}
\right. \, ,
\ee
matched together at a sphere of radius $r_m$.
Here, $\rho_{\rm iso}(\mathbf r)$ is the nonspherical isothermal solution to Eq.~\eqref{eq:Jeans_nonsph}, formally given in Eq.~\eqref{eq:hydro_equilib}, while $\rho_{\cdm}(\mathbf r)$ is the outer profile, taken to be a nonspherical CDM halo as described above.
Generalizing the logic of the spherical Jeans model, we assume that the spherically-averaged density and enclosed mass are matched at $r_m$,
\be 
\label{eq:matching_rm_2}
\int \frac{d\Omega}{4\pi} \, \rho_{\rm iso}(r_m,\theta,\varphi) = \int \frac{d\Omega}{4\pi} \,  \rho_{\cdm}(r_m,\theta,\varphi) \, ,  \qquad
M_{\rm iso}(r_m) = M_{\cdm}(r_m) \, . 
\ee
We also impose the same boundary conditions~\eqref{eq:BC_0} at the origin.
Finally, we require the dark matter potential to be well-behaved (finite) at the origin and at infinity, which in turn fully fixes the boundary conditions for $\phi_{\ell m}$ for $\ell > 0$ without any additional freedom in the model.
We use a relaxation method to solve Eq.~\eqref{eq:main2} in accordance with the discussed boundary conditions.
Further details are given in Appendix~\ref{app:nonspherical}.

\begin{figure}[t]
\centering
\includegraphics[width=0.99\textwidth]{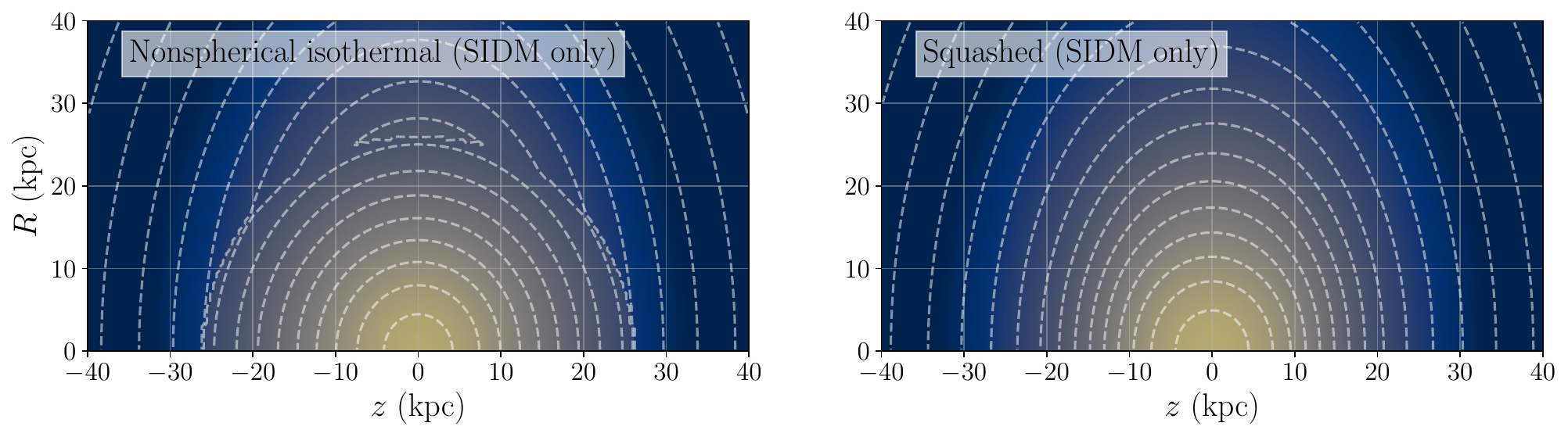}
\caption{\it Dark matter 2D density profiles are shown for the nonspherical isothermal Jeans model with $\ell_{\rm max} = 10$ (left) and squashed Jeans model (right) in cylindrical coordinates $(R,z)$. 
Both panels show an axisymmetric halo with $\Mvir = 2 \times 10^{12} \,\Msun$, $\cvir=10$, $q_0 = 0.6$, $\alpha=0$, $r_m = 26 \; {\rm kpc}$, and no baryons.
Iso-density contours are dashed lines, with light (dark) colors as regions of high (low) density. 
The squashed model yields a gradual rounding of the halo due to self-interactions, while the isothermal model has an abrupt transition at $r_m$.
}
\label{fig:iso_vs_squashed_1}
\end{figure}

Although well motivated, a drawback of this model is that the obtained density profile is not locally continuous across the matching surface at $r_m$, except in the trivial case of spherical symmetry. It may be hoped that with many additional functions $\phi_{\ell m}$, we are free to choose the boundary conditions at $r_m$ such that the density profile is locally continuous for sufficiently high $\ell_{\rm max}$.
We discuss in Appendix~\ref{app:nonspherical} that this is not possible while also preserving a potential that does not blow up at the origin or infinity.
It may be possible to fix this issue by matching on a nonspherical boundary surface, but we do not consider this idea further.

To illustrate the matching discontinuity, Figure~\ref{fig:iso_vs_squashed_1} (left) shows a representative SIDM halo from the nonspherical isothermal Jeans model with no baryons, where the outer halo is an NFW profile \eqref{eq:nfw}, with virial mass $\Mvir = 2 \times 10^{12} \; M_\odot$, concentration $c_{200}=10$, and constant DM density axial ratio $\qcdm =q_0= 0.6$.
The inner halo is matched at $r_m = 26 \; {\rm kpc}$, which corresponds to $\sigma/m = 1 \cmg$.
The contours and shaded coloring show the 2D dark matter density profile as a function of azimuthal radius $R$ and height $z$.
There is an obvious discontinuous transition at the matching radius between the nonspherical outer halo and the nearly round inner halo, as the model abruptly enforces hydrostatic equilibrium within $r_m$.

\subsection{Squashed Jeans model}
\label{sec:squashed}

The nonspherical isothermal Jeans model discussed above has two main flaws. 
First, the model yields density profiles that are discontinuous at $r_m$, which is a consequence of imposing finite asymptotic behavior for the dark matter potential (see  Appendix~\ref{app:nonspherical}).
Second, as we show below, halos tend to be too round compared to those obtained from the simulations.

Both issues can be fixed by modifying the transition between self-interacting and collisionless regions across $r_m$, as a function of the expected number of scatters per particle, $\N$.
This approach is motivated by $N$-body simulations, which show that multiple scatterings are required to erase ellipticity in SIDM halos~\cite{Peter:2012jh}.
The collisionless regime corresponds to $\N \ll 1$, while only for $\N \gg 1$ do we expect to reach hydrostatic equilibrium.
In Appendix~\ref{app:squashed}, we provide an analytic argument to interpolate between the two regimes.

We begin with the spherical isothermal Jeans model discussed in Section \ref{sec:spherical_jeans}.
We denote  $\bar{\rho}_{\rm dm}(r)$ as  the spherical Jeans model profile of Eq.~\eqref{eq:matching}.
Next, assuming axial symmetry, 
we model the nonspherical 2D SIDM halo according to
\be \label{eq:squashed_mapping}
\rho_{\rm dm}(\mathbf r) = \bar{\rho}_{\rm dm}(\rsph) \, ,
\ee
that is, by ``squashing'' iso-density spheres into iso-density spheroids of the same volume, where now the effective spheroidal radius is
\be \label{eq:r_sph}
\rsph(r,\theta) = r \sqrt{ q(\rsph)^{2/3} \sin^2 \theta  + q(\rsph)^{-4/3} \cos^2\theta } \, .
\ee
The shape of the SIDM halo is modeled through the axial ratio $q$, which varies over the halo as a function of $\rsph$. In our setup, $q$ will be a known function, so $\rsph$ can be solved for implicitly from Eq.~\eqref{eq:r_sph} for a given position $(r,\theta)$.

Our ansatz for $q$ is the following\footnote{In this paper, $\log$ refers to the natural logarithm, unless a different base is explicitly indicated.}:
\be \label{eq:q_ansatz_2}
\log q = \log q_{\rm iso} + \log(\qcdm/q_{\rm iso})\, e^{ - k \N} \, ,
\ee
where $q_{\rm iso}$ is the shape for an isothermal halo in hydrostatic equilibrium, $\qcdm$ is the outer halo anisotropy in the collisionless limit, and $\N$ is the number of scatters.
That is, $q=\qcdm$ for negligible scattering ($\N \to 0$), while $q=q_{\rm iso}$ in the collisional limit ($\N \to \infty$). 
In Appendix~\ref{app:squashed}, we derive Eq.~\eqref{eq:q_ansatz_2} from the collisional relaxation rate of an anisotropic plasma in a simplified setup, in which we find $k = \sqrt{2} / 5$.

The radial dependence of $q$ follows from evaluating Eq.~\eqref{eq:q_ansatz_2} at $\rsph$.
The radially-dependent number of scatters is calculated from the spherical Jeans model and rate equation~\eqref{eq:rate}, 
\be\label{eq:Nsolve}
\N(r) = \frac{\N_m \bar\rho(r) v_{\rm rel}(r) }{\bar\rho(r_m) v_{\rm rel}(r_m)} 
\approx 
\frac{\N_m \bar\rho(r) } {\bar\rho(r_m)}\,,
\ee
where the last step follows by approximating the typical scattering velocity to be uniform (corresponding to the assumption that the inner halo is isothermal).
The collisionless shape profile $\qcdm$ is given in Eq.~\eqref{eq:q_cdm}.
The isothermal shape profile $q_{\rm iso}$ depends on the baryon potential and outer halo shapes, as discussed in Sec.~\ref{sec:nonspherical_jeans}.
While in principle this can be obtained from the full numerical solution of the nonspherical isothermal Jeans model, it is also possible to calculate $q_{\rm iso}$ directly semi-analytically (see in particular Eq. \eqref{eq:q_iso_three_terms}).
The latter provides a much faster calculation and is the one used for the squashed Jeans model.
Both methods are discussed and compared in Appendix~\ref{app:nonspherical}.

Returning  to Figure~\ref{fig:iso_vs_squashed_1}, we  contrast the 2D dark matter density profile from the nonspherical isothermal Jeans model (left panel) to that of the squashed Jeans model (right panel).
The latter yields a smooth transition across the matching region from nonspherical outer halo to round inner halo, as is built into the model, unlike the former, which has a discontinuity from nonspherical to (nearly) spherical at $r_m$, as already discussed.

\subsection{Halo shapes and model comparison}\label{sec:halo_shape}

Here, we discuss our method for quantifying and comparing dark matter halo shapes obtained from Jeans modeling and numerical simulations. 
For a given DM density function $\rho_{\rm dm}(\mathbf r)$, we start 
by computing the (non-traceless) quadrupole moment
\be \label{eq:mass_tensor}
M_{ij} = \int_{\rm shell} d^3 r \, r_i r_j \, \rho_{\rm dm}(\mathbf r)\,,
\ee
where $i,j$ label the three Euclidean axes ($x,y,z$).
The integral is performed over a thin spheroidal shell that approximates an iso-density contour for the halo.
The equation for a spheroid is
\be
\frac{R^2}{A^2} + \frac{z^2}{C^2} = 1 \, .
\ee
Next, we define the axis ratio $q=C/A$, where $q < 1$ ($q > 1$) for an oblate (prolate) spheroid, and the effective spheroidal radius
\be\label{eq:sphrhalo}
\rsph = \sqrt{ R^2 q^{2/3} + z^2 q^{-4/3} }  \, ,
\ee
such that the volume of the spheroid is $\frac{4\pi}{3} A^2 C = \frac{4\pi}{3} r_{\rm eff}^3$.
It is also customary to define the semi-major and -minor axes, which here are $a = {\rm max}(A,C)$ and $c = {\rm min}(A,C)$, respectively. 
The ratio of semi-minor to semi-major axes is
\be
c/a = {\rm min}\left(q, q^{-1}\right) \, .
\ee

In the case the system has azimuthal symmetry, $M_{ij}$ can be diagonalized with only two independent nonzero components, $M_{xx} = M_{yy} \equiv M_{RR}$ and $M_{zz}$.
They are computed as
\begin{align} \label{eq:IRR}
M_{RR} &= \pi \int_{\rm shell} dr \,  d\theta \, r^4 \, \sin^3\theta \, \rho_{\rm dm}(r,\theta) \,
= \frac{\pi \Delta r_{\rm eff}}{r_{\rm eff}} \int_0^\pi d\theta \, r(\theta)^5 \, \sin^3 \theta \, \rho_{\rm dm}\big(r(\theta),\theta\big)\,,
\\
M_{zz} &= 2\pi \int_{\rm shell} dr \, d\theta \, r^4 \, \sin\theta \cos^2\theta \, \rho_{\rm dm}(r,\theta) 
= \frac{2\pi \Delta r_{\rm eff}}{r_{\rm eff}}\int_0^\pi d\theta \, r(\theta)^5 \, \sin \theta \, \cos^2 \theta \,  \rho_{\rm dm}\big(r(\theta),\theta\big) \, ,
\label{eq:Izz}
\end{align}
for fixed $r_{\rm eff}$, where $\Delta r_{\rm eff}$ is the thickness of the shell, and the dependence on $q$ is implicit in
\be \label{eq:r_theta}
r(\theta) = \frac{r_{\rm eff}}{\sqrt{\sin^2\theta\,  q^{2/3} + \cos^2\theta \, q^{-4/3} }} \, .
\ee
Finally, the halo shape, at a given $r_{\rm eff}$, is computed from the ratio
\be \label{eq:q_shell}
q = \sqrt{\frac{M_{zz}}{M_{RR}}} \, .
\ee
This equation must be solved implicitly since $q$ enters Eqs.~\eqref{eq:IRR} and \eqref{eq:Izz}, i.e., through the nonsphericity of the shell over which $M_{ij}$ is computed.\footnote{For $N$-body simulations, the integral in Eq.~\eqref{eq:mass_tensor} is replaced by a sum over particles, discussed in Sec.~\ref{sec:sims} below.}

The solution to Eq.~\eqref{eq:q_shell} for $q$ provides exactly the radially-dependent shape of the halo in the case where iso-density contours are spheroidal.
We prove this analytically by taking a spheroidal shell coincident with an iso-density contour, with a given $\rsph$. 
The integrals in Eqs.~\eqref{eq:IRR} and \eqref{eq:Izz} can be performed analytically to yield
\bea
M_{RR} &=& \frac{4\pi}{3} \rho_{\rm dm}(\rsph) \rsph^4 \Delta \rsph q^{-2/3}\,, \\
M_{zz} &=& \frac{4\pi}{3} \rho_{\rm dm}(\rsph) \rsph^4 \Delta \rsph q^{4/3} \, ,
\eea
since $\rho_{\rm dm}$ depends only on $\rsph$ and can be brought outside the integral.
Taking the ratio yields $M_{zz}/M_{RR} = q^2$, satisfying Eq.~\eqref{eq:q_shell}.
Thus, we conclude that the shape profile $q(\rsph)$ introduced for the squashed Jeans model is the same $q$ computed here. 
In general, for the nonspherical isothermal model, the density profile is not spheroidal and the halo shape profile $q(\rsph)$ has to be computed numerically using Eqs.~\eqref{eq:IRR}, \eqref{eq:Izz}, and \eqref{eq:q_shell}.

\begin{figure}[t]
\centering
\includegraphics[width=0.48\textwidth]{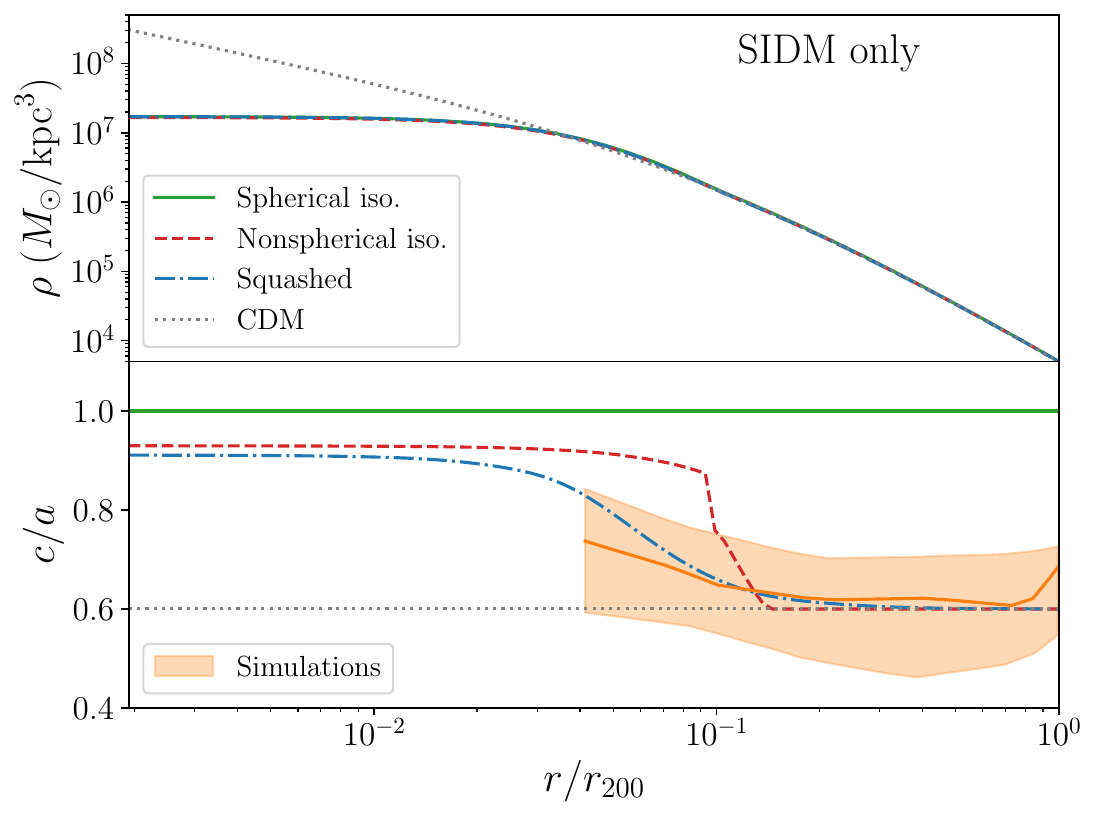}
\includegraphics[width=0.48\textwidth]{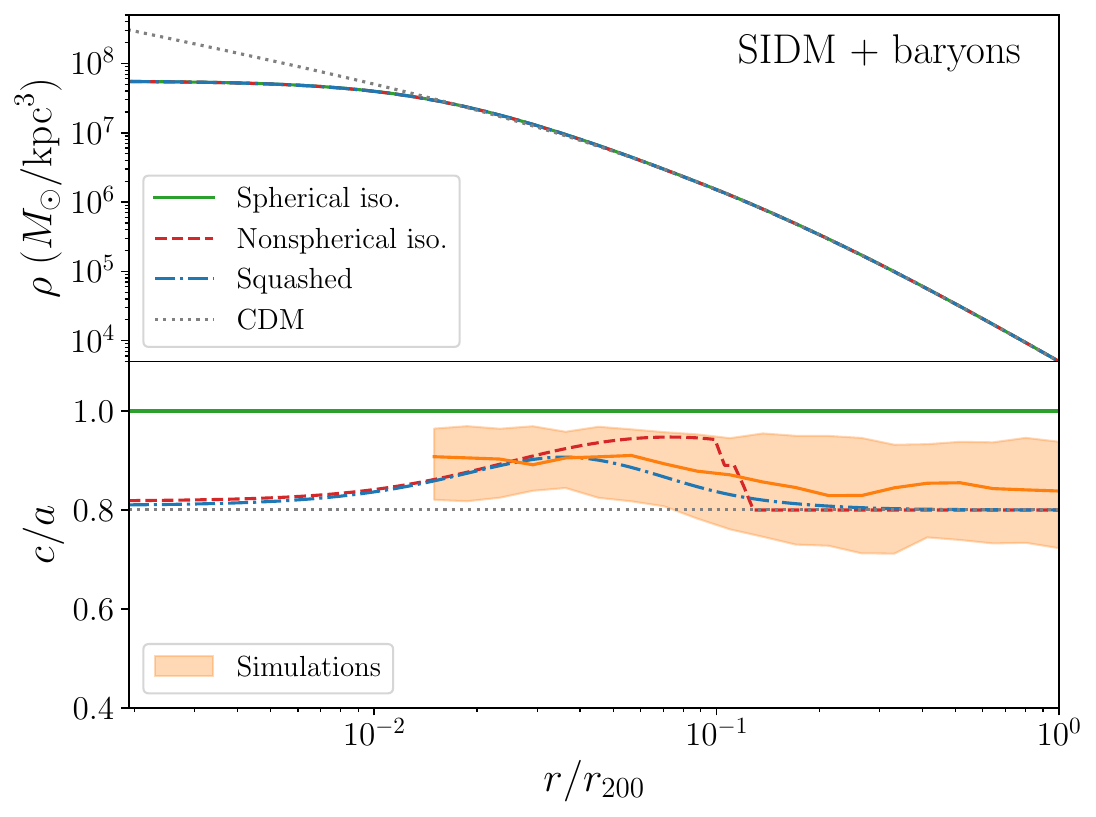}
\caption{\it Spherically-averaged density (upper panels) and halo shape (lower panels) profiles for spherical isothermal (solid lines), nonspherical isothermal with $\ell_{\rm max} = 10$ (dashed), and squashed (dot-dashed) Jeans models for SIDM with $\sigma/m \approx 1 \cmg$, and collisionless CDM (dotted) for a representative $2 \times 10^{12} \Msun$ halo and other model parameters given in the text.
Left side represents the SIDM-only halo with no baryons; right side represents the SIDM halo with a baryonic disk.
Orange bands show shape profiles ($20-80\%$) from $N$-body simulations for SIDM halos with $1 \cmg$ both without baryons  (left) \cite{Peter:2012jh} and with baryons (right) \cite{Robertson:2020pxj}.
}
\label{fig:iso_vs_squashed_2}
\end{figure}

We can now compare the predictions of the different Jeans models discussed so far,
as shown in  Fig.~\ref{fig:iso_vs_squashed_2}.
In the left panel, we have an SIDM-only halo (no baryons) with $\sigma/m = 1 \cmg$ ($r_m = 26 \; \kpc$) and outer halo parameters $c_{200}=10$, $\Mvir = 2 \times 10^{12} \; \Msun$, $\qcdm = 0.6$.
In the right panel, we consider an SIDM halo with baryons, with $\sigma/m = 1 \cmg$ ($r_m = 28 \; \kpc$), and outer halo parameters $c_{200}=10$, $\Mvir = 2 \times 10^{12} \; \Msun$, $\qcdm = 0.8$. 
We take a mildly oblate baryon potential described by a Miyamoto-Nagai disk~\cite{Miyamoto:1975zz} with disk mass $M_{d} = 5 \times 10^{10} \; \Msun$ and scale parameters $(a_{d}, b_{d}) = (2,3) \; {\rm kpc}$.

Fig.~\ref{fig:iso_vs_squashed_2} (upper panels) shows that spherically-averaged density profiles for the nonspherical isothermal (dashed) and squashed (dot-dashed) Jeans models are nearly identical to those of the spherical isothermal Jeans model (solid).
CDM profiles for the outer halo (squashed NFW profiles) are also shown (dotted).
On the left, we have the usual cored profiles of SIDM-dominated halos with modest cross sections~\cite{Spergel:1999mh,Dave:2000ar}.
On the right, we have cuspy SIDM profiles, similar to the collisionless halo, owing to the gravitational potential of baryons~\cite{Kaplinghat:2013xca}.
The conclusion from the upper panels is that nonsphericity in Jeans modeling does not alter the success of the spherical Jeans model in describing the spherically-averaged properties of dark matter halos in $N$-body simulations and observations.

The halo shape profiles, $c/a$, predicted by the two models are shown in Fig.~\ref{fig:iso_vs_squashed_2} (lower panels). The nonspherical isothermal model exhibits an abrupt change from an oblate outer halo to a nearly spherical inner region at the matching radius $r_m$, whereas the squashed model transitions smoothly between the two limits. Physically, a sharp discontinuity in shape is not expected: although a single scattering can isotropize a particle's velocity, the spatial density -- and hence the halo shape -- responds only after many scatterings as particles orbit and redistribute, leading to a gradual rounding of the halo toward smaller radii. This qualitative behaviour is borne out in $N$-body simulations, where the axis ratio $c/a$ typically varies smoothly with radius for individual halos. For comparison, the orange bands in Fig.~\ref{fig:iso_vs_squashed_2} show the median and 20--80\% range of $c/a$ from ensembles of 65 SIDM-only halos~\cite{Peter:2012jh} (left) and 116 SIDM-plus-baryon halos~\cite{Robertson:2020pxj} (right) with $\Mvir \in [10^{12}, 10^{13}] \; \Msun$. Their overall smooth profiles are consistent with the squashed-model prediction and inconsistent with the discontinuous transition of the nonspherical isothermal model. We therefore adopt the squashed Jeans model as our physically motivated description of SIDM halo shapes.

%However, the halo shape profiles, $c/a$, are very different between models, as shown in Fig.~\ref{fig:iso_vs_squashed_2} (lower panels). The nonspherical isothermal model shows an abrupt transition between the oblate outer halo and the rounder inner halo, while the squashed model has a smooth transition. To select between models, we have further compared these Jeans shapes  to halo shape profiles inferred from $N$-body simulations. On the left, we show $c/a$ from SIDM-only simulations with $\sigmam = 1 \cmg$, comprising 65 halos with $\Mvir \in [10^{12}, 10^{13}] \; \Msun$~\cite{Peter:2012jh}, represented by the orange contributions. On the right, we show $c/a$ from SIDM-plus-baryons simulations with $\sigmam = 1 \cmg$ (discussed in Sec.~\ref{sec:sims}), comprising 116 relaxed halos with $\Mvir \in [10^{12}, 10^{13}] \; \Msun$~\cite{Robertson:2020pxj}. The orange bands (lines) show the $20-80\%$ quantiles (medians) for $c/a$ profiles, plotted over the range of resolved radii. Both simulations show that $c/a$ is a smooth function of radius, consistent with the predictions from the  squashed Jeans model. Since simulated halos do not exhibit a sharp transition in $c/a$, we conclude that the nonspherical isothermal Jeans model is unsatisfactory for describing SIDM halo shapes.

\section{N-body simulations}
\label{sec:sims}

To rigorously test the nonspherical Jeans model, we consider halos from the \textsc{Eagle-50} simulations from Ref.~\cite{Robertson:2020pxj} for two models, SIDM with $\sigmam = 1 \cmg$ (\SIDM) and collisionless dark matter with $\sigmam = 0$ (\CDM).
These are high-resolution cosmological simulations for dark matter and baryons, with particle masses $9.7 \times 10^6$ and $1.8 \times 10^6 \, \Msun$, respectively, and spanning a periodic volume of $(50 \; {\rm Mpc})^3$.
Baryonic physics (gas cooling, star formation, stellar feedback, black hole feedback, etc.)~is included following the \textsc{Eagle} galaxy formation model~\cite{2015MNRAS.446..521S, 2015MNRAS.450.1937C}. 
For self-interactions, we consider hard-sphere scattering, i.e., with a differential scattering cross section that is constant in velocity and isotropic in angle.
In the simulations, scattering is treated using a Monte Carlo approach, where at each time step nearby pairs of particles have a probability for scattering given by the rate equation~\cite{2017MNRAS.465..569R}.
Ref.~\cite{2018MNRAS.476L..20R} describes further details of the implementation of SIDM within the \textsc{Eagle} code.

Within the simulation volume, we consider the 250 most massive halos at redshift $z=0$,\footnote{The simulation data described throughout this section is publicly available and can be found at \href{https://doi.org/10.5281/zenodo.16331984}{doi.org/10.5281/zenodo.16331984}} defined using a friends-of-friends algorithm~\cite{1985ApJ...292..371D}, which span a range of virial masses $\Mvir$ between $3.5 \times 10^{11} - 1.6 \times 10^{14} \; \Msun$.
We further require a clear one-to-one correspondence between \CDM\ and \SIDM\ halos, such that the centers of halos (position of the most bound particle) are within 10 kpc and the virial masses are the same within 0.1 dex, which leaves 227 halos.
Lastly, we omit the most massive halo in our sample since it lies above the galactic scales considered in this work, leaving us with a sample of 226 halos spanning $4 \times 10^{11} - 4 \times 10^{13} \Msun$.

\subsubsubsection{2D mass distributions:} We focus on the 2D distribution of dark matter under the assumption of axial symmetry, which is to say that the density depends on cylindrical coordinates $(R,z)$, but is symmetric with respect to the azimuthal angle.
Of course, simulated halos need not follow this symmetry, but we impose this assumption here to enable a comparison to the axisymmetric squashed Jeans model.

We calculate the azimuthally-averaged density distributions from the simulations as follows.
First, we define the halo origin as the position of the most gravitationally bound particle and the $z$-axis as the stellar angular momentum vector, about which we model the distribution as axially symmetric.
Next, we take a regular grid of $R$ and $z$ values over the intervals $[0,100] \; {\rm kpc}$ and $[-100,100] \; {\rm kpc}$, respectively, with grid spacing $\Delta R = \Delta z = 1 \; {\rm kpc}$.
For each particle species (dark matter, stars, gas and black holes), we calculate the azimuthally-averaged density $\rho(R,z)$ as the mass within a given pixel divided by the 3D pixel volume (which is an annular cylinder of mean radius $R$, radial thickness $\Delta R$, and height $\Delta z$). 

{\bf 2D halo shapes:} We calculate azimuthally-averaged halo shapes from our $N$-body simulation data following similar approaches in the literature~\cite{Dubinski:1991bm,Dave:2000ar,Allgood:2005eu,Peter:2012jh}. 
The integral in
Eq.~\eqref{eq:mass_tensor} is estimated by a sum over particles $a$, with mass $m_a$, within a given spheroidal shell
\be \label{eq:mass_tensor_sum}
M_{ij} = \sum_{a \, \in \, \mathrm{shell}} r_{i,a} \, r_{j,a} \, m_a .
\ee
Recalling that we defined $M_{xx} = M_{yy} \equiv M_{RR}$ for the case of azimuthal symmetry, we can define an azimuthally-averaged estimate of $M_{RR}$ as $M_{RR} = \frac{1}{2} (M_{xx} + M_{yy})$. The  axis ratio of an iso-density surface of the halo is estimated as $q = \sqrt{M_{zz} / M_{RR}}$.
However, as described in Sec.~\ref{sec:halo_shape}, the shell within which we include particles in the sum in Eq.~\eqref{eq:mass_tensor_sum} must itself be an approximate iso-density surface, which means it depends on $q$. 
We therefore use an iterative procedure, described as follows. 

First, we consider a shell spanning radii $[r_{\rm min}, r_{\rm max}]$ and with an initial guess for $q$: $q^{(0)}=1$.
That is, the shell for evaluating Eq.~\eqref{eq:mass_tensor_sum} is initially spherical and the sum is evaluated including particles satisfying
$r_{\rm min} < r_a < r_{\rm max}$.
With $M_{zz}$ and $M_{RR}$ calculated from these particles, we obtain a new shape estimate $q^{(1)} = \sqrt{M_{zz} / M_{RR}}$.
The process is further repeated.
For subsequent stages of the iteration, we use the previous shape $q^{(i)}$ to calculate $M_{zz}$ and $M_{RR}$ by including all particles satisfying
\be
r_{\rm min} < \sqrt{R_a^2 \, (q^{(i)})^{2/3} + z_a^2  (q^{(i)})^{-4/3}} < r_{\rm max} \, ,
\ee
which is then used to obtain the next guess $q^{(i+1)}$.
We stop this procedure when subsequent iterations converge within some tolerance, in particular, when $q^{(i+1)}$ and $q^{(i)}$ agree to within 1\%. 
We note that this procedure is similar to that used for fitting 3D ellipsoids to measure halo shapes in $N$-body simulations (such as in Ref. \cite{Allgood:2005eu}), but specialized to the case of azimuthal symmetry about a fixed axis.

Next, the algorithm is applied across a series of radial bins to obtain the shape profile for a given halo. 
For our simulations, we define 35 logarithmically-spaced shells in spheroidal radius $\rsph$, covering a range of scales from 1 kpc to 3.2 Mpc (though no halos extend out this far). 
To make full use of the particles, without introducing strong covariance between the shape measurements in neighboring shells, we use a shell width that is approximately equal to the spacing between shells. 
Our shell spacing leads to neighboring shells with radii that differ by a factor of $f \approx 1.27$.
Therefore, for each bin, we take $r_{\rm min} = \rsph/\sqrt{f} \approx 0.89 \, \rsph$ and $r_{\rm max} = \rsph \sqrt{f} \approx 1.13 \, \rsph$.

In addition to evaluating $q(r_\mathrm{eff})$ from the full particle distribution, we estimate the statistical uncertainty $\sigma_q(r_\mathrm{eff})$ from the standard deviation of $q(r_\mathrm{eff})$ values calculated from bootstrap samples of the particles. For $N$ total particles in the halo, each bootstrap sample of particles is created by selecting $N$ particles (with replacement). 

{\bf Selection of relaxed halos:} To ensure that our model accurately represents the average properties of a halo ensemble, we restrict our analysis to systems that are \textit{relaxed}. We follow the criteria outlined in Ref.~\cite{Neto_2007} for determining relaxed versus unrelaxed systems. The \textit{Neto criteria} evaluate three metrics: (i) $f_\mathrm{sub}$, the mass fraction of substructure whose centers lie within $r_{200}$; (ii) $s$, the normalized offset between the center of mass and the potential center of the halo; and (iii) $2T/|U|$, the virial ratio where $T$ is the kinetic energy and $U$ the gravitational self-potential energy of particles within $r_{200}$. A system is considered relaxed if it satisfies all three of the conditions: $f_\mathrm{sub}<0.1$, $s<0.07$, and $2T/|U|<1.35$. Applying the Neto criteria reduces our simulation suite further to 180 \SIDM\ halos and 176 \CDM\ halos.\footnote{We note that relaxedness is a continuum and there is no bimodal separation between relaxed and unrelaxed systems. 
Moreover, the relaxedness criteria adopted from Ref.~\cite{Neto_2007} relate to the entire halo. 
As such, we find that some halos entering our relaxed sample do exhibit unrelaxed features, especially close to their centers (where our analysis is focused), which in turn impacts our cross-section inferences.}

\begin{figure}[p]
\centering
\includegraphics[width=0.99\textwidth]{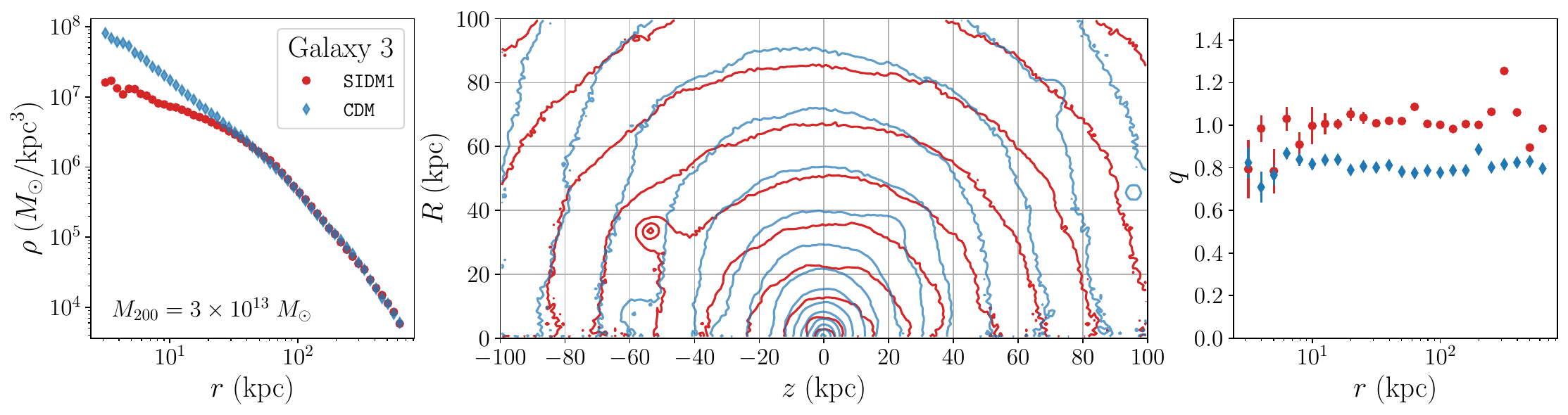}
\includegraphics[width=0.99\textwidth]{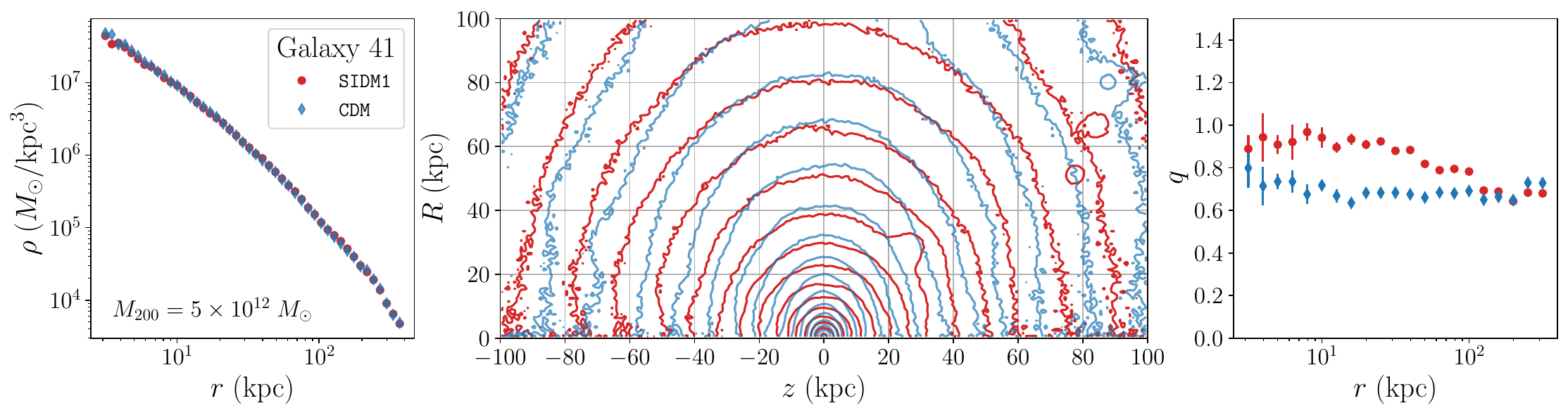}
\includegraphics[width=0.99\textwidth]{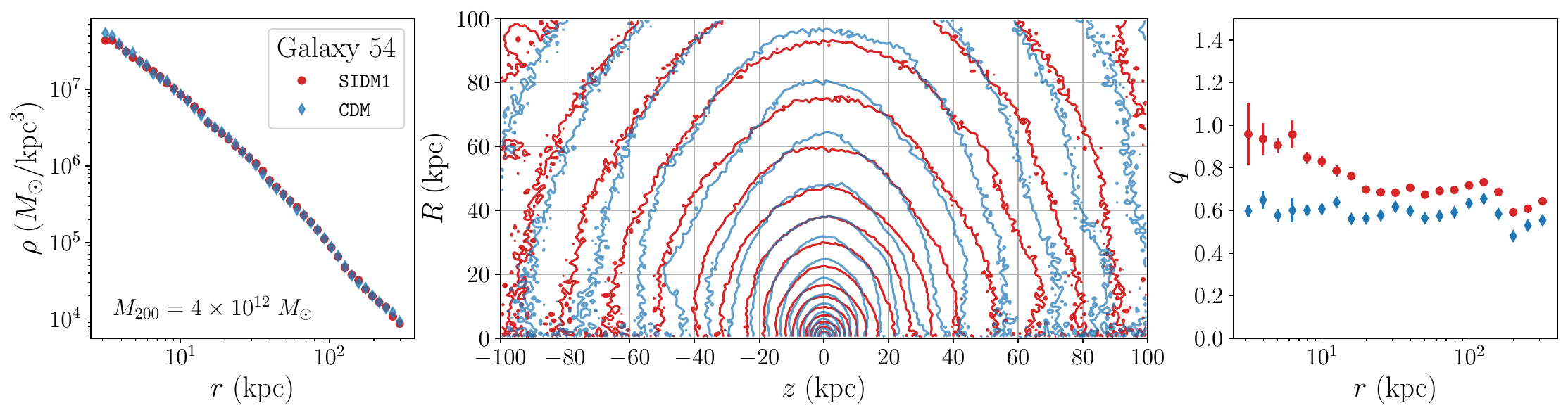}
\includegraphics[width=0.99\textwidth]{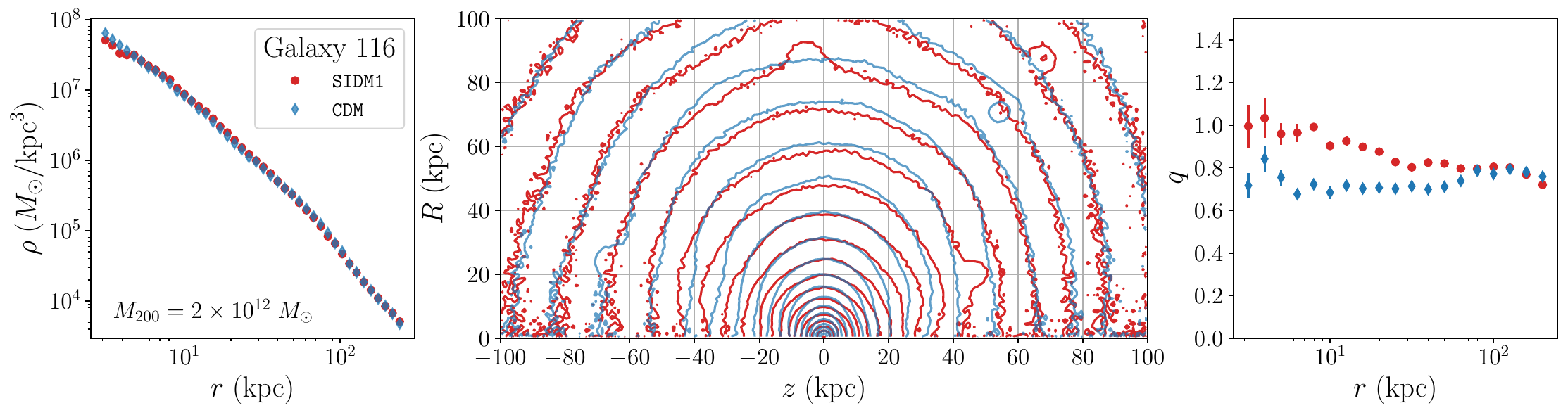}
\includegraphics[width=0.99\textwidth]{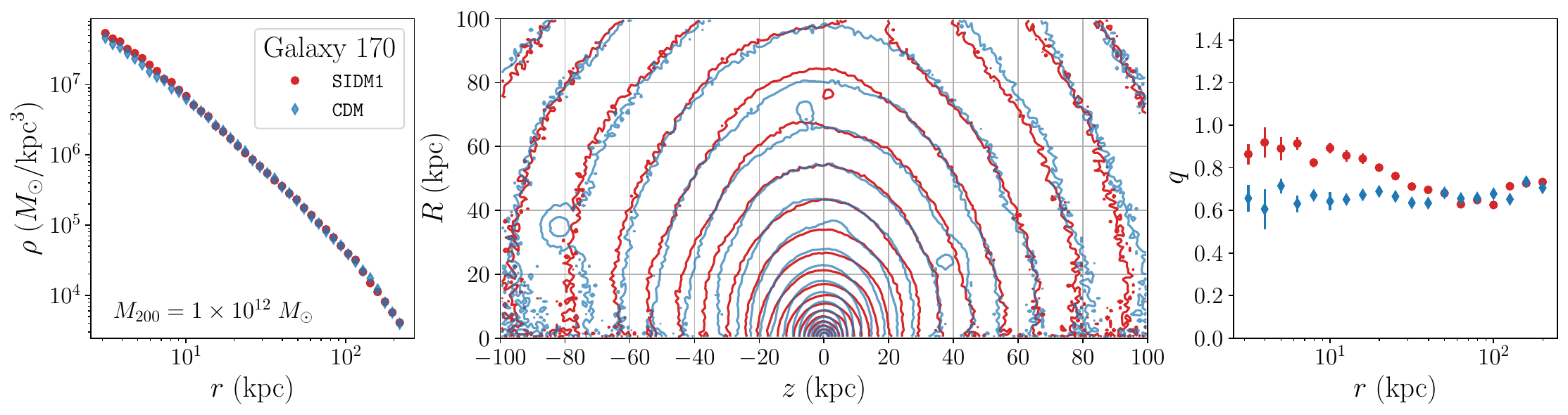}
\caption{\it {\bf Simulated SIDM vs CDM halos}: Spherically-averaged density profiles (left), azimuthally-averaged iso-density contours (center), and halo shape profiles (right) for five systems from \textsc{Eagle-50} simulations for \SIDM\ (red) and \CDM\ (blue). For virial mass in the range $\Mvir \sim 10^{12} - 10^{13} \; \Msun$, SIDM systems can have similar spherically-averaged density profiles, but different halo shapes.
}
\label{fig:cdm_vs_sidm_1}
\end{figure}

\subsection{CDM vs SIDM simulated halos}

First, we discuss how \CDM\ and \SIDM\  simulated halos compare to each other.
Fig.~\ref{fig:cdm_vs_sidm_1} shows the distribution of dark matter for five representative halos, across a range of $\Mvir$ values.
The left panels compare the spherically-averaged density profiles for \SIDM\ (red dots) and \CDM\ (blue diamonds).
For $\Mvir \gtrsim 10^{13} \Msun$, self-interactions typically reduce the central density and flatten the inner slope, which is evident in the uppermost panel (Galaxy 3).
In the remaining panels, for $10^{12} \Msun \lesssim \Mvir \lesssim 10^{13} \Msun$, the spherically-averaged profiles for SIDM and CDM are virtually identical.
This is due to the fluid-like behavior of SIDM responding to the baryon potential to form a cusp~\cite{Kaplinghat:2013xca}.
For the purposes of constraining self-interactions on these intermediate halo mass scales, it is clear that questions of cores versus cusps and the slopes of inner profiles are not sufficient.

Therefore, turning to halo shapes, we ask to what extent the angular distribution of dark matter can be different between SIDM and CDM, even for halos where the spherically-averaged densities look identical.
In Figure~\ref{fig:cdm_vs_sidm_1}, the center panels show the azimuthally-averaged 2D density profiles $\rho_{\rm dm}(R,z)$ for both \SIDM\ (red) and \CDM\ (blue).
By eye, we can see that \SIDM\ and \CDM\ halos have different shapes, with the former appearing somewhat rounder, even in halos below $10^{13} \; \Msun$ where the spherically-averaged profiles are nearly identical.

The difference between \SIDM\ and \CDM\ halos is even more apparent in their shape profiles, shown in the right panels.
Here, we show the radially-dependent axis ratio $q$, along with the shape uncertainties, computed as discussed above.\footnote{To be precise, the abscissa in the right panels is the effective spheroidal radius $\rsph$ defined in Eq.~\eqref{eq:sphrhalo}, not the circular radius $r$.} 
Again, we see that \SIDM\ halos appear rounder than their \CDM\ counterparts, despite similar density profiles below $10^{13} \; \Msun$.

\subsection{Modeling simulated CDM halos}

The outskirts of SIDM halos, where collisions are negligible, should match expectations from cosmological structure formation for CDM.
Since the outer halo profile is a key input for the Jeans model, we test our assumptions for collisionless halos by comparing to CDM simulations.

For modeling CDM halos, we consider two profiles: {\it (i)} an NFW profile, motivated from CDM-only simulations~\cite{Navarro:1995iw,Navarro:1996gj}, and  {\it (ii)} in the presence of baryons,  an NFW profile modified by adiabatic contraction (AC) during galaxy formation, following Cautun et al.~\cite{Cautun:2019eaf}.
In both cases, we allow for nonsphericity with a power-law radial dependence as per Eq.~\eqref{eq:q_cdm}.
Each profile is determined by its virial mass $\Mvir$, concentration $c$, and shape parameters $q_0, \alpha$, as described in Sec.~\ref{sec:CDM}.

We perform Markov Chain Monte Carlo (MCMC) fits to the relaxed 176 galaxy-scale halos from the \CDM\ simulations.
We assume logarithmically-flat priors on $\Mvir, c, q_0$ and a linearly-flat prior on $\alpha$.
To assess the goodness-of-fit, we calculate the total $\chi^2$ between the model and simulation data for both the spherically-averaged density and shape ($q$) profiles given by
\be \label{eq:chi_sq_full}
\chi^2 = \sum_{i \in r \, {\rm bins}} \left( \frac{\rho_i^{\rm sim} - \rho_i^{\rm model}}{\Delta \rho_i^{\rm sim}}\right)^2 + 
\sum_{i \in r \, {\rm bins}} {\rm min} \left[\left( \frac{q_i^{\rm sim} - q_i^{\rm model}}{\Delta q_i^{\rm sim}}\right)^2, \: 9 \right] \, ,
\ee
evaluated for radial bins $i$ spanning $[3 \kpc,\,  r_{200}]$.
(Density and shape profiles have different assumed radial bins.)
For the density profile, we take a $0.1$ dex systematic error, while the statistical error is negligible. 
For the shape profile, we take the statistical error $\sigma_q$ (defined above), due to the finite number of particles in each spheroidal shell, added in quadrature with a systematic error $0.05$ to reflect local small variation from small substructures not captured in the model.
To avoid larger substructures biasing our shape fits, we also truncate outlier contributions to the $\chi^2$ at 3-sigma.

\begin{figure}[t]
\centering
\includegraphics[width=0.99\textwidth]{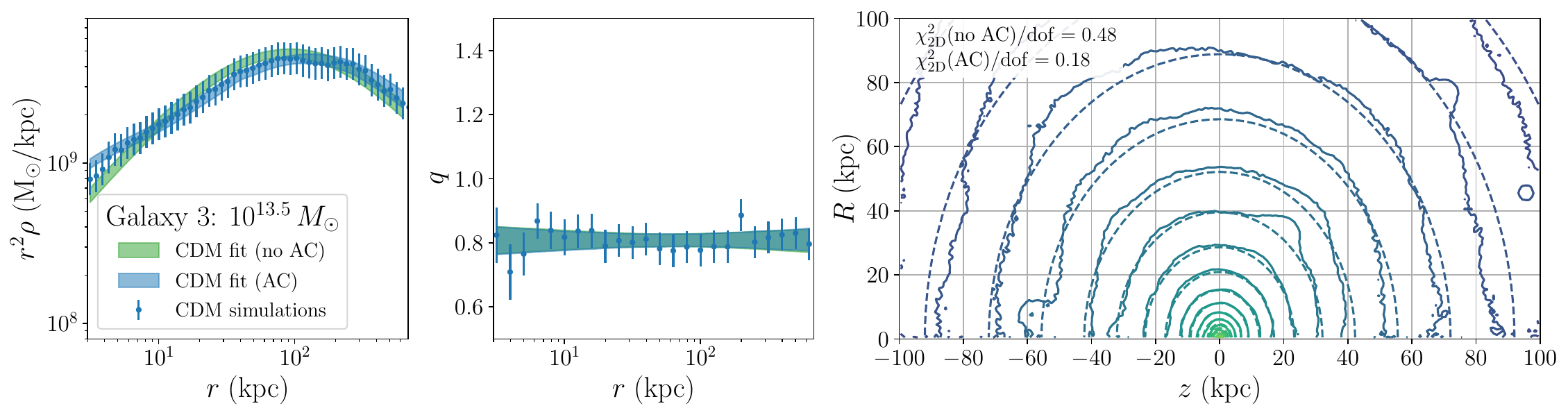}
\includegraphics[width=0.99\textwidth]{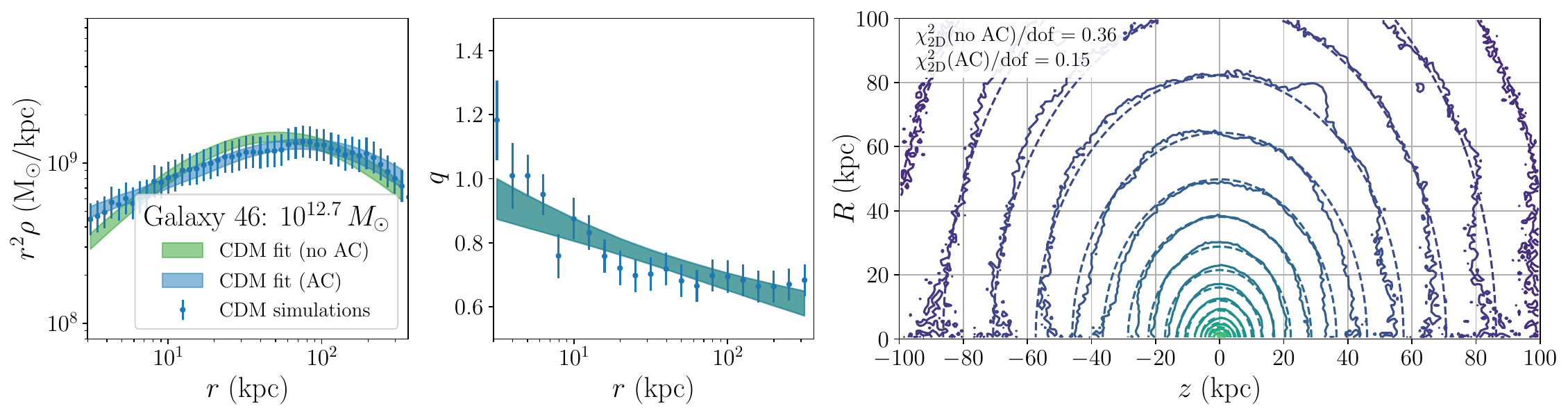}
\caption{\it {\bf CDM halos:} Spherically-averaged density profiles (left), halo shape profiles (center), and azimuthally-averaged iso-density contours (right) for two example systems from the \textsc{Eagle-50} simulation of CDM plus baryons. 
The points show the measured density and shape profiles for the simulated halos, with the error bars denoting the adopted uncertainties, $\Delta \rho$ and $\Delta q$ in equation~\eqref{eq:chi_sq_full}. 
The shaded bands show the 10th - 90th percentiles of the model posterior distributions for $\rho(r)$ and $q(r)$ when fitting to this data, adopting a fit with adiabatic contraction (AC) in blue, and without AC in green. The addition of AC improves the match to $\rho(r)$ at small radii.
Reduced $\chi^2_{\rm 2D}$ values are shown in (right).
}
\label{fig:cdm_fits_1}
\end{figure}

\begin{figure}[t]
\centering
\includegraphics[width=0.99\textwidth]{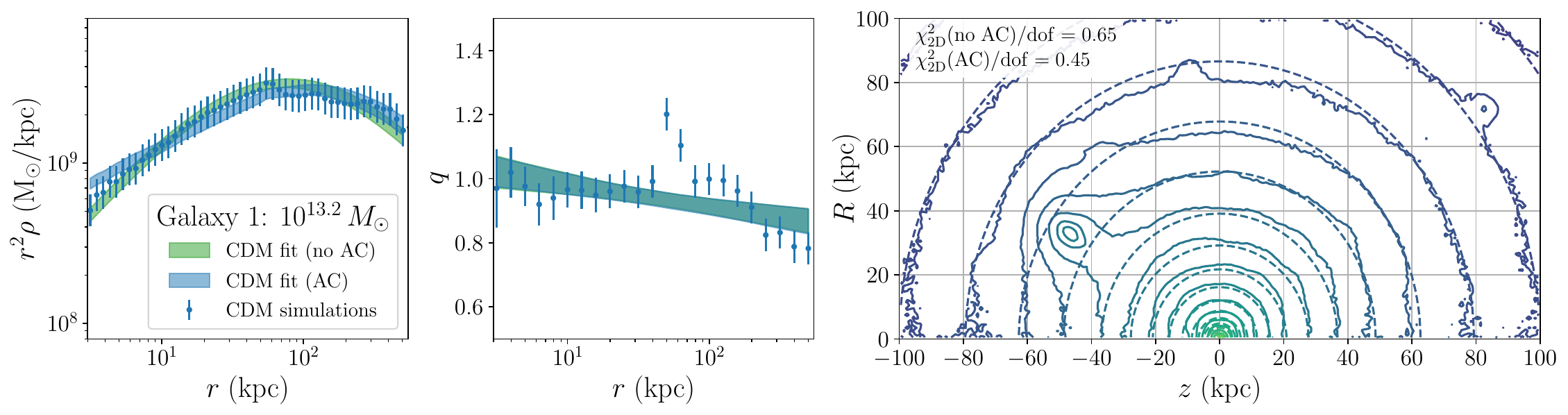}
\includegraphics[width=0.99\textwidth]{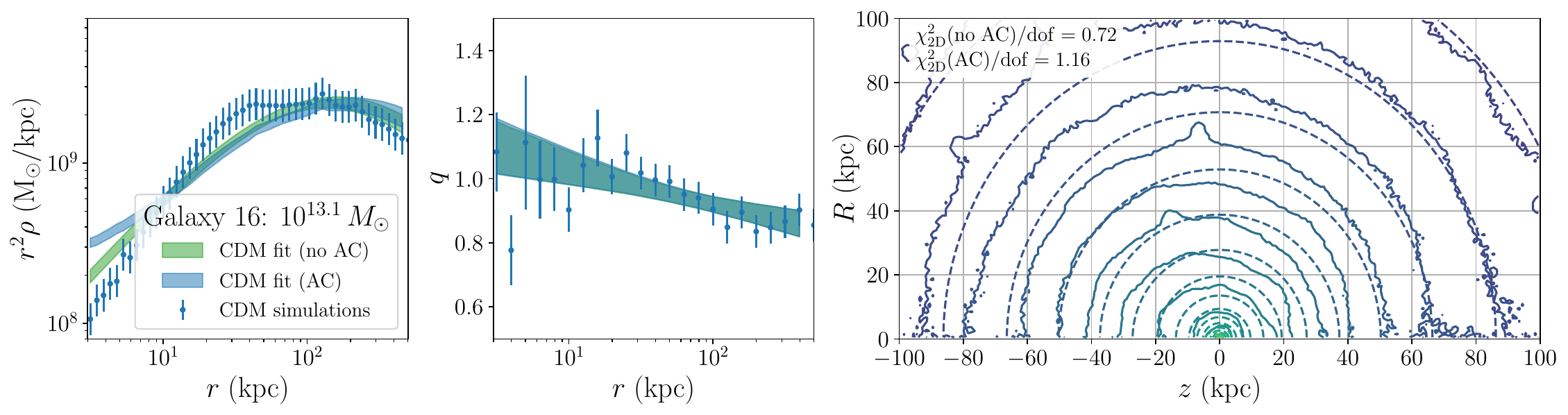}
\includegraphics[width=0.99\textwidth]{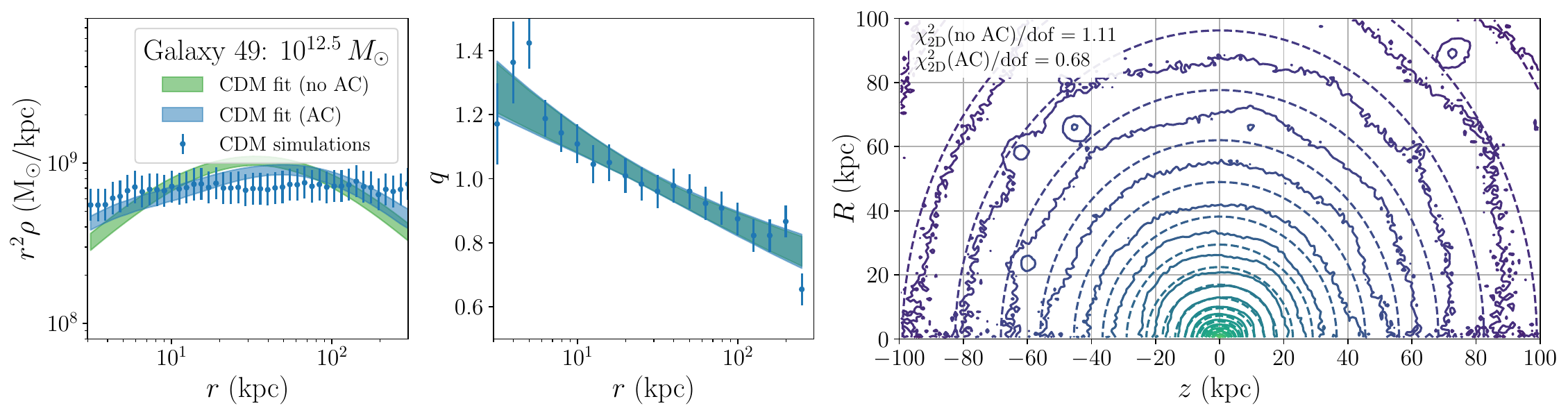}
\includegraphics[width=0.99\textwidth]{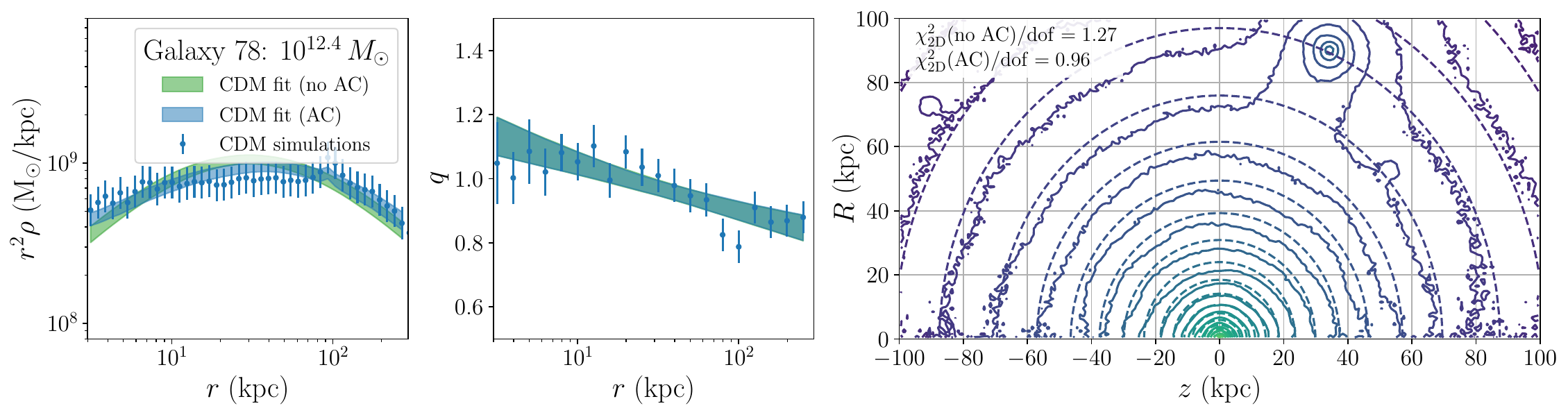}
\caption{\it {\bf CDM halos:} 
Example \CDM\ halos with features not described in our CDM model, i.e., substructure, asymmetries, and steep inner slopes.
Panels as in Fig.~\ref{fig:cdm_fits_1}.
}
\label{fig:cdm_fits_2}
\end{figure}

Figure~\ref{fig:cdm_fits_1} shows representative examples of our fits.
The spherically-averaged density profile (left, shown $r^2$-weighted) and shape profile (center) are the fitted data (blue data points) in our $\chi^2$ function, with error bars representing the total uncertainties assumed in our fit.
The shaded bands represent $80\%$ posterior intervals for our fitted CDM model with (blue) and without (green) AC.
The right panels show the 2D iso-density contours from \CDM\ simulations (solid) and our best-fit model with AC (dashed).
For both systems, our fitting is able to provide a good description of the main halo with AC. 

Although our model is fit to 1D density and shape profiles, we also quantify how well our fit is able to describe the 2D azimuthally-averaged density.
We quantify the 2D goodness-of-fit as
\be \label{eq:chi_sq_2D}
\chi^2_{\rm 2D} = \sum_{i \in R \, {\rm bins}} \sum_{j \in z \, {\rm bins}} \left( \frac{\rho_{ij}^{\rm sim} - \rho_{ij}^{\rm model}}{\Delta \rho_{ij}^{\rm sim}}\right)^2
\ee
where $\rho_{ij}^{\rm sim}$ is the density in a given cylindrical bin $i,j$ in $(R,z)$, spanning $[0,100]$ and $[-100,100]$ ${\rm kpc}$, respectively, $\Delta \rho_{ij}^{\rm sim}$ is taken to be a $0.1$ dex uncertainty for each bin, 
and $\rho_{ij}^{\rm model}$ is the prediction from our best-fit model minimizing Eq.~\eqref{eq:chi_sq_full}.
In Figure~\ref{fig:cdm_fits_1} (right panels), we provide the values of $\chi_{\rm 2D}^2$ per degree of freedom with and without AC.
Thus, we see that the profiles with AC yield better descriptions of the 2D profiles of the relaxed \CDM\ halos, not just their 1D spherically-averaged profiles.

On the other hand, Figure~\ref{fig:cdm_fits_2} shows some of the counter-examples to illustrate the short-comings of our model.
Galaxy 1 is an example with substructure. 
Our fit procedure provides a good description of the overall smooth main halo, although the subhalo around $60 \; \kpc$ is not described in our model.
Galaxy 16 exhibits an obvious asymmetry about the assumed galactic plane ($z=0$), indicative of nonrelaxedness or other systematic mismodeling effects.
Galaxy 49 exhibits a $\sim 1/r^2$ inner density profile that is steeper than can be explained in our AC model.
Galaxy 78 has both a steep inner density slope and an obvious substructure.
These examples illustrate that there are aspects of the simulated halos lying outside the scope of our analysis, for which Jeans modeling may yield incorrect cross-section inferences.
Whether the situation could be improved by improvements in the model or halo selection is deferred to future work.

\begin{figure}[t]
\centering
\includegraphics[width=0.483\textwidth]{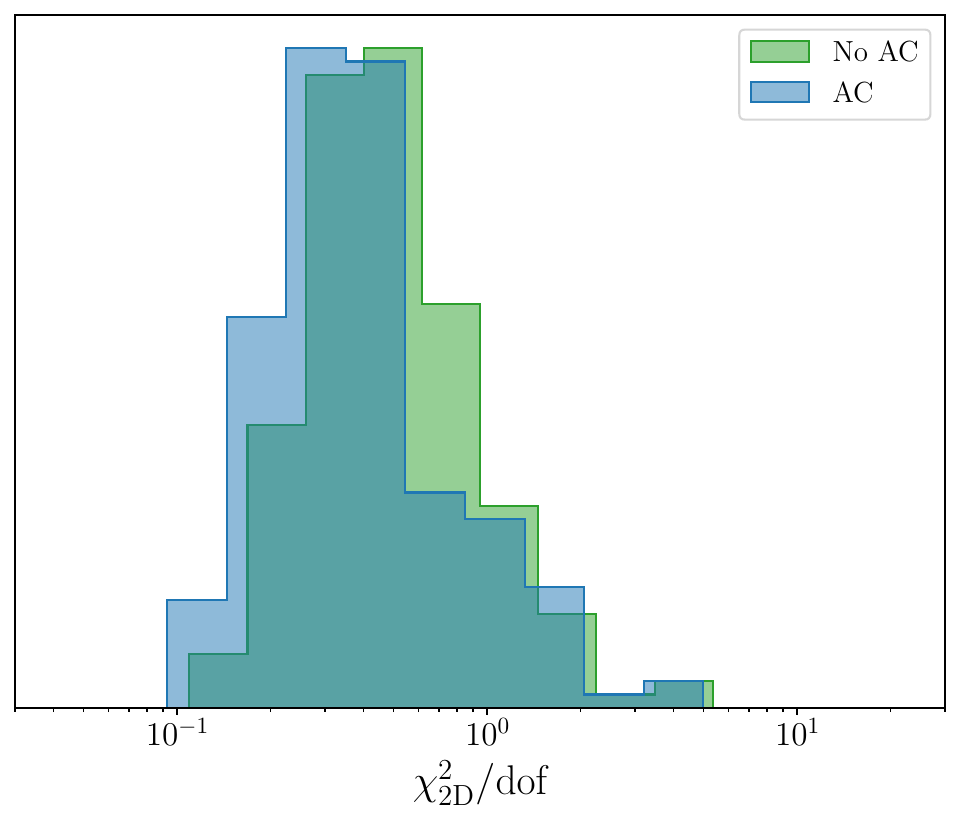}
\includegraphics[width=0.507\textwidth]{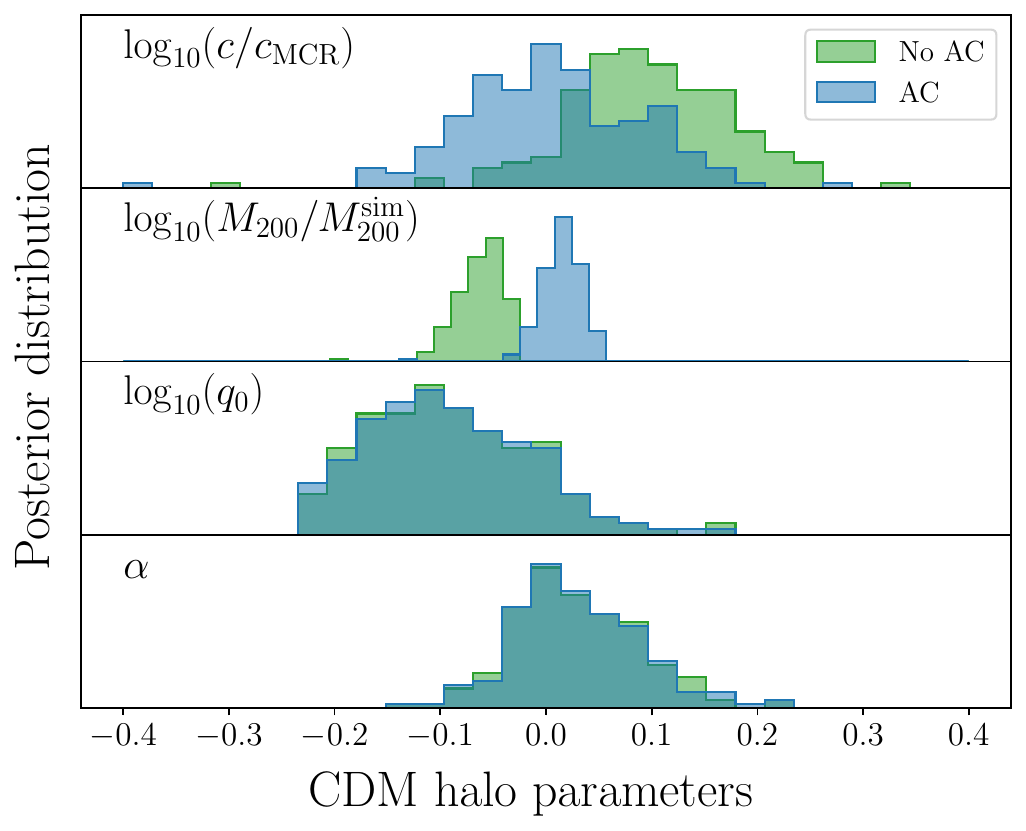}
\caption{\it
Comparison of \CDM\ halo fits with and without adiabatic contraction (AC) for 176 relaxed systems. \textbf{Left:} Distribution of reduced $\chi^2_{\rm 2D}$ values computed from comparing the azimuthally averaged $\rho(R,z)$ from the simulations with the best-fitting model; AC models (blue) systematically yield lower $\chi^2_{\rm 2D}/{\rm dof}$ than no–AC models (green). \textbf{Right:} Distributions of best–fit halo parameters for the same sample. We show $\log_{10}(c/c_{\rm MCR})$, $\log_{10}(M_{200}/M_{200}^{\rm sim})$, $\log_{10}(q_0)$, and the shape parameter $\alpha$. AC fits (blue) bring $c$ and $M_{200}$ into closer agreement with the physically motivated values ($c_{\rm MCR}$ and $M_{200}^{\rm sim}$) than the no–AC fits (green).}
\label{fig:cdm_fits}
\end{figure}

On the whole, spheroidal NFW halos with AC provide a good description of azimuthally-averaged 2D density profiles. 
The left panel of Figure~\ref{fig:cdm_fits} shows the distribution of (reduced) $\chi^2_{\rm 2D}$ values for all 176 relaxed \CDM\ halos. 
Halo models with AC provide better fits compared to uncontracted halos.
In Figure~\ref{fig:cdm_fits} (right), we show the distribution of best-fit model parameters both without AC (green) and with AC (blue) for the 176 relaxed \CDM\ systems.
The means and standard deviations for each parameter and model are quoted in Table~\ref{tab:CDM_params}.
We normalize $c$ relative to the value obtained from the mass-concentration relation, $c_{\rm MCR}$~\cite{Ludlow:2013vxa}, and $\Mvir$ to the virial mass directly inferred from the simulations, $\Mvir^{\rm sim}$.
The values for $c,\Mvir$ obtained from our AC model fits are in much better agreement with these physically motivated values compared to our fits without AC.
Thus, our AC model for CDM is the clear winner, and we adopt it hereafter as the outer halo in the Jeans model.

\begin{table}[b]
\begin{tabular}{c|c|c}
\hline
parameter & no AC & with AC \\
\hline
$\log (c/c_{\rm MCR})$  & $\hphantom{-}0.10 \pm 0.08$ & $\hphantom{-}0.01 \pm 0.08$  \\
$\log (\Mvir/\Mvir^{\rm sim})$    & $-0.06 \pm 0.02$ & $\hphantom{-}0.02 \pm 0.02$  \\
$\log q_0$    & $-0.09 \pm 0.08$ & $-0.09 \pm 0.08$  \\
$\alpha$    & $\hphantom{-}0.03 \pm 0.06$ & $\hphantom{-}0.03 \pm 0.06$  \\
\hline
\end{tabular}
\caption{Mean and standard deviations of best-fit model parameters for cases with and without AC.}
\label{tab:CDM_params}
\end{table}

\subsection{Baryon potential}
\label{sec:baryon_potential}

Lastly, we discuss the baryon potential $\Phi_b$ for the simulated halos, which is also a required  input for the Jeans model.
For the spherical Jeans model, we use the spherically-symmetric baryon potential $\Phi_b(r)$ computed from the spherically-averaged density of stars and gas.
For the nonspherical Jeans model, since we focus on the axisymmetric case, we calculate the baryon potential $\Phi_b(R,z)$ from the azimuthally-averaged mass distribution for stars and gas. 
This calculation effectively treats particles as rings, making use of the analytic form for the gravitational potential of a uniform density ring (see, e.g.~\cite{2009EJPh...30..623C}). 

The gravitational potential of a uniform density ring, with total mass $M$ and radius $R_r$, centered at the origin and lying in the $z=0$ plane, is  
\be
\label{eq:potential_from_ring}
\Phi_{\rm ring}(R,z) = - \frac{2 G M}{\pi \sqrt{(R + R_r)^2 + z^2}} \, K \left( \frac{4 R R_r}{(R + R_r)^2 + z^2}\right),
\ee
expressed in terms of the complete elliptic integral of the first kind
\be
\label{eq:ellipk}
K(m) = \int_0^{\pi/2} \frac{d\theta}{\sqrt{1 - m \sin^2(\theta)}} \, .
\ee
The inner baryon potential of a given halo is computed as a sum over contributions from star and gas particles $i$, each with mass $m_i$ and located at position $(R_i,z_i)$, within a radius 100 kpc of the halo center. Thus, the total baryon potential is given by
\be \label{eq:Phi_b_rings}
\Phi_b^{\rm in}(R,z) = - \sum_i \frac{2 G m_i}{\pi \sqrt{(R + R_i)^2 + (z-z_i)^2}} \, K \left( \frac{4 R R_i}{(R + R_i)^2 + (z- z_i)^2}\right) \, .
\ee
Note that by construction, this potential satisfies Poisson's equation for baryons, as required by the Jeans model. 
We evaluate Eq.~\eqref{eq:Phi_b_rings} on the same pixel grid given above, from which we make a cubic interpolation function in $(R,z)$ to evaluate the potential at any position $r < 100 \; {\rm kpc}$.

Our choice of 100 kpc is motivated by being far larger than the baryonic scale radii of our simulations. 
However, it is necessary to extrapolate $\Phi_b$ beyond our grid since for large cross sections, we may have $r_m > 100 \; {\rm kpc}$.
For this, we adopt a multipole expansion for the outer baryon potential (truncated at 10th order).
In the general case, this is
\be
\Phi_b^{\rm out}(\mathbf r) = - \frac{G M_b(r)}{r} + \sum_{\ell=1}^{10} \sum_{m=-\ell}^{+\ell} \frac{c_{\ell m} Z_{\ell m}(\theta,\varphi)}{r^{\ell+1}} \,,
\ee
where $Z_{\ell m}(\theta,\varphi)$ are real-valued spherical harmonics discussed in  Appendix \ref{app:nonspherical}, and  the multipole coefficients are matched to the inner potential  by imposing continuity of the potential at $r=100\,{\rm kpc}$,
\be
c_{\ell m} = \left. r^{\ell + 1} \int d\Omega \, Z_{\ell m}(\theta, \varphi) \, \Phi_b^{\rm in}(\mathbf r) \right|_{r = 100 \; {\rm kpc}} \, \, .
\ee
With azimuthal symmetry, only $m=0$ terms contribute.
Our complete formula for the baryon potential is thus
\be
\Phi_b(\mathbf r) = \left\{ \begin{array}{ll} \Phi_b^{\rm in}(\mathbf r) + \Delta \Phi\,, & r \le 100 \; {\rm kpc} \,,\\
\Phi_b^{\rm out}(\mathbf r)\,, & r > 100 \; {\rm kpc} \,,\end{array} \right.
\ee
where we include an additional constant
\be
\Delta \Phi = \left. - \frac{G M_b(r)}{r} - \int \frac{d\Omega}{4\pi} \, \Phi_b^{\rm in}(\mathbf r) \right|_{r = 100 \; {\rm kpc}}\,,
\ee
to ensure a continuous monopole term at $r = 100 \; {\rm kpc}$.

\section{Jeans model halo shapes}
\label{sec:shapes}

To confront the spherical and squashed Jeans model against data from simulated halos, we determine the halo-model parameters by sampling the posterior with MCMC using the Metropolis algorithm.
Our fits comprise all relaxed galaxy-sized halos in our sample from both \SIDM\ and \CDM\ simulations.
The goal is to determine to what extent Jeans modeling can infer the ``true'' self-interaction cross section $\sigmam$ in both cases, and whether the inference is improved by shape information. Simultaneously,  we also test to what extent the minimal shape function ansatz (Eq. \ref{eq:q_ansatz_2}) can successfully recover the 2D shapes of \SIDM\ halos.

For the spherical Jeans model, there are three parameters: $(\Mvir, c, r_m)$.
The virial mass $\Mvir$ and concentration $c$ fix the outer collisionless halo profile, assumed to be an adiabatically contracted NFW profile, which is matched at $r_m$ to determine the collisional inner profile using the ``outside-in'' procedure, described in Sec.~\ref{sec:spherical_jeans}.
We take logarithmically-flat priors on these parameters. 
Since the baryon potential is calculated on a discrete spatial grid of spacing $1 \kpc$, we impose an additional prior $r_m > 1 \kpc$ to avoid modeling below this scale.

For the squashed Jeans model, there are five input parameters: $(\Mvir,c,r_m ,q_0 ,\alpha)$, where $q_0$ and $\alpha$ describe the outer halo shape normalization and slope as per Eq.~\eqref{eq:q_cdm}.
Guided by the expectation that the outer halo shape for SIDM halos should be similar to CDM halos, we adopt Gaussian priors on $\log_{10} q_0, \alpha$ to match the means and widths inferred from our \CDM-fits (see Table~\ref{tab:CDM_params}): $\log_{10} q_0 =-0.09 \pm 0.08$ and $\alpha = 0.03 \pm 0.06$.
From these parameters, the full halo shape profile $q(r)$ is computed following the discussion surrounding Eq.~\eqref{eq:q_ansatz_2}.

To quantify the  goodness-of-fit relative to the simulations, we compute the total $\chi^2$  for both spherically-averaged density and shape profiles following Eq.~\eqref{eq:chi_sq_full}.
(For the spherical Jeans model, the shape term is neglected since it contributes an overall constant.)
Lastly, we determine $\sigmam$ from our posterior distributions and the rate equation~\eqref{eq:rate}.

\subsection{Squashed vs spherical Jeans model} 

\subsubsection{SIDM halos}

\begin{figure}[p]
\centering
\includegraphics[width=0.99\textwidth]{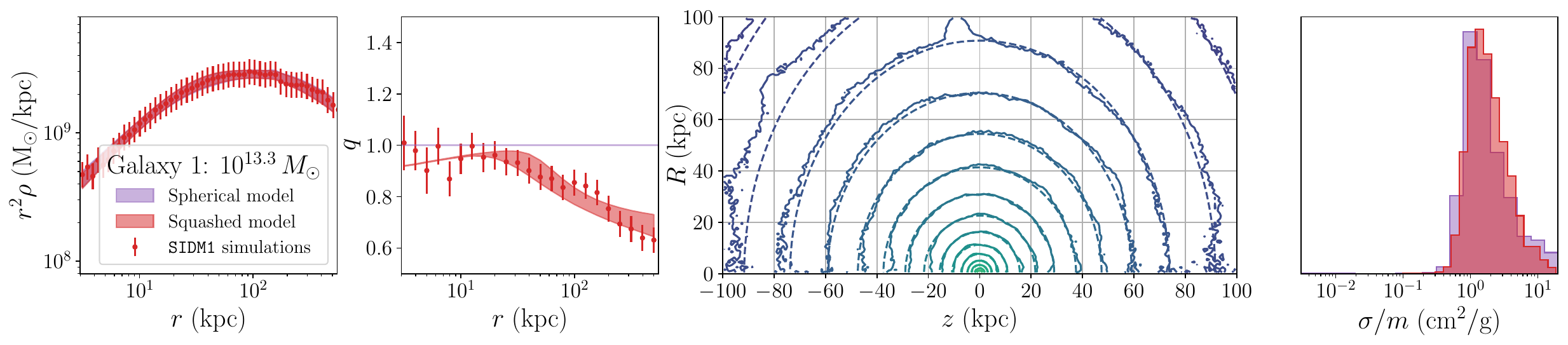}
\includegraphics[width=0.99\textwidth]{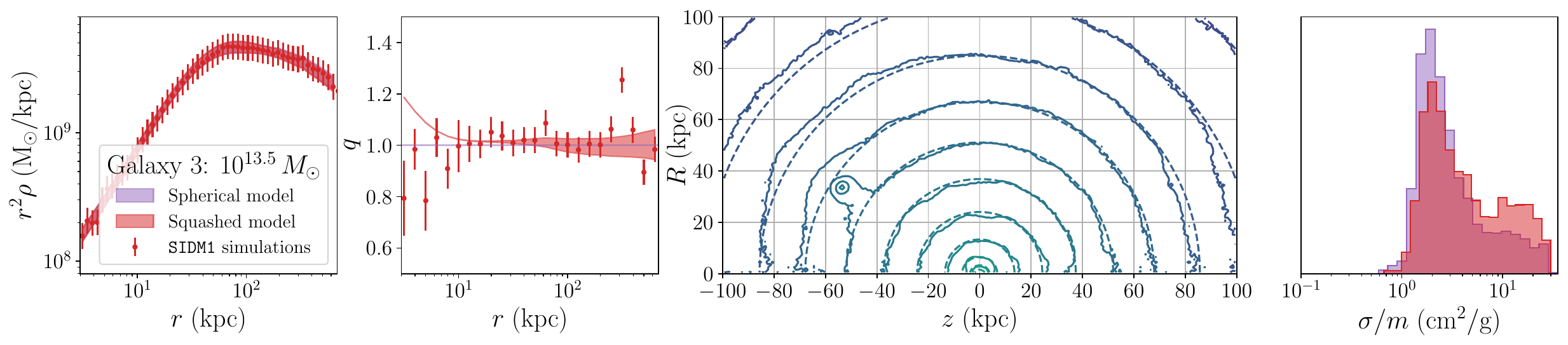}
\includegraphics[width=0.99\textwidth]{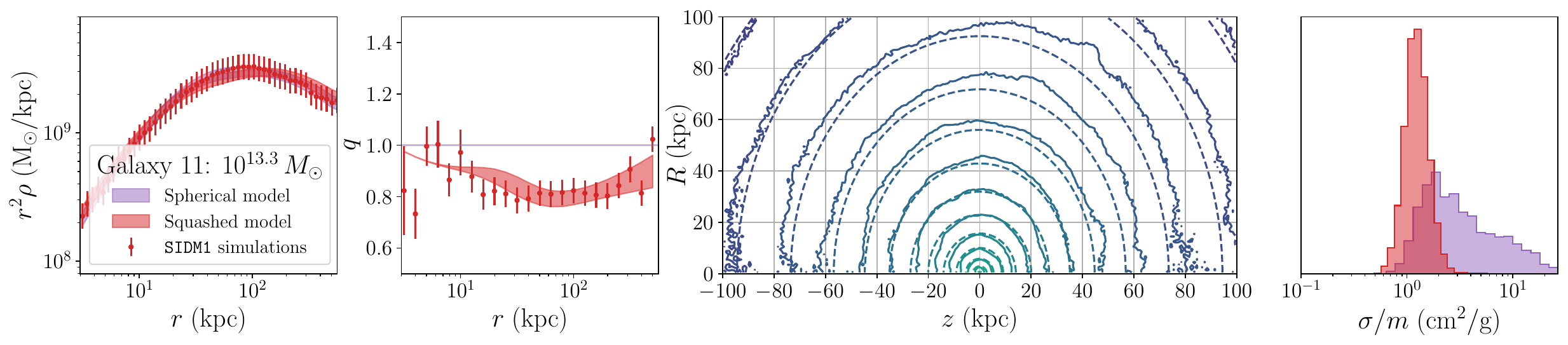}
\includegraphics[width=0.99\textwidth]{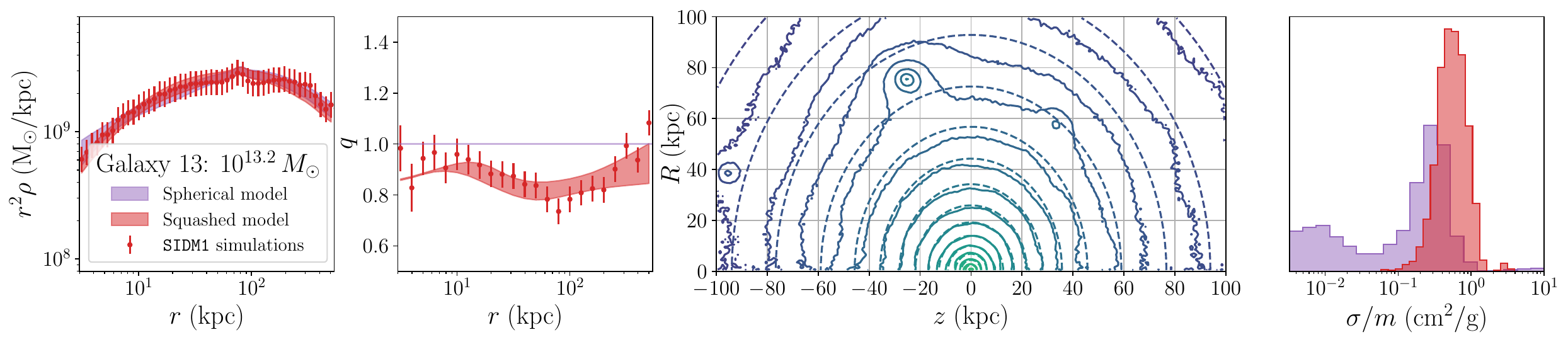}
\includegraphics[width=0.99\textwidth]{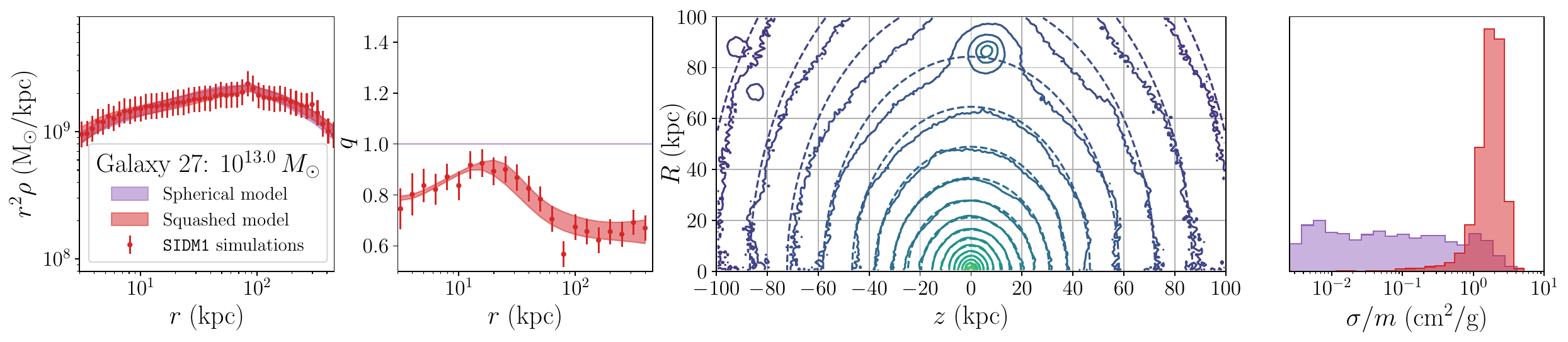}
\caption{\it \textbf{High–mass SIDM halos: squashed vs.\ spherical Jeans fits.} Comparison of \SIDM\ simulations to squashed and spherical Jeans fits for higher-mass systems ($\Mvir > 10^{13} M_\odot$).
Panels (left-to-right) show (a) spherically-averaged density and (b) halo shape profiles, where red (purple) bands denote 80\% confidence region from our nonspherical (spherical) fits;
(c) azimuthally-averaged 2D iso-density contours for \SIDM\ simulations (solid) and our best-fit squashed Jeans model profile (dashed), and (d) red (purple) posterior distribution for $\sigma/m$ for squashed (spherical) model.}
\label{fig:Squashed_fit_plots_SIDM1_high_mass}
\end{figure}

\begin{figure}[p]
\centering
\includegraphics[width=0.99\textwidth]{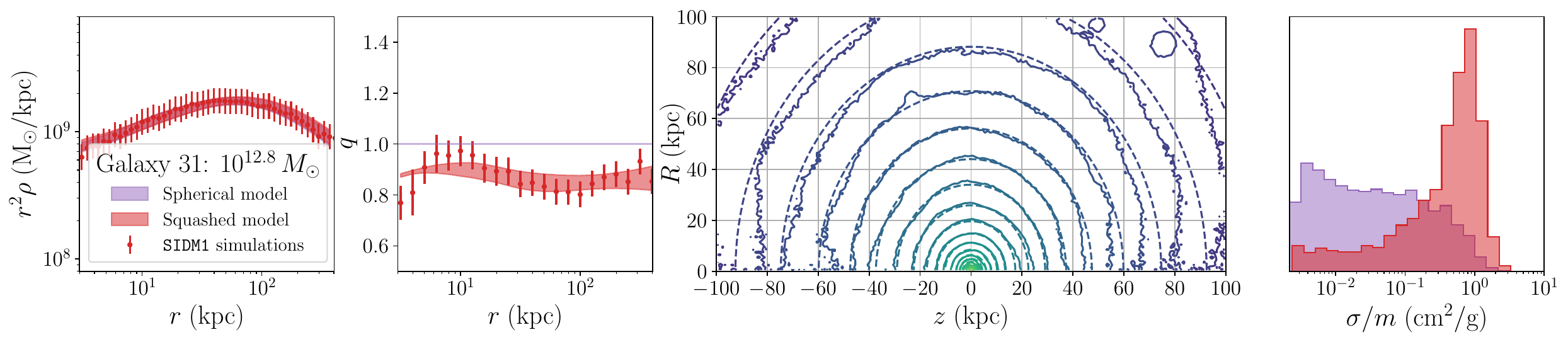}
\includegraphics[width=0.99\textwidth]{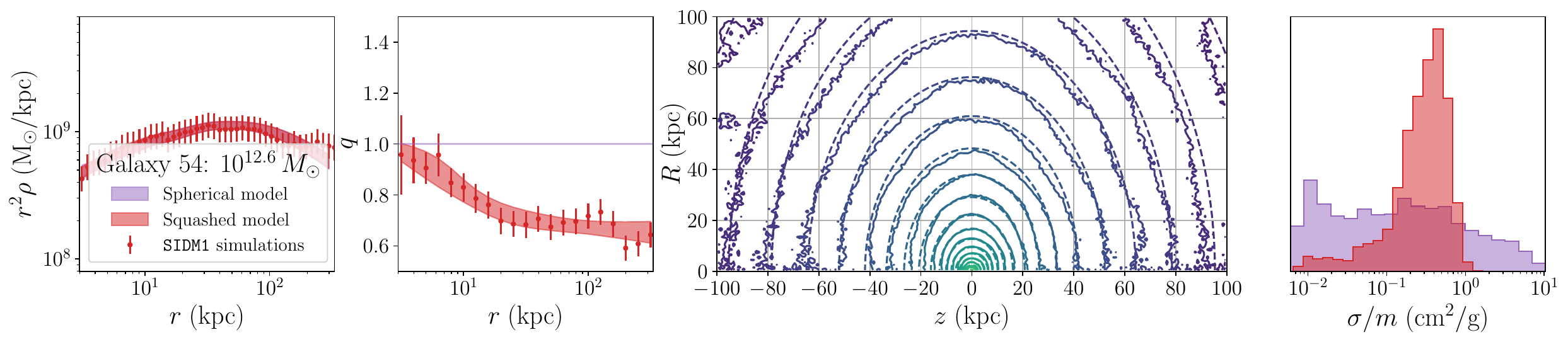}
\includegraphics[width=0.99\textwidth]{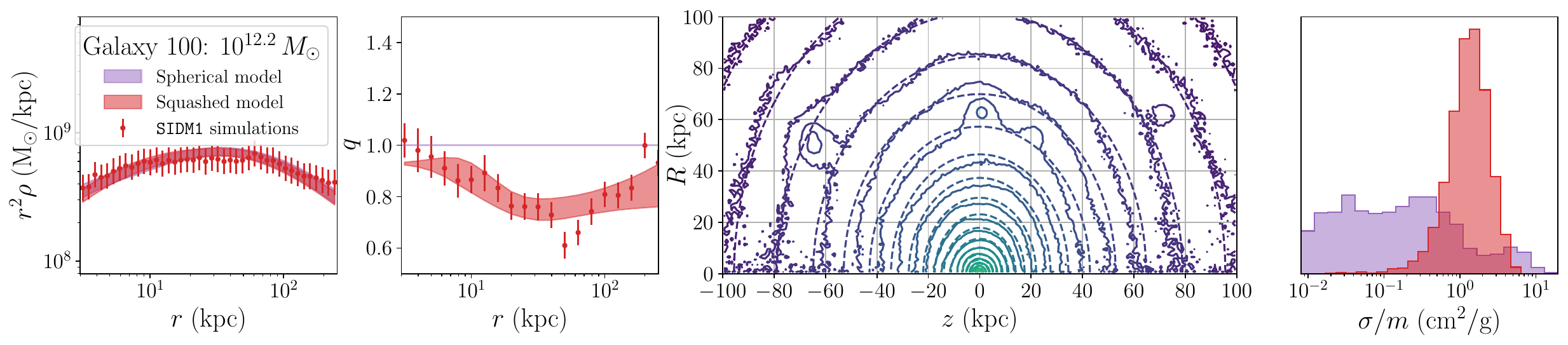}
\includegraphics[width=0.99\textwidth]{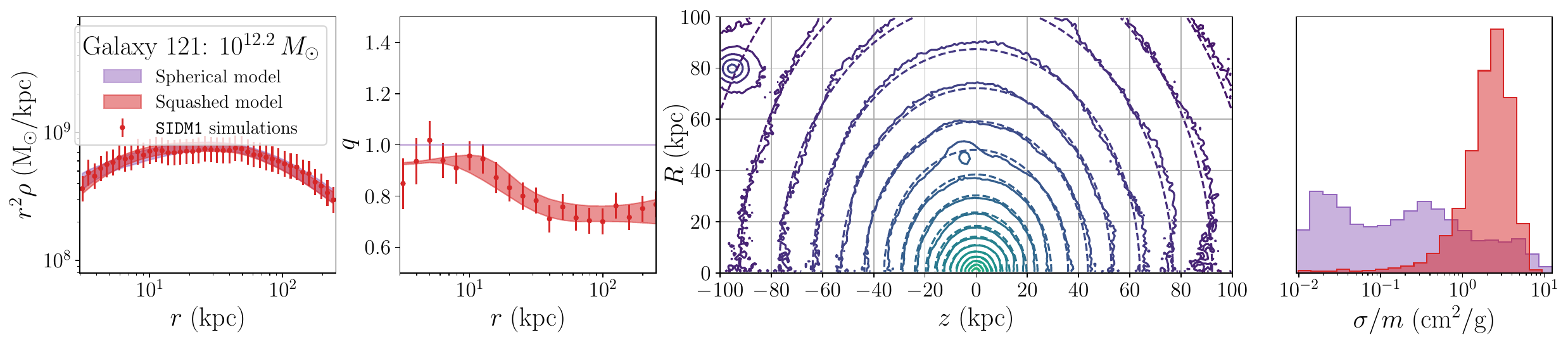}
\includegraphics[width=0.99\textwidth]{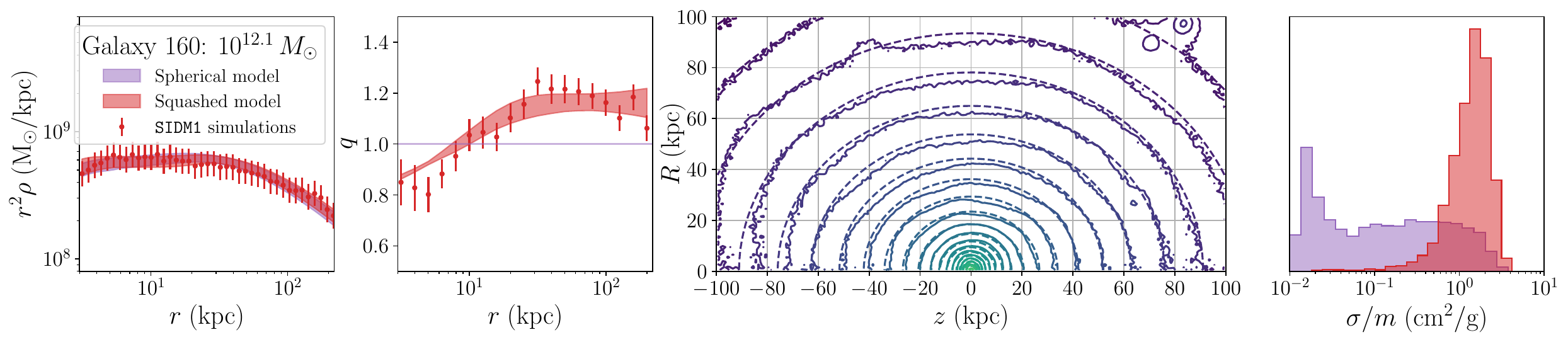}
\includegraphics[width=0.99\textwidth]{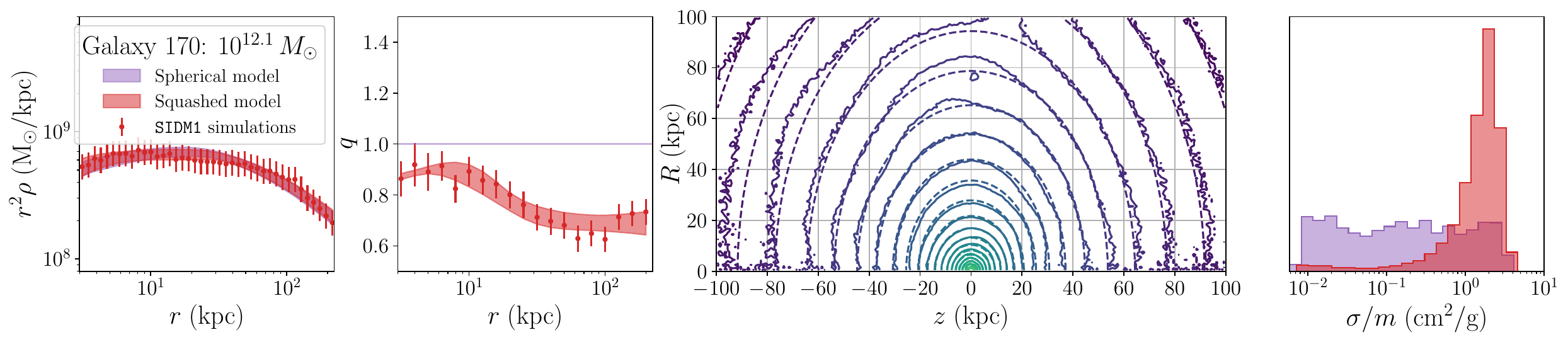}
\caption{\it \textbf{Example lower-mass SIDM systems with informative shapes.} Comparison of \SIDM\ simulations to squashed and spherical Jeans model fits for lower-mass systems ($\Mvir < 10^{13} M_\odot$).
Panels as in Figure~\ref{fig:Squashed_fit_plots_SIDM1_high_mass}.
}
\label{fig:Squashed_fit_plots_SIDM1_low_mass}
\end{figure}

\begin{figure}[p]
\centering
\includegraphics[width=0.99\textwidth]{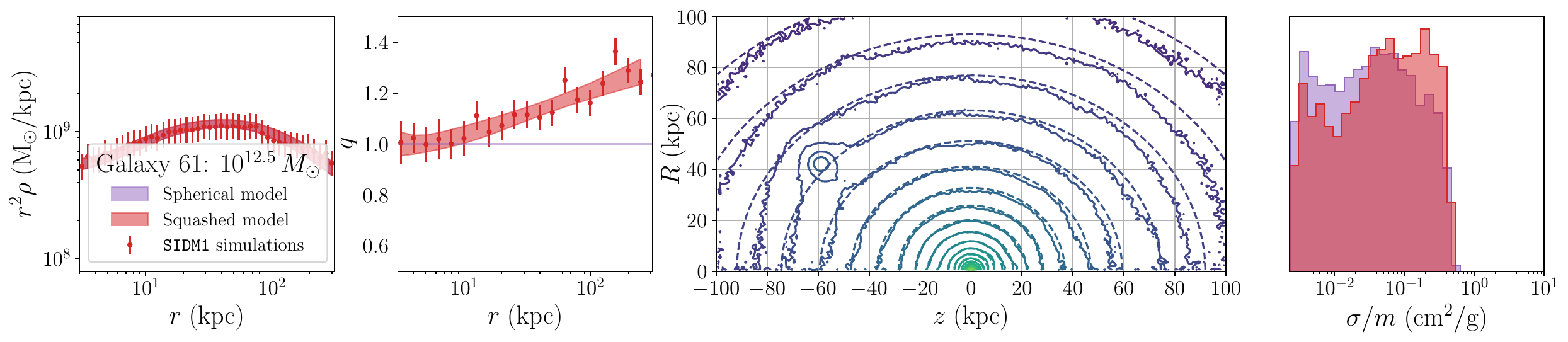}
\includegraphics[width=0.99\textwidth]{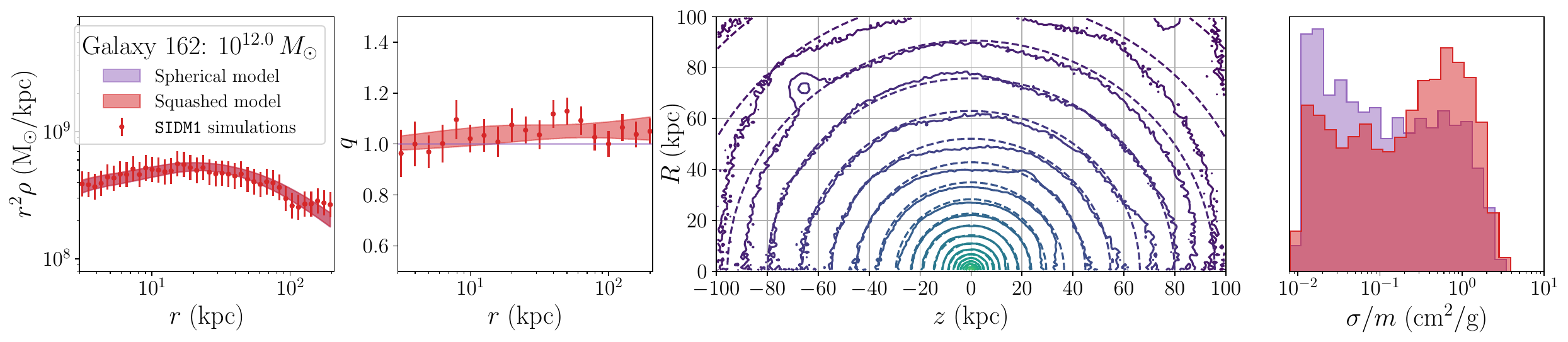}
\includegraphics[width=0.99\textwidth]{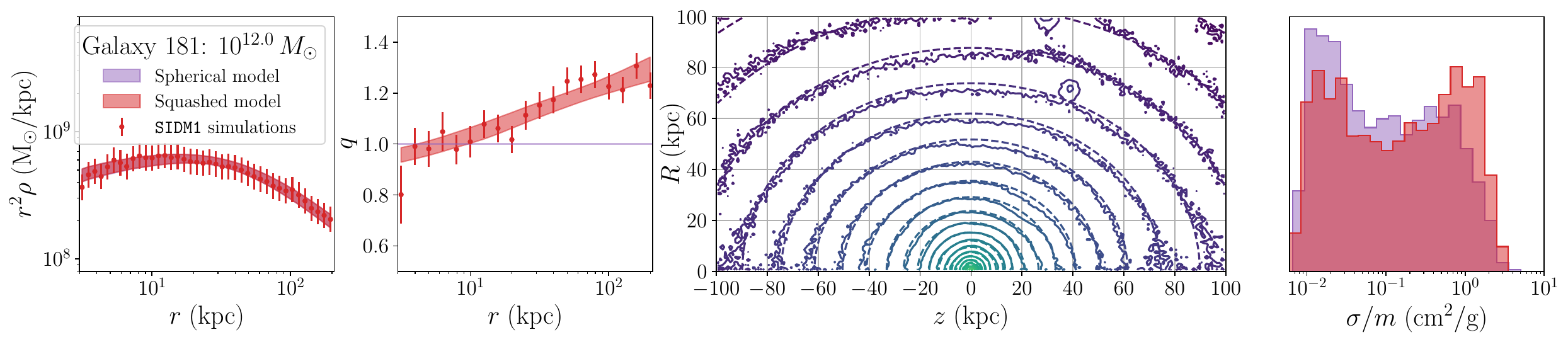}
\includegraphics[width=0.99\textwidth]{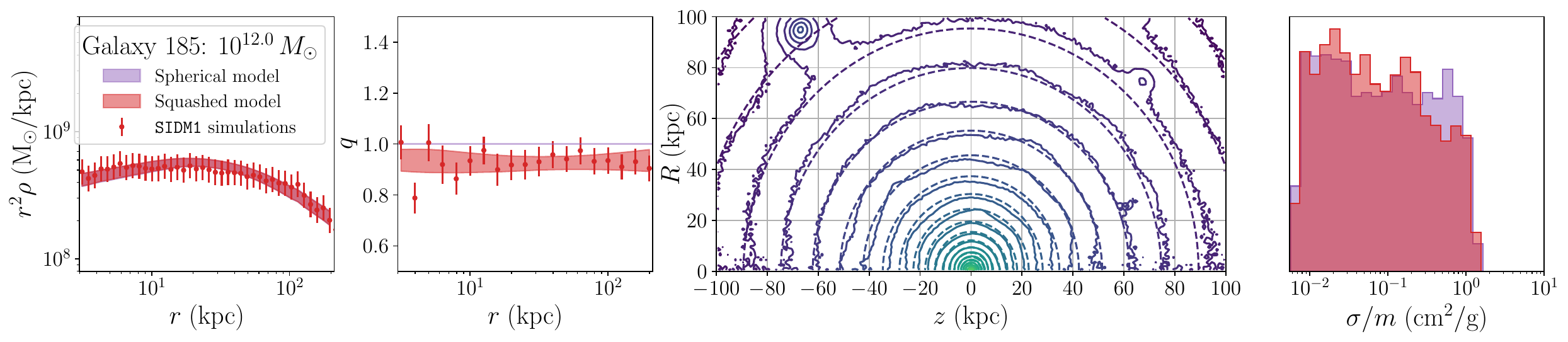}
\includegraphics[width=0.99\textwidth]{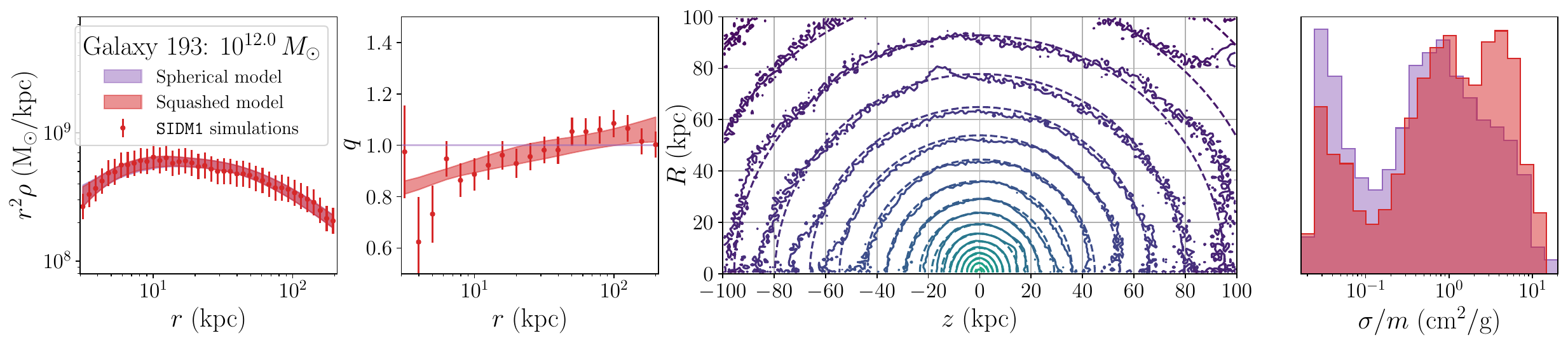}
\caption{\it \textbf{Example lower-mass SIDM systems with uninformative shapes.} Comparison of \SIDM\ simulations to squashed and spherical Jeans model fits for lower-mass systems ($\Mvir < 10^{13} M_\odot$).
Panels as in Figure~\ref{fig:Squashed_fit_plots_SIDM1_high_mass}.
}
\label{fig:Squashed_fit_plots_SIDM1_low_mass_2}
\end{figure}

In Figures~\ref{fig:Squashed_fit_plots_SIDM1_high_mass}-\ref{fig:Squashed_fit_plots_SIDM1_low_mass_2}, we show the results of our Jeans model fits to several representative \SIDM\ halos among our sample of 180 relaxed systems. 
Each row corresponds to one system, and the panel columns are as follows, from left to right:
\begin{enumerate}[label=(\alph*)]
    \item {\it Density profiles ($r^2$-weighted):} Data points represent the spherically-averaged density from \SIDM\ simulations, with our assumed uncertainties. The bands are profiles from our fits (80\% quantiles). 
    Squashed and spherical Jeans models yield nearly identical profiles, except for some systems that show a small difference at the innermost radii.
    \item {\it Halo shape profiles:} Data points represent the spheroidal axis ratio $q$ from \SIDM\ simulations, with our assumed uncertainties. 
    The squashed model band (80\% quantile) from our nonspherical fits describes the general trend of the radially-dependent shape, but not outlier points from substructure. 
    The spherical model has a trivial shape ($q=1$).
    \item {\it Density contours:} Solid lines show the 2D azimuthally-averaged iso-density contours from \SIDM\ simulations, while the dashed lines show those for the best-fit squashed Jeans model. By eye, we see generally good agreement, except for occasional substructures that are not captured in our model. 
    \item {\it Scattering cross section:} Histograms show the marginalized posteriors for $\sigmam$ for squashed and spherical Jeans models. We expect the posterior to be localized around the true value, $\sigmam = 1 \cmg$.
\end{enumerate}

First, Figure~\ref{fig:Squashed_fit_plots_SIDM1_high_mass} shows our results for five representative higher-mass halos, between $(1-4) \times 10^{13}\, \Msun$.
We find that our spherical fit is able to infer a nonzero cross section for most higher-mass systems (Galaxies 1, 3, 11), but a few systems (Galaxies 13, 27) are also consistent with $\sigmam \ll 1$, i.e., CDM. 
Our nonspherical fit improves our fits in some cases (Galaxies 11, 13, 27), with $\sigmam$ more peaked around $\sigmam = 1 \cmg$, while the others (Galaxy 1, 3) show no improvement with shape data.

Next, we turn to lower-mass halos, between $10^{12} - 10^{13} \, \Msun$.
Here, CDM and SIDM halos have similar spherically-averaged profiles, making it difficult to distinguish between them using spherical data alone (see Figure~\ref{fig:cdm_vs_sidm_1}).
However, CDM and SIDM halos have distinct shape profiles, and our nonspherical analysis can distinguish between models for some but not all systems.
Figure~\ref{fig:Squashed_fit_plots_SIDM1_low_mass} shows five successful examples along these lines. 
For these halos, our spherical fit yields $\sigmam$ posteriors consistent with a wide range of cross sections, including CDM-like halos with $\sigmam \ll 1 \cmg$.
However, shape information breaks the degeneracy. 
Our squashed fits favor SIDM, with posteriors consistent with $\sigmam = 1 \cmg$.

On the other hand, some systems do not benefit from a nonspherical analysis, such as those shown in Figure~\ref{fig:Squashed_fit_plots_SIDM1_low_mass_2}. 
These systems are consistent with a wide range of cross sections in both spherical and nonspherical fits.
Evidently, not all shape profiles (even those that are radially dependent) escape the degeneracy between CDM and SIDM.
For such cases, only an upper limit on $\sigmam$ is possible.

\begin{figure}[t]
\centering
\includegraphics[width=0.483\textwidth]{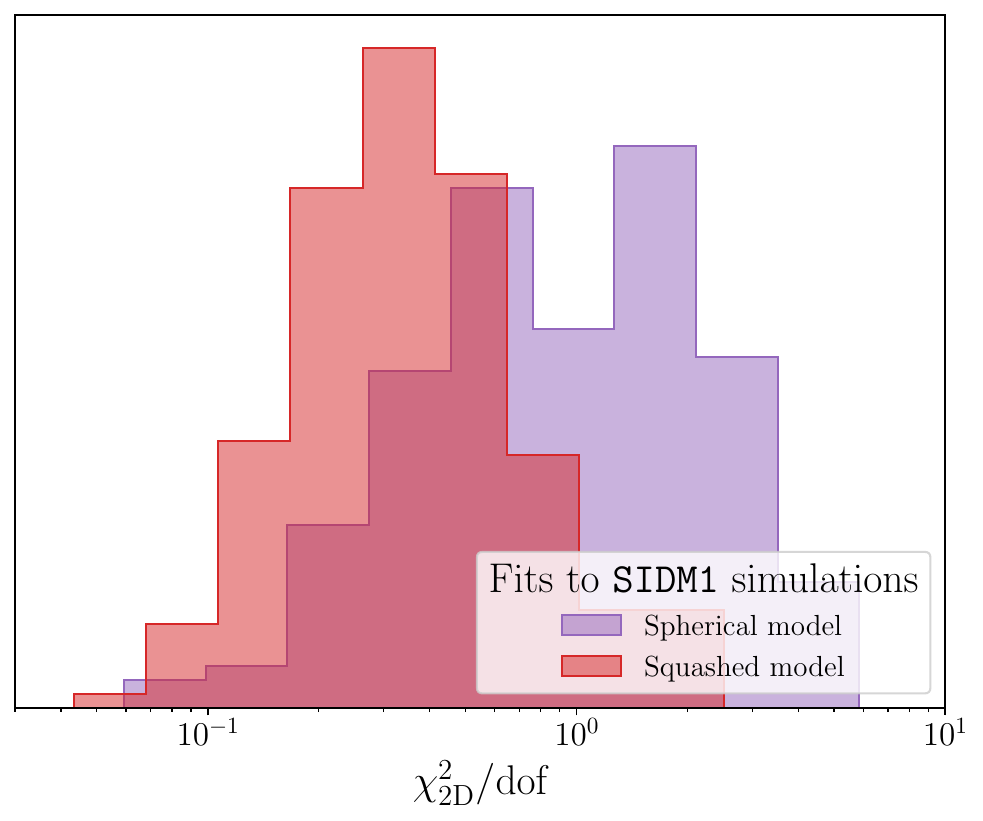}
\caption{\it Comparison of $\chi^2_{\rm 2D}$ per degree-of-freedom (dof) values for best-fit spherical and squashed models from our MCMC for \SIDM\ halos. 
}
\label{fig:sidm_fits}
\end{figure}

To compare the spherical and squashed Jeans model fits across all 180 relaxed SIDM1 halos, we calculate $\chi_{\rm 2D}$, comparing the azimuthally-averaged 2D density profiles for our models and simulations, following Eq.~\eqref{eq:chi_sq_2D}.
Figure~\ref{fig:sidm_fits} shows the distribution of reduced $\chi_{\rm 2D}$ values evaluated for each of our best-fit MCMC points for each system.
The squashed models provide lower values of $\chi_{\rm 2D}$ compared to spherical models.
Our conclusion is that despite being fit to 1D density and shape profiles, the squashed model does provide an improvement in modeling the true 2D density of the halos as expected.

\subsubsection{CDM halos}

\begin{figure}[t]
\centering
\includegraphics[width=0.99\textwidth]{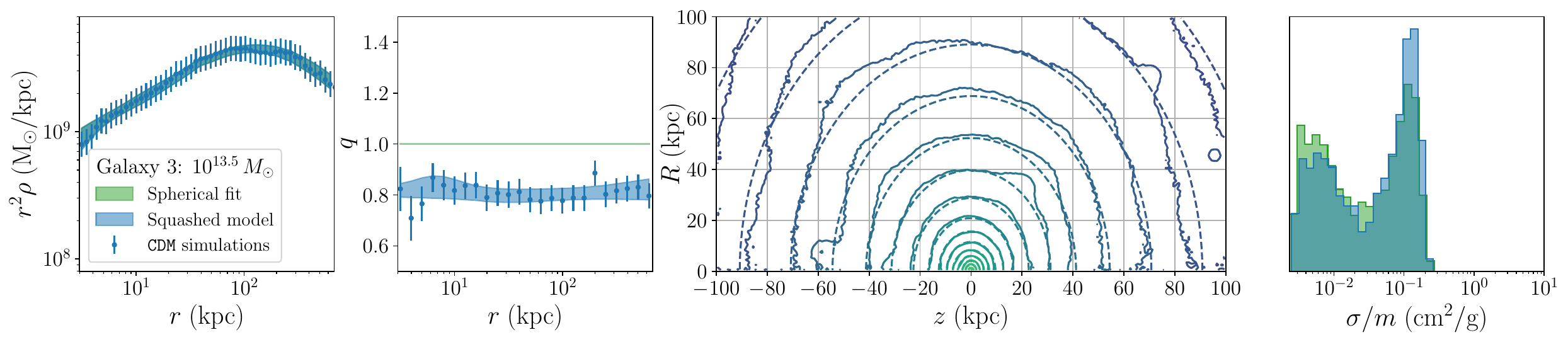}
\includegraphics[width=0.99\textwidth]{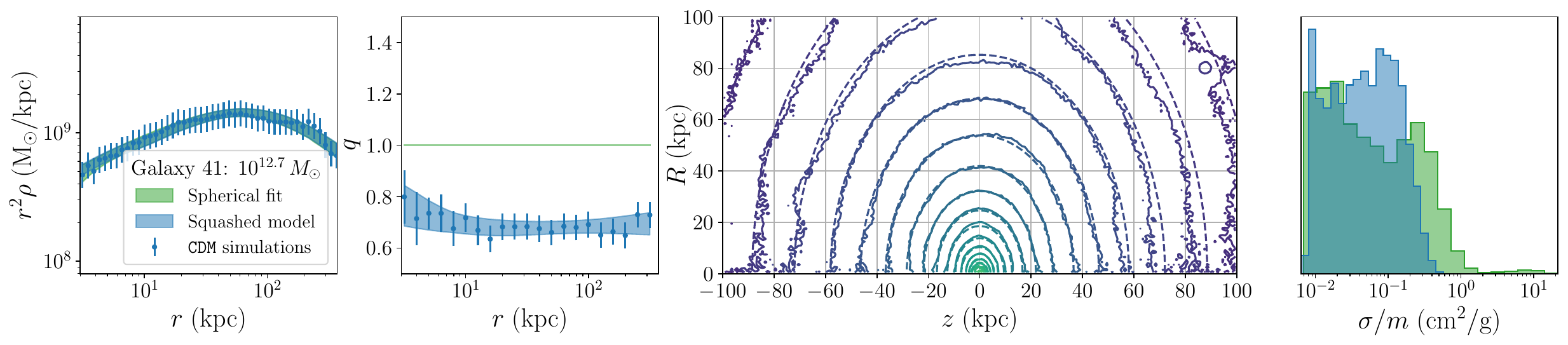}
\includegraphics[width=0.99\textwidth]{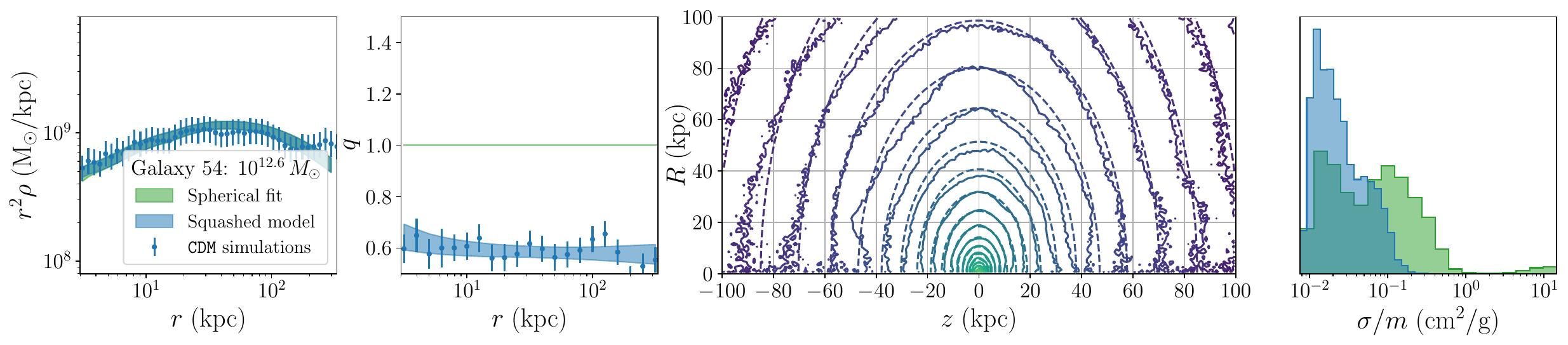}
\includegraphics[width=0.99\textwidth]{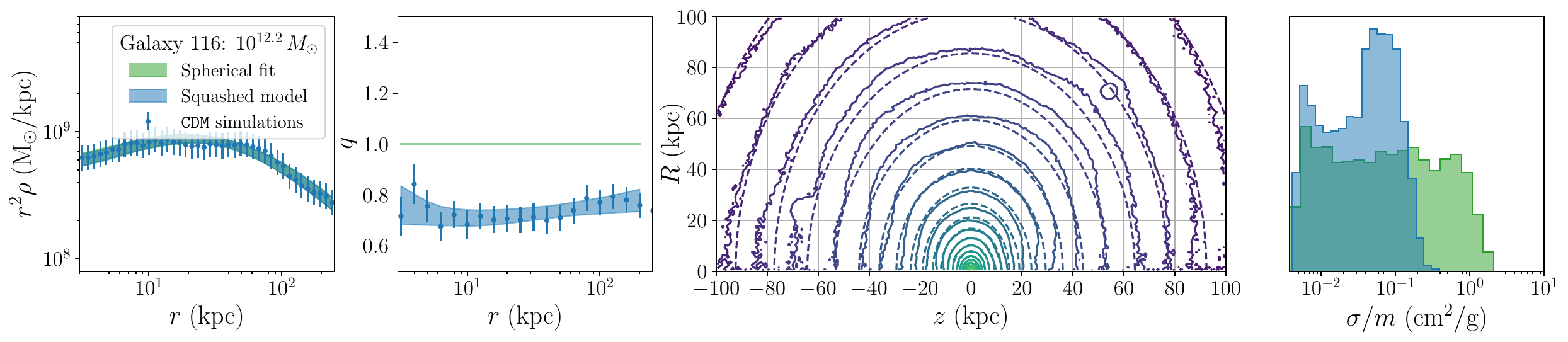}
\includegraphics[width=0.99\textwidth]{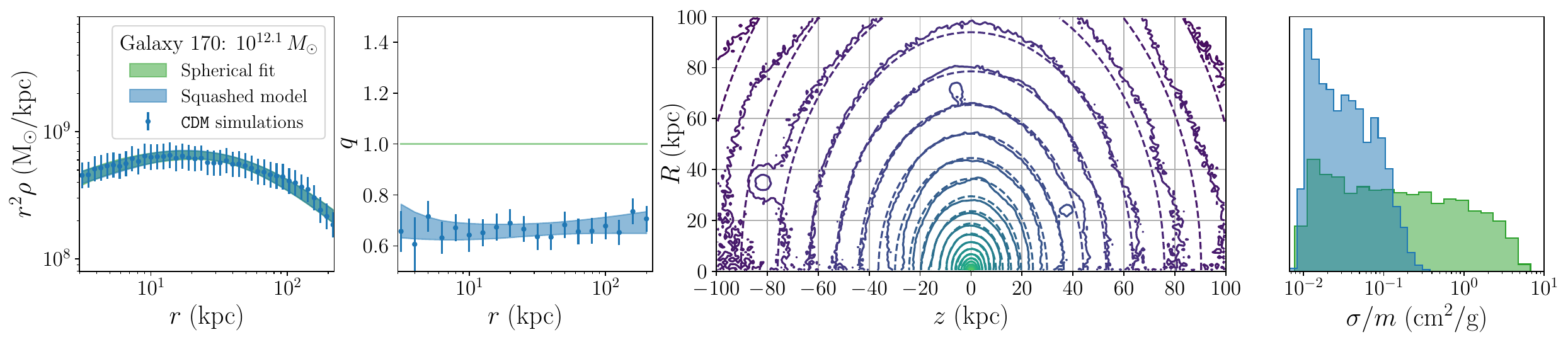}
\caption{\it {\bf Example CDM systems with upper limits on $\sigma/m$.} Comparison of \CDM\ simulations to squashed and spherical Jeans model fits.
Panels (left-to-right) show (a) spherically-averaged density and (b) halo shape profiles, with blue (green) bands denoting 80\% confidence region from our nonspherical (spherical) fits;
(c) azimuthally-averaged 2D iso-density contours from \CDM\ simulations (solid) and our best-fit squashed Jeans model profile (dashed), and (d) posterior distribution for $\log\sigma/m$ (in $\cmg$) for squashed (blue) and spherical (green) Jeans models.
}
\label{fig:Squashed_fit_plots_CDM_1}
\end{figure}

\begin{figure}[t]
\centering
\includegraphics[width=0.99\textwidth]{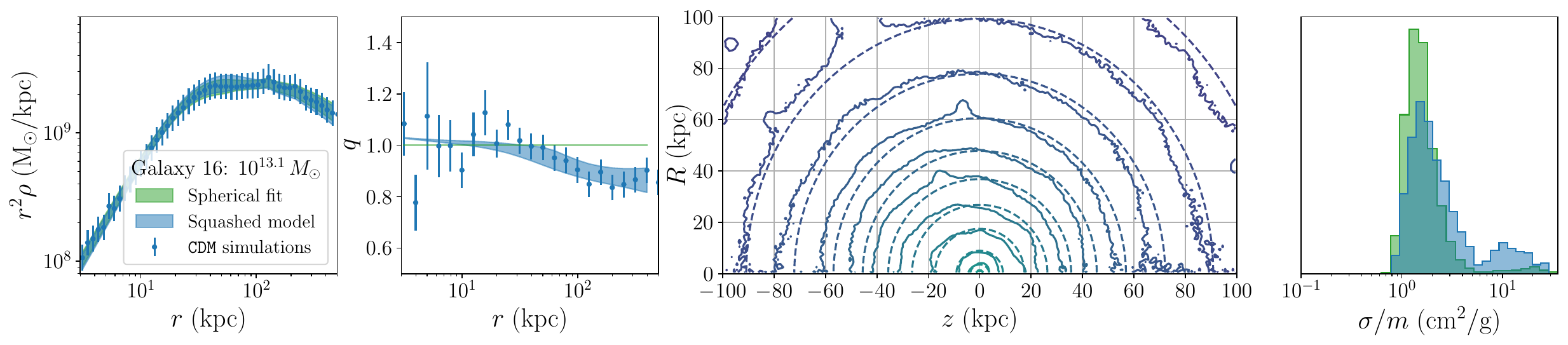}
\includegraphics[width=0.99\textwidth]{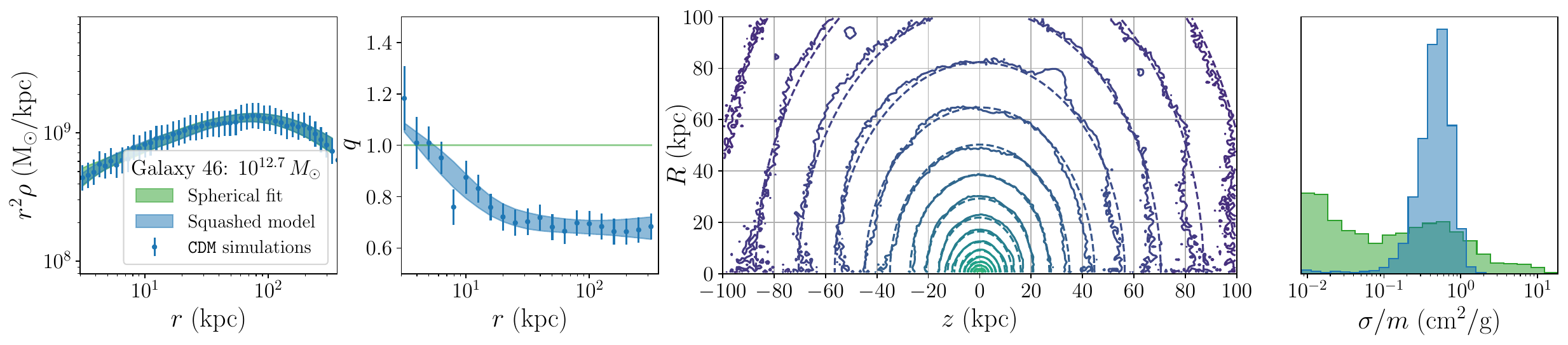}
\includegraphics[width=0.99\textwidth]{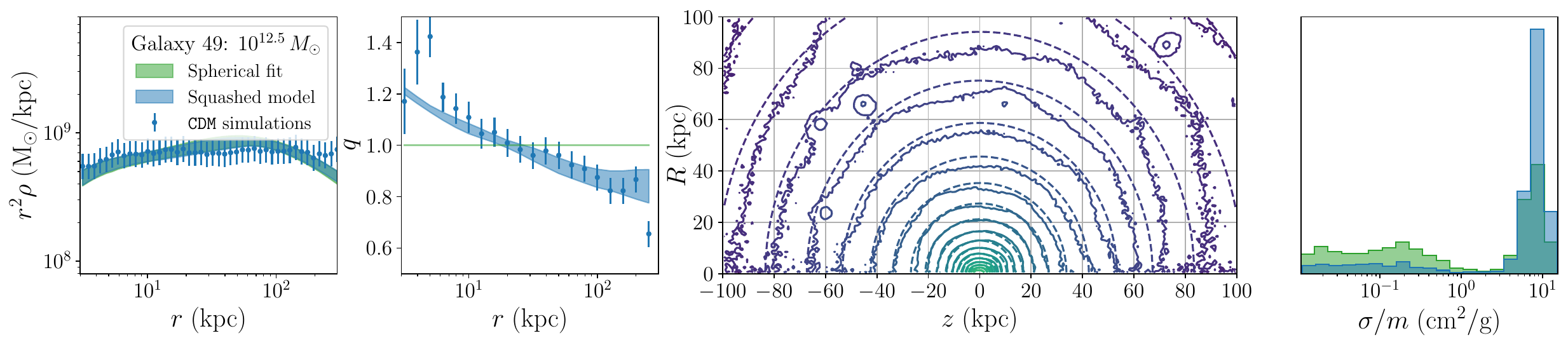}
\includegraphics[width=0.99\textwidth]{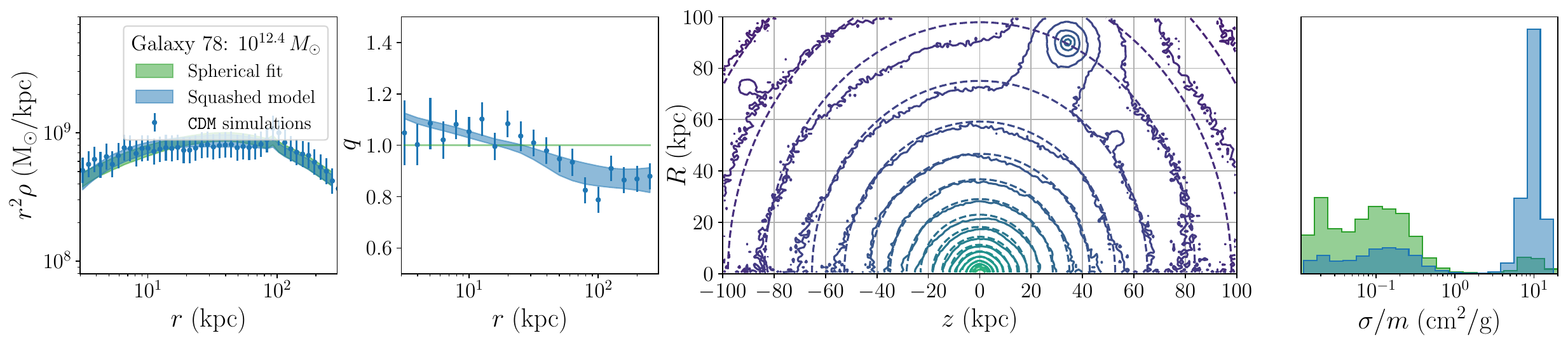}
\caption{\it {\bf Example CDM systems with spurious inference for $\sigma/m$.} 
Comparison of \CDM\ simulations to squashed and spherical Jeans model fits, for example, relaxed halos for which the squashed Jeans model infers a nonzero value of $\sigmam$.
Panels as in Figure~\ref{fig:Squashed_fit_plots_CDM_1}.
Compare to Figs.~\ref{fig:cdm_fits_1} and \ref{fig:cdm_fits_2}.}
\label{fig:Squashed_fit_plots_CDM_2}
\end{figure}

Next, in Figures~\ref{fig:Squashed_fit_plots_CDM_1} and \ref{fig:Squashed_fit_plots_CDM_2}, we show the results of our Jeans model fits to several representative systems among our sample of 176 relaxed \CDM\ halos. 
The panels are as in the preceding Figures~\ref{fig:Squashed_fit_plots_SIDM1_high_mass}-\ref{fig:Squashed_fit_plots_SIDM1_low_mass_2} for \SIDM\ halos.
Here, however, the data is from collisionless \CDM\ simulations and we expect inferred cross sections $\sigmam \ll 1 \cmg$.
(We note that posterior distribution for $\sigmam$ does not extend to zero due to our prior $r_m > 1 \kpc$.)

Figure~\ref{fig:Squashed_fit_plots_CDM_1} recalls the same \CDM\ halos shown previously in Figure~\ref{fig:cdm_vs_sidm_1}. 
There, we argued that higher-mass halos above $10^{13} \, \Msun$ (Galaxy 3) exhibit different density profiles between SIDM and CDM halos, while lower-mass halos between $10^{12}-10^{13}\, \Msun$ have degenerate density profiles but different shapes (Galaxies 41, 54, 116, 170).
The implication for the Jeans model is as follows: 
\begin{itemize}
\item For higher-mass \CDM\ halos, the spherical Jeans model is sufficient to constrain SIDM below $1 \cmg$. 
For example, for Galaxy 3, our spherical fit yields $\sigmam < 0.14 \cmg$ (90\%), while including shape information does not much impact the constraint.
\item For lower-mass \CDM\ halos, including shape data dramatically improves our constraints on $\sigmam$ compared to a spherical analysis. For Galaxies 41, 54, 116, 170, for example, our nonspherical analysis yields upper limits spanning $0.07-0.17 \cmg$ (90\%), compared to upper limits $0.3 - 1.8 \cmg$ from our spherical analysis. 
\end{itemize}

Lastly, shown in Figure~\ref{fig:Squashed_fit_plots_CDM_2}, there are a few systems that are ``false positives.'' 
For these halos, contrary to expectations, a sizable nonzero cross section is inferred from \CDM\ data in our nonspherical analysis. 
Galaxy 16 is asymmetric about the $z=0$ plane, while Galaxies 49 and 78 have prominent substructures around $r \approx 150$ and $100 \kpc$, respectively. 
Moreover, in Figure~\ref{fig:cdm_fits_2}, we see that none of their density profiles are well-fit by NFW profiles with or without AC.
This may indicate that Galaxies 16, 49, and 78 are not sufficiently relaxed, despite passing the criteria of Neto et al~\cite{Neto_2007}.
On the other hand, for Galaxy 46, none of these explanations seems to apply. 
This halo appears smooth by eye, has a low degree of nonrelaxedness according to the Neto et al~\cite{Neto_2007} metrics, and moreover its spherically-averaged density is well-fit by an NFW profile with AC as for most relaxed halos in our sample (see Figure~\ref{fig:cdm_fits_1}).
Evidently, the azimuthally-averaged shape profile for this halo is different than our ansatz for CDM in such a way as to mimic an SIDM halo.

\subsection{Inference for cross section and other parameters}

\begin{figure}[t]
\centering
\includegraphics[width=0.99\textwidth]{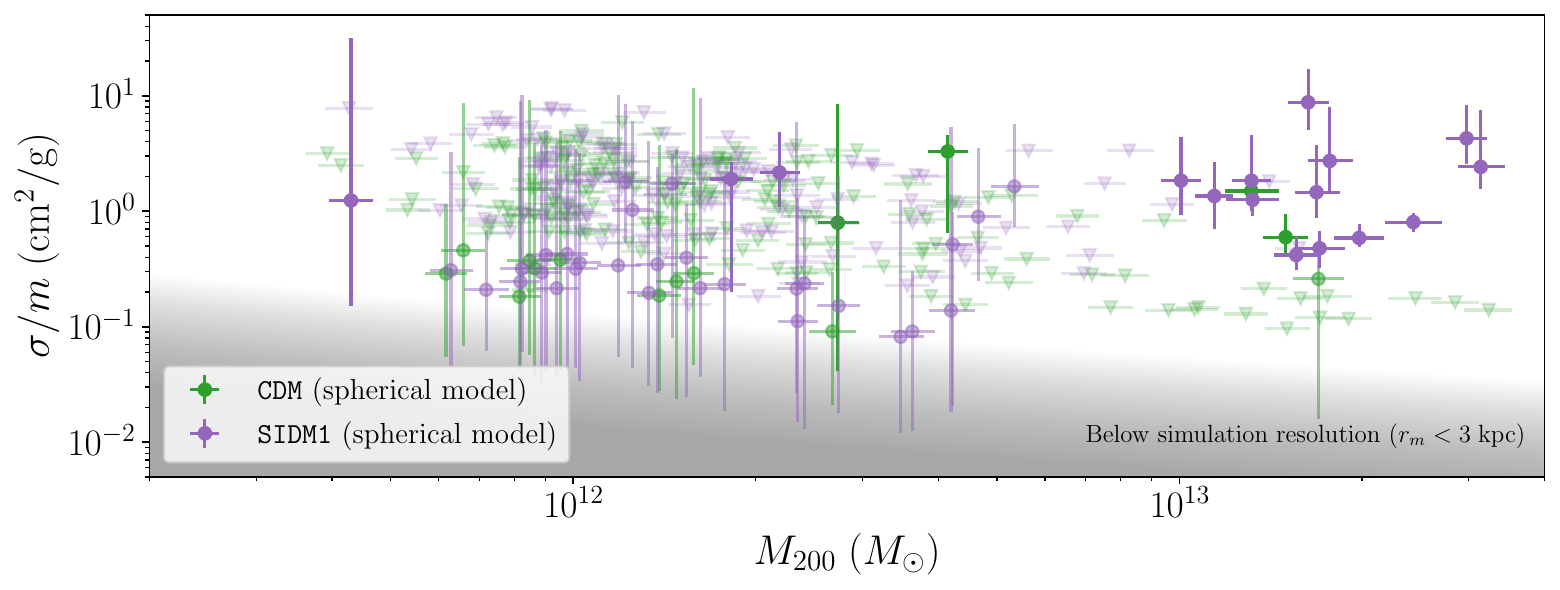}
\includegraphics[width=0.99\textwidth]{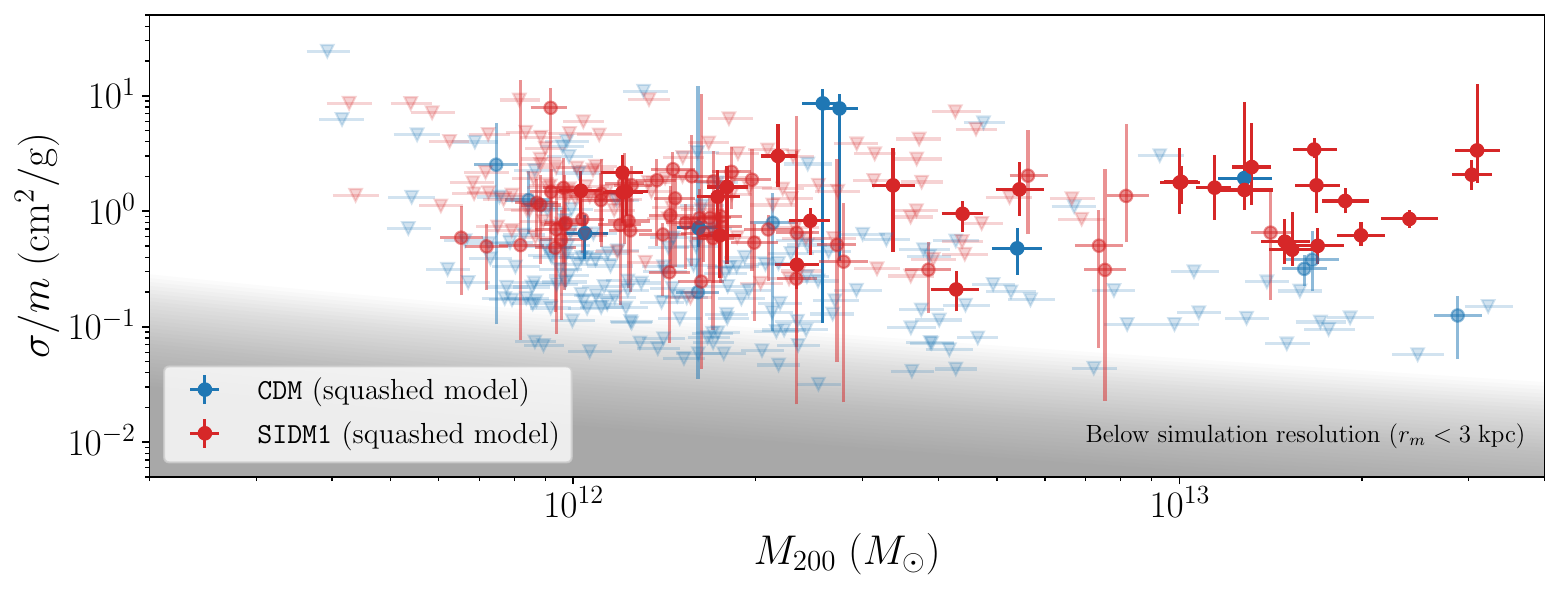}
\caption{\it \textbf{Inferred SIDM cross section vs.\ halo mass from Jeans modeling.} Top: Spherical Jeans model fits for relaxed \SIDM\ (purple) and \CDM\ (green) halos. Bottom: squashed Jeans fits for the same sample of halos (\SIDM\ in red; \CDM\ in blue). Points show the posterior median of $\sigma/m$ with central 68\% credible intervals; darker symbols indicate stronger cross section inferences. Down-pointing triangles mark 90\% upper limits when a nonzero cross section is not favored. Adding in shape information significantly improves how often we can tightly constrain the cross section in systems with $M_{200} = 10^{12} - 10^{13} \mathrm{M_\odot}$.}
\label{fig:cross_sections}
\end{figure}

Here we assess how well Jeans modeling is able to infer the true cross section from simulation data, with our results shown in Figure~\ref{fig:cross_sections}. 
The upper panel shows our spherical Jeans model results for \SIDM\ (purple) and \CDM\ (green) relaxed halos. 
The lower panel shows our nonspherical results using the squashed Jeans model, combining both density and shape profiles, again for both \SIDM\ (red) and \CDM\ (blue) relaxed halos.

Each data point represents 68\% posterior values for $\sigmam$ and $\Mvir$ for a given halo.
Bold (faint) data points indicate halos with a relatively strong (weak) inference on $\sigmam$. 
We have quantified our confidence in our cross section inference by comparing the maximum likelihood $\mathcal{L}$ calculated in our MCMC algorithm for samples with $r_m > 3 \kpc$ against that with $r_m < 3 \kpc$, which are effectively CDM for the purposes of our fit.
Defining the difference in maximum log-likelihoods
\be
\Delta \chi^2 = 
2\log \mathcal{L}_{\rm max}(r_m > 3 \kpc) - 2\log \mathcal{L}_{\rm max}(r_m > 3 \kpc) \, , 
\ee
we take $\Delta \chi^2 > 9$ to be a strong inference and $4 < \Delta \chi^2 < 9$ to be a weak inference.
For $\Delta \chi^2 < 4$, we obtain no cross section inference. 
Such halos are indicated as down-pointing triangles representing 90\% upper limits on $\sigmam$. 
The shaded gray region shows the $3 \kpc$ resolution limit in our simulated halos, expressed in the $\Mvir$-$\sigmam$ plane.
This approximate boundary is calculated using the spherical Jeans model without baryons with an outer halo satisfying the mean mass-concentration relation~\cite{Dutton:2014xda} and modifying these assumptions can shift its position.

First, we focus on higher-mass halos, $\Mvir > 10^{13} \, \Msun$, and compare the upper and lower panels of Figure~\ref{fig:cross_sections}.
\begin{itemize}
\item The spherical analysis (upper panel) infers a nonzero $\sigmam$ for 13/16 simulated \SIDM\ halos, with median values that span $0.4-8.8 \cmg$.
For \CDM\ halos, only upper limits on $\sigmam$ are inferred for most systems, as expected. These constraints are at or below $0.3 \cmg$ ($90\%$) for 13/17 systems.
\item The nonspherical analysis (lower panel) infers a nonzero $\sigmam$ for 16/16 simulated \SIDM\ halos, with median values that span $0.5-3.4 \cmg$.
For \CDM\ halos, again, most systems yield only upper limits on $\sigmam$, which are at or below $0.3 \cmg$ ($90\%$) for 12/17 systems.
\end{itemize}
Our conclusion for higher-mass halos is that the spherical Jeans model generally does well in distinguishing CDM versus SIDM at $1 \cmg$, albeit with a large scatter in inferred values of $\sigmam$ for SIDM.
The squashed Jeans model provides a modest improvement, narrowing the posterior distributions on $\sigmam$ for some \SIDM\ systems (see Figure~\ref{fig:Squashed_fit_plots_SIDM1_high_mass}), and reducing the overall scatter in $\sigmam$ across halos.
For high-mass \CDM\ halos, the spherical and squashed models yield similar results.
We note there is one false positive \CDM\ halo, Galaxy 16, where we strongly infer $\sigmam \approx 1 \cmg$ in both spherical and nonspherical fits.
Shown in Figure~\ref{fig:cdm_fits_2}, this halo is asymmetric about the $z=0$ plane and its spherically-averaged density is not well fit by NFW profiles with or without AC, which could signal nonrelaxedness beyond the criteria we have assumed.

Next, we turn to lower-mass halos, $\Mvir < 10^{13} \, \Msun$, again comparing the upper and lower panels of Figure~\ref{fig:cross_sections}.
\begin{itemize}
\item In the spherical analysis, only 3/164 (29/164) simulated \SIDM\ halos yield a strong (weak) inference for nonzero $\sigmam$.
For \CDM\ halos, only 15/159 systems yield stringent upper limits at the level of $0.3 \cmg$ (90\%) or less, but 123/159 systems have relatively worse upper limits above $0.5 \cmg$ (90\%) that would not exclude SIDM.
\item Our nonspherical analysis yields a strong (weak) inference for $\sigmam$ for 14/164 (50/164) simulated \SIDM\ halos.
For \CDM\ halos, 94/159 systems have 90\% upper limits at or below $0.3 \cmg$, while 34/159 halos have 90\% upper limits above $0.5 \cmg$.
\end{itemize}
We conclude that the spherical Jeans model offers little discriminating power between \SIDM\ and \CDM\ halos below $10^{13} \, \Msun$, but the situation is much improved with the squashed Jeans model.
Including halo shape constraints yields nonzero $\sigmam$ inferences for more \SIDM\ halos (see Figure~\ref{fig:Squashed_fit_plots_SIDM1_low_mass}) and strengthens constraints from \CDM\ halos (see Figure~\ref{fig:Squashed_fit_plots_CDM_1}).
However, even after including shape data, many \SIDM\ halos yield no inference in our nonspherical analysis.
This is to be expected since many SIDM halos are either spherical or have shape profiles consistent with our CDM shape profile ansatz.

\begin{figure}[t]
\centering
\includegraphics[width=0.99\textwidth]{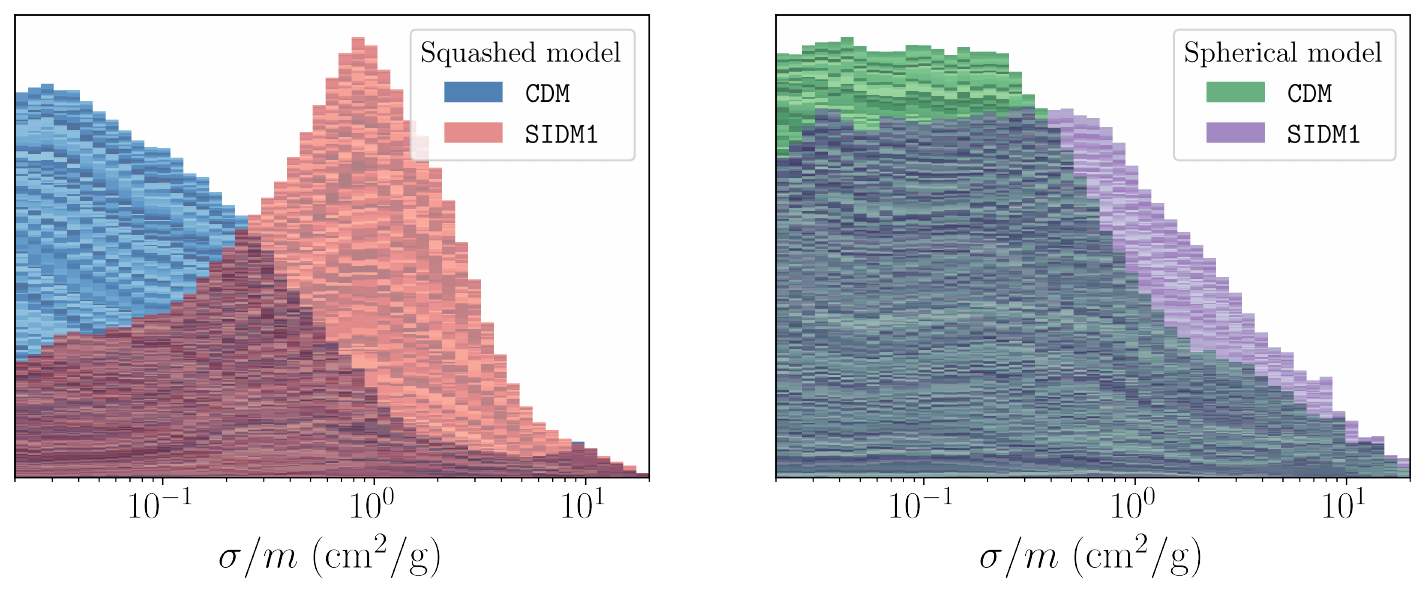}
\caption{\it \textbf{Stacked histogram for low-mass halos.} Left panel: Stacked histogram of posterior distributions for $\sigmam$ for all \SIDM\ (red) and \CDM\ (blue) halos below $10^{13} \Msun$ from the squashed Jeans model. Right panel: Similar for low-mass \SIDM\ (purple) and \CDM\ (green) halos with the spherical Jeans model.}
\label{fig:stacked_cross_sections}
\end{figure}

Figure~\ref{fig:stacked_cross_sections} further illustrates the role of shape data in constraining $\sigmam$ for low-mass halos.
The left panel shows a stacked histogram of our posterior distributions for all \SIDM\ (red) and \CDM\ (blue) halos below $10^{13} \Msun$ from the squashed Jeans model.\footnote{Different shades within the histogram represent individual posteriors in the stack among the sample of 164 \SIDM\ and 159 \CDM\ halos.}
The right panel shows the same plot for \SIDM\ (purple) and \CDM\ (green) halos using the spherical Jeans model.
Comparing these plots, it is clear that the squashed Jeans model is superior at distinguishing $\sigmam \approx 1 \cmg$ for \SIDM\ halos and $\sigmam \ll 1 \cmg$ for \CDM\ halos compared to the spherical model.

\begin{figure}[t]
\centering
\includegraphics[width=0.99\textwidth]{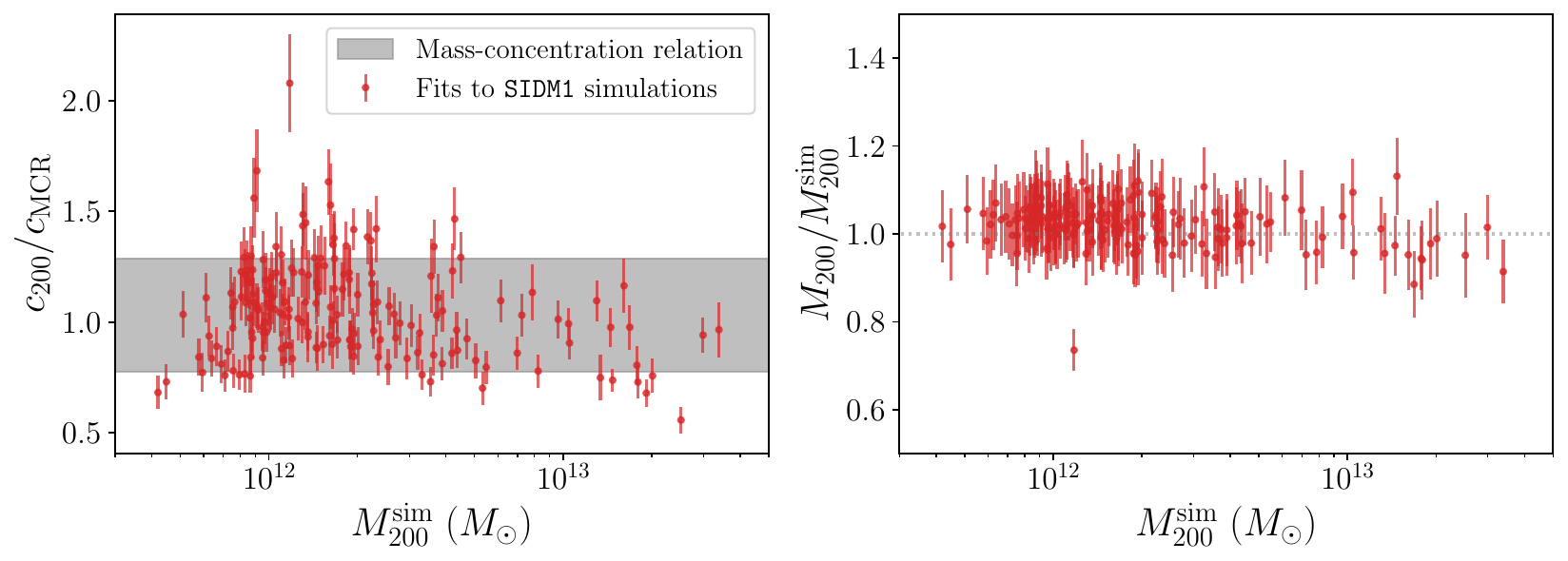}
\caption{\it Left panel shows the ratio of concentration parameters inferred for 180 \SIDM\ halos in our squashed model fit (red points) relative to the mean value from the cosmological mass-concentration relation (scatter in the latter is shown by the gray band) from Ref.~\cite{Dutton:2014xda}.
Right panel shows the corresponding ratio of virial masses $\Mvir$ from our fits relative to those directly computed from simulations, $\Mvir^{\rm sim}$.
}
\label{fig:mass_concentration}
\end{figure}

Lastly, we consider other parameters inferred from our nonspherical fits.
While we have not adopted priors on $(\cvir, \Mvir)$, we now check a posteriori whether our inferred values are reasonable.
The expectations are that {\it (i)} our inferred $\Mvir$ will be consistent with values directly measured from simulations, $\Mvir^{\rm sim}$, and {\it (ii)} our inferred concentration $\cvir$ will be consistent with the mass-concentration relation, $c_{\rm MCR}$, since SIDM halos follow the same process of cosmological structure formation as CDM halos.
Figure~\ref{fig:mass_concentration} (left) shows the ratio between our inferred $\cvir$ and the mean value of $c_{\rm MCR}$, evaluated at $\Mvir^{\rm sim}$, from Ref.~\cite{Dutton:2014xda}.
The gray band represents the 68\% cosmological scatter in $c_{\rm MCR}$~\cite{Dutton:2014xda}.
Most systems in our analysis (red data points) are consistent with this expectation.
Figure~\ref{fig:mass_concentration} (right) shows the ratio between our inferred $\Mvir$ and $\Mvir^{\rm sim}$.
Nearly all systems in our analysis (red data points) are consistent with the expectation $\Mvir/\Mvir^{\rm sim} = 1$ (dashed line).
These checks lend credence to the robustness of our approach.

\section{Conclusions}
\label{sec:conclude}

We have introduced a new semi-analytical framework, the squashed Jeans model, to describe the shape profiles of SIDM halos.
The standard lore is that SIDM halos are rounder compared to their collisionless counterparts.
However, quantitative predictions for SIDM halo shapes depend on the baryon distribution, as well as on residual triaxiality from the dark matter halo. Both effects are included in our model.

Our work extends the spherical isothermal Jeans model, widely used in the literature to describe the spherically-averaged properties of SIDM halos.
While previous attempts to generalize the Jeans model beyond spherical symmetry maintain isothermality and hydrostatic equilibrium, we have shown these assumptions lead to halos that {\it (i)} are too round compared to $N$-body simulations, and {\it (ii)} cannot be easily embedded within a collisionless outer halo without a discontinuous density profile.

In contrast, our setup is based on the fact (borne from $N$-body simulations) that multiple self-interactions are required to impact halo shapes. 
The transition between collisional and collisionless regions of a halo is a smooth function of the mean number of scatters per particle, $\mathcal{N}(r)$, at a given radius.
The inner halo reaches the isothermal limit for $\mathcal{N} \gg 1$, not where $\mathcal{N} = 1$ as typically assumed, while the outer halo reduces to the collisionless limit for $\mathcal{N} \ll 1$.
In addition, we provide an efficient prescription for calculating the shape profile in the isothermal limit that is accurate, derived from first principles, and avoids the computationally expensive step of solving a multivariate Poisson's equation.

In this work, we have specialized to the case of azimuthal symmetry, though the method can be generalized to relax this assumption, as we discuss below.
In our systems of study (from \textsc{Eagle-50} simulations), the axis of symmetry is defined by the stellar angular momentum vector.
In this case, halo shapes are described by the axis ratio profile $q(r)$, where $q > 1$ ($q < 1$) is prolate (oblate).
\\

Here we give a step-by-step description for how the  squashed Jeans model is implemented. We also highlight the differences with the (usual) spherical Jeans model.
\begin{enumerate}[topsep=0pt]
\item \underline{Model inputs:} There are four quantities that must be fixed as initial inputs.
    \begin{itemize}[noitemsep, topsep=0pt]
      \item $\bar{\rho}_\mathrm{cdm}(r)$ is the spherically-averaged density profile of a collisionless halo, i.e., in the absence of self-interactions. This could be an NFW profile, an adiabatically-contracted NFW profile, an Einasto profile, etc.
      \item $q_{\rm cdm}(r)$ is the axis ratio profile the halo would have in the absence of self-interactions. We assumed a power-law profile $q_{\rm cdm}(r) \propto q_0 \, r^\alpha$ in this work.
      \item $\rho_\mathrm{b}(\mathbf r)$ is the (nonspherical) baryon density, from which we calculate the baryon potential $\Phi_b$. These quantities are azimuthally averaged in this work.
      \item $r_m$ is the matching radius delineating the inner and outer regions, corresponding to a given value of $\sigma/m$.
    \end{itemize}
These inputs are identical to those in the spherical Jeans model, except allowing for nonsphericity in the outer halo and baryon distribution.
\item \underline{Solve the spherical Jeans model:} 
We follow the usual approach established in the literature.
\begin{itemize}[noitemsep, topsep=0pt]
\item Solve the spherically-symmetric isothermal Jeans equation~\eqref{eq:Jeans_sph} in the region $0 \leq r \leq r_m$ to obtain the inner SIDM density profile, $\bar{\rho}_\mathrm{iso}(r)$, subject to boundary conditions $\bar{\rho}_\mathrm{iso}(r_m) = \bar{\rho}_\mathrm{cdm}(r_m)$, $M_\mathrm{iso}(r_m) = M_\mathrm{cdm}(r_m)$, with the spherically-averaged baryon density. 
We use the relaxation method (see Appendix~\ref{app:spherical}), which also determines the central density $\bar{\rho}_\mathrm{iso}(0) = \rho_0$ and 1D velocity dispersion $\sigma_0$.   
\item The prediction for the full spherically-averaged density profile, $\bar{\rho}_\mathrm{dm}(r)$, is then $\bar{\rho}_\mathrm{iso}(r)$ for $r < r_m$ and $\bar{\rho}_\mathrm{cdm}(r)$ for $r \geq r_m$.
\item The number of scatters per particle, as a function of radius, is determined by $\mathcal{N}(r) \approx \bar\rho_{\rm dm}(r) / \bar\rho_{\rm dm}(r_m)$ according to the rate equation~\eqref{eq:rate}, which also fixes $\sigma/m$ corresponding to $\mathcal{N} = 1$.
\end{itemize}  
\item \underline{Determine the isothermal shape profile:} We calculate the axis ratio profile $q_\mathrm{iso}(r)$ for an isothermal halo in hydrostatic equilibrium for radii $r < r_m$. 
Here we have explored two methods.
\begin{itemize}[noitemsep, topsep=0pt]
\item The computationally expensive way to determine $q_{\rm iso}$ is a brute-force multivariate solution to Poisson's equation~\eqref{eq:main}, which we have done in Appendix~\ref{app:nonspherical}.
\item Our preferred method is to calculate $q_\mathrm{iso}(r)$ directly following Eq.~\eqref{eq:q_iso_three_terms}, which is derived via an expansion in deviations from spherical symmetry.
This approach is accurate and far more efficient numerically.
\end{itemize}
\item \underline{Construct the halo shape profile:}
The full axis ratio profile for the entire halo, $q_\mathrm{dm}(r)$, is determined by -- at each radius -- interpolating between $q_{\rm cdm}(r)$ and $q_{\rm iso}(r)$ using Eq.~\eqref{eq:q_ansatz_2}. This interpolation is based upon the local scattering rate, with $q_\mathrm{dm} \to q_{\rm cdm}$ when $\mathcal{N} \ll 1$ and $q_\mathrm{dm} \to q_{\rm iso}$ when $\mathcal{N} \gg 1$.  
\item \underline{Squashed profiles:} 
Lastly, the full nonspherical dark matter halo profile $\rho_\mathrm{dm}(\mathbf r)$ is determined from $\bar{\rho}_\mathrm{dm}(r)$ and $q_\mathrm{dm}(r)$. This is done by ``squashing'' iso-density spheres into iso-density spheroids of the same volume, following Eqs.~\eqref{eq:squashed_mapping} and \eqref{eq:r_sph}. 
\end{enumerate}
Ultimately, the method we have implemented is not much more numerically intensive than the usual spherical Jeans model.
\\

Within the nonspherical squashed Jeans model developed in this work, a diverse family of two-dimensional halo shape profiles naturally arises, offering a richer structural description of DM systems compared to the spherical models. For CDM halos, the shape profile follows a simple power-law dependence on radius, while SIDM halos require a more flexible---yet still semi-analytically tractable---shape function to capture their radial evolution. 
This allows our framework to successfully reproduce both SIDM and CDM halo shapes
across a wide range of halo masses.

Halo shape information can be used to infer the scattering cross section and physical halo parameters.
For higher-mass halos in our sample ($\Mvir > 10^{13} \Msun$), the squashed Jeans model using halo shape data provides only modest improvement relative to the spherical Jeans model, although with the posterior distributions for $\sigmam$ less spread for the former.
However, for lower-mass halos ($\Mvir \sim 10^{12} - 10^{13} \Msun$), 
halo shapes provide a new discriminant between CDM and SIDM scenarios, while spherically averaged density profiles of the two models are otherwise indistinguishable.
For many relaxed systems, the inferred cross sections are consistent with the true simulated values, $\sigmam \approx 1 \cmg$ for SIDM and $\sigmam \ll 1 \cmg$ for CDM.
On the other hand, not all systems show an improved inference with shape data.

In this work, we have used the squashed Jeans framework to model 2D axisymmetric halo shapes with constant self-interaction cross sections. 
We defer to future work the consideration of velocity-dependent scattering, as well as the generalization to the 3D case, which follows straightforwardly by taking iso-density ellipsoids to model triaxiality.
We also defer to future work the application of the squashed Jeans model to analyze observational data, such as halo shape measurements from gravitational lensing or X-ray observations, as a fast alternative to $N$-body simulations.

\section*{Acknowledgments} 

This research was enabled in part by support provided by SciNet (\url{scinethpc.ca}) and the Digital Research Alliance of Canada (\url{alliance.can.ca}).
AS and ST thank the support of the Natural Sciences and Engineering Research Council of Canada (NSERC). The work of
YFB has been partially supported by the European Research Council under Advanced Investigator Grants ERC–AdG–885414 and  ERC–AdG–101200505. 
Lastly, we thank Brian Colquhoun for collaboration in the early stages of this project.

\clearpage

\appendix

\section{ Relaxation method  }
\label{app:relaxation}
In this appendix we describe the implementation of a general relaxation method for solving a system of coupled first order ordinary differential equations (ODEs). This method will  then be used to solve the   systems of ODEs presented in the main text.   Consider a general system of coupled ODEs taking the form
\be \label{eq:ODEs}
\mathbf{y}^\prime(x) + \mathbf{F}(\mathbf y, x, \mathbf p) = 0\,,
\ee
where $\mathbf{y}^\prime = \frac{d\mathbf y}{dx}$, and 
\be
\mathbf{y} = \left( \begin{array}{c} y_1(x) \\ \vdots \\ y_n(x) \end{array} \right)\,,
\ee
is a vector of $n$ functions, $y_1(x), \, ..., \, y_n(x)$, that we wish to solve for as a boundary value problem over the domain $x_{\rm min} \le x \le x_{\rm max}$.
Here $\mathbf{F}$ represents a similar vector, with each element an expression involving $y_1(x), \, ..., \, y_n(x)$ as well as $m$ constant parameters which we collect in the vector
\be
\mathbf{p} = \left( \begin{array}{c} p_1 \\ \vdots \\ p_m \end{array} \right) \, .
\ee
We suppose that $\mathbf{p}$ is unknown and is to be solved for along with the $y_i(x)$ functions themselves.
Hence, we require imposing $n + m$ boundary conditions on our system.

First, we define a grid of $N$ points
\be
x_1 \equiv x_{\rm min}, \, x_2, \, \dots, \, x_{N-1} , \, x_N \equiv x_{\rm max} \,,
\ee
with the boundary of our domain at the end points.
The spacing on the grid need not be uniform.
We also discretize the functions to be solved for as
\be
\mathbf{y}_i = \mathbf{y}(x_i) \, .
\ee
After discretization, we have an algebraic problem to solve for $N \times n + m$ unknowns: $\mathbf{y}_1, \, \dots,  \, \mathbf{y}_N, \, \mathbf{p}$.

Next, we write the original differential equation \eqref{eq:ODEs} in the discretized form as
\be \label{eq:ODE_disc_1}
\mathbf{E}_{i}(\mathbf{y}_{i+1}, \mathbf{y}_{i},\mathbf p) = 
\mathbf{y}_{i+1} - \mathbf{y}_i + (x_{i+1} - x_i) \,  \mathbf{F}(\bar{\mathbf{y}}_{i}, \bar{x}_{i}, \mathbf{p})
\ee
where we denote the mean values between grid points using the notation
\be
\bar{\mathbf{y}}_{i} = \frac{\mathbf{y}_{i+1} + \mathbf{y}_i}{2} \, , \quad
\bar{x}_{i} = \frac{x_{i+1} + x_i}{2} \, .
\ee
We seek the solution where $\mathbf{E}_1 = \dots = \mathbf{E}_{N-1} = 0$, which gives us $(N-1) \times n$ equations.
With $n+m$ boundary conditions, we are able (in principle) to solve for all $N \times n + m$ unknowns.
For linear equations, we are able solve the system by simple matrix inversion.
Here, however, we consider $\mathbf{F}$ to be in general a  nonlinear function of $\mathbf{y}$ and $\mathbf{p}$.

In the relaxation method proposed here, we follow an iterative approach to solving the nonlinear system of equations.
First, we take initial guess for the unknown variables $\mathbf{y}_i$, $\mathbf{p}$. 
Next, we update our guess
\be
\mathbf{y}_i \longrightarrow \mathbf{y}_i^{\rm new} = \mathbf{y}_i + \boldsymbol{\Delta} \mathbf{y}_i \, , \quad
\mathbf{p} \longrightarrow \mathbf{p}^{\rm new} = \mathbf{p} + \boldsymbol{\Delta} \mathbf{p} \, .
\ee
For suitably-chosen $\boldsymbol{\Delta} \mathbf y_i$ and $\boldsymbol{\Delta} \mathbf p$, our unknown functions and parameters relax toward their true solutions.
The process is then repeated until the desired level of convergence is achieved.

The remaining task is to determine the shifts $\boldsymbol{\Delta} \mathbf y_i$ and $\boldsymbol{\Delta} \mathbf p$ for each relaxation step.
While we cannot determine the optimal shifts
that satisfy $\mathbf{E}_i(\mathbf{y}_{i+1}^{\rm new}, \mathbf{y}_{i}^{\rm new}, \mathbf{p}^{\rm new}) = 0$ exactly, due to the nonlinear nature of the system, we can approximate them by first linearizing in the shifts.
Here it is helpful to write out all indices explicitly:
\bea
E_{ai}(\mathbf{y}_{i+1}^{\rm new}, \mathbf{y}_{i}^{\rm new}, \mathbf{p}^{\rm new}) &\approx& 
E_{ai}(\mathbf{y}_{i+1}, \mathbf{y}_{i}, \mathbf{p}) + \sum_{b,j} \frac{\partial E_{ai}}{\partial y_{bj}} \Delta y_{bj} + 
\sum_{q} \frac{\partial E_{ai}}{\partial p_{q}} \Delta p_{q} \notag \\
&\approx& 
E_{ai}(\mathbf{y}_{i+1}, \mathbf{y}_{i}, \mathbf{p}) + \Delta y_{a,i+1} - \Delta y_{ai} \notag \\
&& \qquad + \; (x_{i+1} - x_i) \, \left( \sum_{b} \frac{\partial F_{a}}{\partial y_{b}} \frac{\Delta y_{b,i+1} + \Delta y_{bi}}{2} + 
\sum_{q} \frac{\partial F_{a}}{\partial p_{q}} \Delta p_{q}  \right) \, ,
\label{eq:ODE_disc}
\eea
where $a,b = 1, \dots, n$ labels the components of $\boldsymbol{\Delta}\mathbf{y}_i$, $\mathbf{F}$, etc., and $q = 1, \dots, m$ labels the parameters, i.e., the components of $\mathbf{p}$.
Imposing Eq.~\eqref{eq:ODE_disc} to be zero, the shifts are determined by solving a linear system of equations
\be \label{eq:big_matrix_1}
\left( \begin{array}{ccccccc}
-\mathbb{1}_n + \mathbb{A}_1 & \mathbb{1}_n + \mathbb{A}_1 & \mathbb{0}_{n\times n} & \dots & \mathbb{0}_{n\times n} & \mathbb{0}_{n\times n}& \mathbb{B}_1 \\
\mathbb{0}_{n\times n} & -\mathbb{1}_n + \mathbb{A}_2 & \mathbb{1}_n + \mathbb{A}_2 & \mathbb{0}_{n\times n} & \dots & \mathbb{0}_{n\times n} & \mathbb{B}_2 \\ 
 & & \ddots & \ddots & & &  \vdots \\
 \mathbb{0}_{n\times n} & \mathbb{0}_{n\times n} &  \dots & \mathbb{0}_{n\times n} & -\mathbb{1}_n + \mathbb{A}_{N-1} & \mathbb{1}_n + \mathbb{A}_{N-1} & \mathbb{B}_{N-1} 
\end{array} \right)
\boldsymbol{\Delta}\mathbf{Y} =
- \left( \begin{array}{c}
\mathbf{E}_{1} \\
\vdots \\
\mathbf{E}_{N-1}
\end{array} \right) .
\ee
We have defined $n \times n$ matrix blocks
\be \label{eq:derivative1}
(\mathbb{A}_{i})_{ab} = \frac{1}{2} (x_{i+1} - x_i) \left. \frac{\partial F_a}{\partial y_b} \right|_{\substack{\mathbf{y} = \bar{\mathbf{y}}_{i} \\ x = \bar{x}_{i} }} 
\ee
and $n \times m$ matrix blocks
\be \label{eq:derivative2}
(\mathbb{B}_{i})_{aq} = (x_{i+1} - x_i )  \left. \frac{\partial F_a}{\partial p_q} \right|_{\substack{\mathbf{y} = \bar{\mathbf{y}}_{i} \\ x = \bar{x}_{i} }}  \, ,
\ee
and $\mathbb{1}_n$, $\mathbb{O}_{n\times n}$ are the $n \times n$ identity and zero matrices, respectively.
The shifts to be solved for are organized into a single $N\times n + m$ component vector
\be
\boldsymbol{\Delta}\mathbf{Y} =
\left( \begin{array}{c}
\boldsymbol{\Delta}\mathbf{y}_{1} \\
\vdots \\
\boldsymbol{\Delta}\mathbf{y}_{N} \\
\boldsymbol{\Delta}\mathbf{p} \\
\end{array} \right) \, .
\ee
However, it is not yet possible to solve Eq.~\eqref{eq:big_matrix_1} since we have not yet included the boundary conditions and the matrix shown is not a square matrix.

Finally, let us include the boundary conditions, expressed as
\be \label{eq:BC_form}
\mathbf{H}(\mathbf y_1, \mathbf{y}_N, \mathbf p) = 0 \, ,
\ee
where $\mathbf{H}$ is an $(n+m)$-component vector that depends only on $\mathbf{y}(x)$ at the end points.
Following a similar procedure of linearizing in the shifts, we have
\be
H_\alpha(\mathbf y_1^{\rm new}, \mathbf y_N^{\rm new}, \mathbf p^{\rm new}) \approx
H_\alpha(\mathbf y_1, \mathbf y_N, \mathbf p)  + \sum_a \left( \frac{\partial H_\alpha}{\partial y_{a1}} \Delta y_{a1} + \frac{\partial H_\alpha}{\partial y_{aN}} \Delta y_{aN} \right) + \sum_q \frac{\partial H_\alpha}{\partial p_q} \Delta p_q  \label{eq:disc_BCs}
\ee
where $\alpha = 1, \dots, n+m$ labels the components of $\mathbf{H}$.
We define the matrix blocks
\be \label{eq:disc_BCs_blocks}
(\mathbb{C}_1)_{\alpha a} =  \frac{\partial H_\alpha}{\partial y_{a1}} \, , 
\quad
(\mathbb{C}_N)_{\alpha a} = \frac{\partial H_\alpha}{\partial y_{aN}} \, ,
\quad
(\mathbb{D})_{\alpha q} = \frac{\partial H_\alpha}{\partial p_q} \, .
\ee
where $\mathbb{C}_{1,N}$ are $(n+m) \times n$ and $\mathbb{D}$ is $(n+m) \times m$.
At last, we can incorporate the boundary conditions into Eq.~\eqref{eq:big_matrix_1}, writing the full system of equations as
\be
\mathbbm{M} \, \boldsymbol{\Delta}\mathbf{Y} = - \mathbf{E}\,,
\ee
where we now have an $(nN+m) \times (nN + m)$ square matrix
\bea 
\label{eq:big_matrix_2}
\mathbbm{M} &=& \left( \begin{array}{ccccccc}
-\mathbb{1}_n + \mathbb{A}_1 & \mathbb{1}_n + \mathbb{A}_1 & \mathbb{0}_{n\times n} & \dots & \mathbb{0}_{n\times n} & \mathbb{0}_{n\times n}& \mathbb{B}_1 \\
\mathbb{0}_{n\times n} & -\mathbb{1}_n + \mathbb{A}_2 & \mathbb{1}_n + \mathbb{A}_2 & \mathbb{0}_{n\times n} & \dots & \mathbb{0}_{n\times n} & \mathbb{B}_2 \\ 
 & & \ddots & \ddots & & &  \vdots \\
 \mathbb{0}_{n\times n} & \mathbb{0}_n &  \dots & \mathbb{0}_{n\times n} & -\mathbb{1}_n + \mathbb{A}_{N-1} & \mathbb{1}_n + \mathbb{A}_{N-1} & \mathbb{B}_{N-1}\\
\mathbb{C}_1 & \mathbb{0}_{(n+m) \times n} & \dots & \dots & \mathbb{0}_{(n+m) \times n} & \mathbb{C}_N & \mathbb{D} 
\end{array} \right)
\begin{array}{l} 
\} n \; {\rm rows} \\[1pt]
\} n \; {\rm rows} \\[1pt]
\vdots \\[1pt]
\} n \; {\rm rows} \\[1pt]
\} n+m \; {\rm rows} \\[1pt]
\end{array} \\
&& \qquad \;
\underbrace{\qquad}_{n \, {\rm cols}} \quad \;\;\;\;
\underbrace{\qquad}_{n \, {\rm cols}} \quad \;\;\;\;\;
\underbrace{\qquad}_{n \, {\rm cols}} \quad
\;\;\dots\;\; \quad \;\;\;
\underbrace{\qquad}_{n \, {\rm cols}} \quad \;\;\;\;\;\;\;\;
\underbrace{\qquad}_{n \, {\rm cols}} \quad \;\;\;\;\;
\underbrace{\qquad}_{m \, {\rm cols}} 
\notag
\eea
and the $(nN + m)$-component vector
\be \label{eq:big_E}
\mathbf{E} =
\left( \begin{array}{c}
\mathbf{E}_{1} \\
\vdots \\
\mathbf{E}_{N-1} \\
\mathbf{H} \\
\end{array} \right) \, .
\ee

For each iteration, one takes an initial guess for $\mathbf{y}_i$, $\mathbf{p}$ and computes $\mathbbm{M}$ and $\mathbf{E}$.
Then, the shifts are determined by
\be \label{eq:update}
\boldsymbol{\Delta} {\mathbf Y} = - \mathbbm{M}^{-1} \mathbf{E} \,,
\ee
and the solution is updated.
The process is repeated until convergence is achieved. 
If overshooting is a concern, Eq.~\eqref{eq:update} can be modified to $\boldsymbol{\Delta} {\mathbf Y} = - \kappa \mathbbm{M}^{-1} \mathbf{E}$
with $\kappa < 1$.
In our work, we find $\kappa = 1$ suffices.

\section{Spherical isothermal Jeans model}

\label{app:spherical}

Here we discuss our numerical implementation of the spherical Jeans model, both with and without baryons using the relaxation method described in Appendix \ref{app:relaxation} to solve the respective differential equations. 

\subsection{Case without baryons}

A spherically-symmetric dark matter-only profile can be solve numerically and  serves therefore as a initial guess for  relaxation profiles in the case in which baryons and nonspherical effects are introduced. The Jeans and Poisson's equations 
Eqs.~\eqref{eq:Jeans_sph} and \eqref{eq:Poisson} respectively,  yield two first-order equations
\be \label{eq:jeans_no_baryons}
\frac{ \partial \rho_{\rm dm} }{\partial r} = - \frac{G \rho_{\rm dm} }{ \sigma_0^2 r^2} M_{\rm dm}  \, , \quad
\frac{\partial M_{\rm dm}}{\partial r} = 4\pi r^2 \rho_{\rm dm} \, ,
\ee
for the dark matter density $\rho_{\rm dm}(r)$ and enclosed mass $M_{\rm dm}(r)$.
We can further  express Eq.~\eqref{eq:jeans_no_baryons} as a system of first-order equations for two new functions $h(x)$, $j(x)$ by a change of variables
\be
\rho_{\rm dm}(r) = \rho_0 e^{h(x)} \, ,
\quad 
M_{\rm dm}(r) = \frac{4\pi}{3} \rho_0 r^3 e^{j(x)} \, ,
\quad
x = r/r_0 \, ,
\ee
with central density $\rho_0 = \rho_{\rm dm}(0)$ and characteristic radius $r_0 = \sigma_0/\sqrt{4\pi G \rho_0}$.
We have then
\be \label{eq:jeans_no_baryons_2}
\frac{\partial h}{\partial x} + \frac{x}{3} e^{j(x)} = 0 \, , \quad 
\frac{\partial j}{\partial x} + \frac{3}{x} \left( 1 - e^{h(x) - j(x)}\right) = 0 \, ,
\ee
subject to the boundary conditions $h(0) = j(0) = 0$.\footnote{In practice, it is useful to implement the boundary conditions at a small but nonzero value $x$. Noting the asymptotic solutions $h(x) \approx - \tfrac{1}{6} x^2$ and $j(x) \approx - \tfrac{1}{10} x^2$ for $x \ll 1$, we set $h(\epsilon) = - \tfrac{1}{6} \epsilon^2$ and $j(\epsilon) = - \tfrac{1}{10} \epsilon^2$ for $x = \epsilon \ll 1$.}
Solutions to Eq.~\eqref{eq:jeans_no_baryons_2} are universal in that they do not depend on additional inputs for $\rho_{\rm dm}$ and $M_{\rm dm}$, and it suffices to tabulate numerical values for $h(x)$, $j(x)$ once and for all.

Next, we translate back into the physical profiles $\rho_{\rm dm}(r)$, $M_{\rm dm}(r)$.
If $\rho_0$ and $r_0$ (or $\sigma_0$) are given, this is a straightforward task.
Here, however, we prefer to implement an ``outside-in'' matching by fixing the boundary conditions at the matching radius $r_m$\footnote{To avoid notational confusion, we define $r_m$ to be the matching radius that is also denoted $r_1$ in the literature. Here we use $r_1$ to label the first grid point in the relaxation method.}
\be
\rho_{\rm dm}(r_m) = \rho_m \, , \quad
M_{\rm dm}(r_m) = M_m \, ,
\ee
where $\rho_m = \rho_{\rm CDM}(r_m)$, $M_m = M_{\rm CDM}(r_m)$ are given, determined by the CDM profile of the outer halo.
Defining the ratio
\be
\mathcal{R} = \frac{M_m}{4\pi r_m^3 \rho_m} \, ,
\ee
the matching condition at $r_m$ can be expressed as a single condition
\be \label{eq:ratio_matching}
\frac{M_{\rm dm}(r_m)}{4\pi r_m^3 \rho_{\rm dm}(r_m)} = \frac{1}{3} e^{j(x_m) - h(x_m)} = \mathcal{R} \, ,
\ee
where $x_m = r_m/r_0$. 
Eq.~\eqref{eq:ratio_matching} is solved numerically to determine $x_m$.\footnote{For some values of $\mathcal{R}$, multiple solutions can be found.  In this case, we take the smallest value of $x_m$.  
For $\mathcal{R} > \mathcal{R}_{\rm max} \approx 1.2615$, no solution for $x_m$ can be found, i.e., the matching conditions cannot be satisfied.  
In this case, we replace $\mathcal{R} \to \mathcal{R}_{\rm max}$ in Eq.~\eqref{eq:ratio_matching}, which yields $x_m \approx 22.544$, for the purposes of obtaining an initial guess for the relaxation method.}
Finally, we fix the unknown parameters according to
\be
r_0 = r_m/x_m \, , \quad
\rho_0 = \rho_m e^{-h(x_m)} \, , \quad
\sigma_0 = \sqrt{4\pi G \rho_0 r_0^2} \, .
\ee

\subsection{Case with baryons}
As a warm up, we illustrate the relaxation method described in Appendix \ref{app:relaxation} by considering the Jeans model with baryons under the assumption of spherical symmetry.
The equations to be solved are obtained again from Eqs.~\eqref{eq:Jeans_sph} and \eqref{eq:Poisson}: 
\be \label{eq:jeans_sph_baryons}
\frac{ \partial \rho_{\rm dm} }{\partial r} = - \frac{G \rho_{\rm dm} }{ \sigma_0^2 r^2} \left( M_{\rm dm} + M_b \right) \, , \quad
\frac{\partial M_{\rm dm}}{\partial r} = 4\pi r^2 \rho_{\rm dm} \, ,
\ee
where we take the baryon enclosed mass profile $M_b(r)$ to be a known function.
If the baryons are spherically symmetric, their potential $\Phi_b$ is a function of $r$ only and is determined by $\frac{\partial \Phi_b}{\partial r} = G M_b/r^2$ up to an integration constant.
It is again useful to make a change of variables
\be
\rho_{\rm dm}(r) = \rho_0 e^{-\phi_{\rm dm}(r)- \phi_b(r)} \, , \quad
M_{\rm dm}(r) = \frac{4\pi}{3} \rho_0 r^3 e^{-\eta_{\rm dm}(r)}\,,
\ee
to define two new functions $\phi_{\rm dm}(r)$, $\eta_{\rm dm}(r)$, as well as the normalized baryon potential $\phi_b(r) = (\Phi_b(r) - \Phi_b(0))/\sigma_0^2$.

More generally, one can apply the spherical Jeans model to {\it nonspherical} baryon distributions as via the spherical averaged of the baryon potentials, that is, 
the baryon enclosed mass profile is related to the angular average of the potential
\be
\frac{\partial}{\partial r} \int \frac{d\Omega}{4\pi} \, \Phi_b(\mathbf r) = \frac{G M_b(r)}{r^2} \, .
\ee
Hence, one can  take the normalized spherically-averaged  baryon distribution to be
\be
\phi_b(r) =\int \frac{d\Omega}{4\pi} \left( \frac{\Phi_b(\mathbf r) - \Phi_b(0)}{\sigma_0^2} \right) \, .
\ee

After some manipulation, we write Eq.~\eqref{eq:jeans_sph_baryons} in the general form~\eqref{eq:ODEs} with
\be
\mathbf{y} = \left( \begin{array}{c} \phi_{\rm dm} \\ \eta_{\rm dm} \end{array} \right) \, , \quad
\mathbf{F}(\mathbf y,r,\mathbf p) = \left( \begin{array}{c} - \frac{r}{3 r_0^2} e^{-\eta_{\rm dm}} \\ - \frac{3}{r}(1 - e^{\eta_{\rm dm} - \phi_{\rm dm} - \phi_b}) \end{array} \right),
\ee
with unknown parameters
\be
\mathbf{p} = \left( \begin{array}{c} r_0^2 \\ \sigma_0^2 \end{array} \right)\,,
\ee
that we will determine along with $\phi_{\rm dm}$ and $\eta_{\rm dm}$. (Note that $\sigma_0^2$ appears implicitly through $\phi_b$.)
To determine our two unknown functions $\phi_{\rm dm}$, $\eta_{\rm dm}$ and two parameters, we impose four boundary conditions. 
At
$r=0$, we have
\be\label{eq:bc1}
\rho_{\rm dm}(0) = \rho_0 \, , \quad
M_{\rm dm}(0) = 0 \, ,
\ee
which becomes $\phi_{\rm dm}(0) = \eta_{\rm dm}(0) = 0$. At a given matching radius $r_m$, we have
\be\label{eq:bc2}
\rho_{\rm dm}(r_m) = \rho_m \, , \quad
M_{\rm dm}(r_m) = M_m \, ,
\ee
where $\rho_m$, $M_m$ are assumed to be known, determined by fixing the outer halo.

To implement the relaxation method, we take a discrete grid of $N$ points $r_i$, spanning from $r_1 = 0$ to $r_N = r_m$.
We define our unknown functions on the grid
\be
\phi_i = \phi_{\rm dm}(r_i) \, , \quad 
\eta_i = \eta_{\rm dm}(r_i) \, ,
\ee
dropping the ``${\rm dm}$'' subscript for simplicity.
As our initial guess, we take the density and enclosed mass profiles, as well as the parameters, to be as found from the spherical Jeans model without baryons
\be
\phi_i = - h(r_i/r_0) - \phi_b(r_i) \, , \quad
\eta_i = - j(r_i/r_0) \, ,
\ee
where the functions $h(x)$, $j(x)$ are solutions to Eq.~\eqref{eq:jeans_no_baryons_2}.

Next, to iterate our solution, we construct the matrix $\mathbbm{M}$ and vector $\mathbf{E}$, defined in Eqs.~\eqref{eq:big_matrix_2} and \eqref{eq:big_E}, respectively.
These are built from the matrix blocks~\eqref{eq:derivative1} and \eqref{eq:derivative2}
\bea
\mathbb{A}_i &=& \frac{1}{2} (r_{i+1} - r_i) \left( \begin{array}{cc} 0 & 
\frac{\bar{r}_{i}}{3 r_0^2} e^{-\bar{\eta}_{i}} \\
- \frac{3}{\bar{r}_{i}} e^{\bar{\eta}_{i} - \bar{\phi}_{i} - \phi_b(\bar{r}_{i})} &
\frac{3}{\bar{r}_{i}} e^{\bar{\eta}_{i} - \bar{\phi}_{i} - \phi_b(\bar{r}_i)} \end{array} 
\right) \,,\\
\mathbb{B}_i &=& (r_{i+1} - r_i) \left( \begin{array}{cc}
\frac{\bar{r}_i}{3 r_0^4} e^{-\bar{\eta}_{i}} &
0 \\
0 &
\frac{3}{\bar{r}_{i}\sigma_0^2} \phi_b(\bar{r}_i) e^{\bar{\eta}_{i} - \bar{\phi}_{i} - \phi_b(\bar{r}_i)} \end{array} \right) \, ,
\eea
and the vectors~\eqref{eq:ODE_disc_1}
\be
\mathbf{E}_i = 
\left( \begin{array}{c} \phi_{i+1} - \phi_i 
- (r_{i+1} - r_i) \frac{\bar{r}_i}{3 r_0^2} e^{-\bar{\eta}_{i}} \\
\eta_{i+1} - \eta_i - (r_{i+1} - r_i) \frac{3}{\bar{r}_i}(1 - e^{\bar\eta_{i} - \bar\phi_{i} - \phi_b(\bar r_i}) 
\end{array} \right) \, ,
\ee
for $i = 0, \dots, N-1$.
We also put the four boundary conditions in Eqs. \eqref{eq:bc1} and \eqref{eq:bc2}, in  the form \eqref{eq:BC_form}
\be
\mathbf{H}(\mathbf y_1, \mathbf y_N, \mathbf p) = \left( \begin{array}{c} \phi_1  \\ \eta_1 \\
\rho_0 e^{-\phi_N - \phi_b(r_N)} - \rho_m \\
\frac{4\pi}{3} \rho_0 r_N^3 e^{-\eta_N} - M_m \end{array} \right) = 0\,,
\ee
where $\rho_0 = \sigma_0^2/(4\pi G r_0^2)$ is a function of $\mathbf{p}$.
The corresponding $4 \times 2$ matrix blocks~\eqref{eq:disc_BCs_blocks} are
\bea
\mathbb{C}_1 &=& \left( \begin{array}{c} \mathbb{1}_{2} \\ \mathbb{0}_{2 \times 2} \end{array} \right) \, , \quad 
\mathbb{C}_N = \left( \begin{array}{c} \mathbb{0}_{2 \times 2} \\ \hline 
\begin{array}{cc} - \rho_0 e^{-\phi_N - \phi_b(r_N)} & 0 \\
0 & - \frac{4\pi}{3} \rho_0 r_N^3 e^{-\eta_N} \end{array} \end{array} \right)\,, \\
\mathbb{D} &=& \left( \begin{array}{c} \mathbb{0}_{2 \times 2} \\ \hline 
\begin{array}{cc} - \frac{\rho_0}{r_0^2} e^{-\phi_N - \phi_b(r_N)} & \frac{\rho_0}{\sigma_0^2} (1+\phi_b(r_N) ) \, e^{-\phi_N - \phi_b(r_N)} \\
- \frac{4\pi\rho_0 r_N^3}{3r_0^2} e^{-\eta_N}  &  \frac{4\pi\rho_0 r_N^3}{3\sigma_0^2} e^{-\eta_N} \end{array} \end{array} \right) \, ,
\eea
each of which we have divided into two $2\times 2$ subblocks with a horizontal line for clarity.

\section{Nonspherical isothermal Jeans model}
\label{app:nonspherical}

In this appendix we present our method for solving the nonspherical isothermal Jeans model, which describes a dark matter halo in hydrostatic equilibrium within a given matching radius $r_m$.
As discussed in the main text, this model has issues. Namely, it predicts a sharp feature in the halo shape at $r_m$, as hydrostatic equilibrium is strictly enforced within this radius, whereas numerical simulations for SIDM do not exhibit this feature.
This suggests that hydrostatic equilibrium is reached gradually across the matching radius, which is built into the preferred squashed Jeans model described in the main text.

Despite its failure, it is useful to provide our numerical method for the nonspherical Jeans model, taking advantage of the relaxation method introduced in Appendix \ref{app:relaxation}, since,  first, it represents the asymptotic limit for the squashed Jeans model, which is used as an input, and second, this model connects to previous work in the literature for modeling isothermal dark matter halos.

This appendix is divided in three parts: First we describe the setup and equations to be solved, continuing the discussion from Sec.~\ref{sec:nonspherical_jeans}.
Then, we discuss our numerical algorithm using the relaxation method.
Finally, we present an analytic formula to describe the isothermal halo shapes in the limit of hydrostatic equilibrium.

\subsection{Setup and boundary conditions}

For nonspherical halos, the dark matter density for the inner profile is given by the solution to the Jeans equation~\eqref{eq:hydro_equilib}
\be \label{eq:jeans_solution}
\rho_{\rm dm}(\mathbf r) = \rho_0 \, e^{- \phi_{\rm dm}(\mathbf r) - \phi_b(\mathbf r)} \, ,
\ee
expressed in terms of the dimensionless rescaled potentials for dark matter and baryons~\eqref{eq:dimless_pot}.
The former is expanded as
\be \label{eq:phi_dm_sph_exp}
\phi_{\rm dm}(\mathbf r)  = \sum_{\ell=0}^{\ell_{\rm max}} \sum_{m=-\ell}^\ell \phi_{\ell m}(r) \, Z_{\ell m}(\theta,\varphi) \, ,
\ee
truncated at a fixed order $\ell_{\rm max}$, in terms of 
real-valued spherical harmonics, (also known as tesseral harmonics) $Z_{\ell m}(\theta,\varphi)$. These are defined by 
\be
Z_{\ell m}(\theta,\varphi) = \left\{ \begin{array}{ll} 
\frac{1}{\sqrt{2}} \big( Y_\ell^{-m}(\theta,\varphi) + (-1)^m Y_\ell^m(\theta,\varphi) \Big) & m>0 \,,\\
Y_\ell^0(\theta,\varphi) & m=0 \,,\\
\frac{i}{\sqrt{2}} \big( Y_\ell^m(\theta,\varphi) - (-1)^m Y_\ell^{-m}(\theta,\varphi) \big) & m < 0\,,
\end{array}\right. 
\ee 
where $Y_{\ell}^m$ are the scalar spherical harmonics.
They are orthonormal functions
\be
\int d\Omega \, Z_{\ell m}(\theta,\varphi) Z_{\ell^\prime m^\prime}(\theta,\varphi) = \delta_{\ell \ell^\prime} \delta_{m m^\prime} \, ,
\ee
and can be expressed as
\be
Z_{\ell m}(\theta,\varphi) = \left\{ \begin{array}{ll} \sqrt{ \frac{2(2\ell + 1)}{4\pi}  } \sqrt{ \frac{(\ell - m)!}{(\ell+m)!}} P_\ell^m (\cos\theta) \cos( m \varphi) & m>0\,, \\
 \sqrt{ \frac{ 2\ell + 1}{4\pi}  } P_\ell^0(\cos\theta) & m=0\,, \\
\sqrt{ \frac{2(2\ell + 1)}{4\pi}  } \sqrt{ \frac{(\ell - |m|)!}{(\ell+|m|)!}} P_\ell^{|m|} (\cos\theta) \sin( |m| \varphi) & m<0\,,
\end{array} \right. \label{Zlm}
\ee
where the associated Legendre polynomials are
\be
P_\ell^m(x) =  (1-x^2)^{m/2} \frac{d^m}{{dx}^m} P_\ell(x)\,.
\ee
where $P_\ell$ are the standard Legendre polynomials. Note we have defined $P_\ell^m$ without the Condon-Shortley phase $(-1)^m$, which is a more standard convention. In that case, we must include an additional factor of $(-1)^m$ in Eq.~\eqref{Zlm}.

In terms of the spherical harmonics, Poisson's equation~\eqref{eq:main2} is
\be
\frac{\partial^2 \phi_{\ell m}}{\partial r^2} + \frac{2}{r} \frac{ \partial \phi_{\ell m}}{\partial r} - \frac{\ell (\ell + 1)}{r^2} \phi_{\ell m} 
=  \left( \frac{4\pi G \rho_0}{\sigma_0^2} \right) \int d\Omega \, Z_{\ell m}(\theta,\varphi) \,  e^{-\phi_b(\mathbf r)} \, e^{-\phi_{\rm dm}(\mathbf r)} \, .\label{eq:mainPDE2}
\ee
Thus we arrive at system of nonlinear second-order ordinary differential equations for $\phi_{\ell m}(r)$, with two additional parameters $\rho_0$ and $\sigma_0$.

To implement the relaxation method discussed in Appendix \ref{app:relaxation}, we require a system of first-order equations. 
Let us define
\be 
\mu_{\ell m}(r) = r^2 \frac{\partial \phi_{\ell m}}{\partial r} \, .
\ee
We note that $\mu_{00}$ is proportional to the enclosed mass within a sphere of radius $r$,
\be
\mu_{00}(r) = \frac{\sqrt{4\pi} G}{\sigma_0^2} M_{\rm dm}(r) \, .
\ee
Then Eq.~\eqref{eq:mainPDE2} can be rewritten in terms of first order ODEs only,  
\bea \label{eq:dphi_dr}
\frac{\partial \phi_{\ell m}}{\partial r} + \mathcal{G}_{\ell m}(\mathbf{y}, r, \mathbf p) &=& 0 \,,\\
\frac{\partial \mu_{\ell m}}{\partial r} + \mathcal{F}_{\ell m}(\mathbf{y}, r, \mathbf p) &=& 0 \,  ,\label{eq:dmu_dr}
\eea
where the functions $\mathcal{F}_{\ell m}$ and $\mathcal{G}_{\ell m}$ are 
\bea
\mathcal{F}_{\ell m}(\mathbf{y}, r, \mathbf p) &=& - \ell(\ell+1) \phi_{\ell m} - \frac{r^2}{r_0^2} \int d\Omega \, Z_{\ell m}(\theta,\varphi) \,   e^{-\phi_b(\mathbf r)} \, e^{-\phi_{\rm dm}(\mathbf r)}\,, \label{eq:F}\\
\mathcal{G}_{\ell m}(\mathbf{y}, r, \mathbf p) &=& - \frac{\mu_{\ell m}}{r^2} \, ,
\label{eq:G}
\eea
with $r_0 = \sigma_0/\sqrt{4\pi G \rho_0}$. We have organized the unknown functions and parameters into two vectors
\be
\mathbf{y} = \left( \begin{array}{c} \phi_{00} \\ \phi_{10} \\ \vdots \\ \mu_{00} \\ \mu_{10} \\ \vdots \end{array} \right) \, , \quad \mathbf{p} = \left(\begin{array}{c} r_0^2 \\ \sigma_0^2 \end{array} \right) \, .
\ee
In total, we have $2 n_{\ell m}$ functions to solve for, where $n_{\ell m}$ is the number of $(\ell,m)$-modes, along with two parameters $r_0^2$, $\sigma_0^2$. (Note $\sigma_0^2$ appears implicitly in Eq.~\eqref{eq:F} through $\phi_b$.)
Therefore, a total of $2 n_{\ell m} + 2$ boundary conditions are required.
In cases with azimuthal symmetry, it suffices to neglect $m\ne 0$ modes and so $n_{\ell m} = \ell_{\rm max} + 1$, while in the general case $n_{\ell m} = (\ell_{\rm max} + 1)^2$.

Next, we turn to the boundary conditions, beginning with the higher harmonics $\ell > 0$.
It is well-known that the solution to Poisson's equation in spherical coordinates can be expressed in a  multipole decomposition.
In terms of the functions defined here, this is
\bea
\phi_{\ell m}(r) &=& - \frac{4\pi G}{\sigma_0^2} \frac{1}{2\ell + 1} \left( \frac{1}{r^{\ell+1}} \int_0^r dx \, x^{\ell+ 2} \rho_{\ell m}(x) + r^{\ell} \int_r^\infty dx \, x^{1-\ell} \rho_{\ell m}(x) \right) 
\label{eq:phi_lm}\\
\mu_{\ell m}(r) &=& \frac{4\pi G}{\sigma_0^2} \frac{1}{2\ell + 1} \left( \frac{\ell + 1}{r^\ell} \int_0^r dx \, x^{\ell+ 2} \rho_{\ell m}(x) - \ell r^{\ell + 1} \int_r^\infty dx \, x^{1-\ell} \rho_{\ell m}(x) \right) \, ,
\label{eq:mu_lm}
\eea
where $\rho_{\ell m}(r) = \int d\Omega \, Z_{\ell m}(\theta,\varphi) \, \rho_{\rm dm}(\mathbf r)$.
This is merely a formal solution since $\rho_{\rm dm}$ itself depends on $\phi_{\ell m}$ according to Eqs.~\eqref{eq:jeans_solution} and~\eqref{eq:phi_dm_sph_exp}.
The key point is these solutions have physically sensible asymptotic behaviors, scaling like $\phi_{\ell m} \propto r^\ell$ at the origin and $r^{-(\ell+1)}$ at infinity, so that the potential does not diverge in both limits.
We therefore require 
\be \label{eq:boundarycondition_0}
\phi_{\ell m}(0) = 0 \, ,
\ee
as well as $\phi_{\ell m}(\infty) = 0$.
The latter is satisfied automatically in Eqs.~\eqref{eq:phi_lm}, which we can impose on our solutions as a boundary condition.
It is useful take a linear combination Eqs.~\eqref{eq:phi_lm} and \eqref{eq:mu_lm}, 
evaluated at $r=r_m$,
\be \label{eq:boundarycondition_rm}
\mu_{\ell m}(r_m) + (\ell+1) r_m \phi_{\ell m}(r_m) = - \frac{4\pi G}{\sigma_0^2} r_m^{\ell + 1} J_{\ell m} \, ,
\ee
where $J_{\ell m} = \int_{r_m}^\infty dx \, x^{1-\ell} \rho_{\ell m}(x)$.
Since the integral $J_{\ell m}$ goes over $r > r_m$ only, it is fixed given the outer halo profile $\rho_{\rm cdm}$.
We emphasize that the two conditions \eqref{eq:boundarycondition_0} and \eqref{eq:boundarycondition_rm} are essential.
Different prescriptions not satisfying these conditions will lead to solutions differing by homogeneous terms $A r^\ell + B/r^{\ell+1}$, which would have the incorrect asymptotic behavior.\footnote{Boundary condition~\eqref{eq:boundarycondition_rm} implies that the potential is continuous between the inner and outer halos, i.e., everywhere at the matching sphere of radius $r_m$. 
However, the dark matter density $\rho_{\rm dm}(\mathbf r)$ is {\it not} continuous at $r_m$; only the spherically-averaged profile is continuous according to Eq.~\eqref{eq:matching_rm_2} 
We have attempted an alternative boundary condition imposing that the dark matter density is continuous at the matching sphere. However, this choice is not physical, as it leads to potentials with divergent asymptotic behavior as $r \to \infty$ from additional homogeneous terms $\sim r^{\ell}$.}

Lastly, we turn to the $\ell=m=0$ mode.
In the same spirit as the spherical Jeans model, we 
impose $\rho_{\rm dm}(0) = \rho_0$ and $M_{\rm dm}(0) = 0$, which give boundary conditions
\be \label{eq:boundary_condition_00}
\phi_{00}(0) = 0, \quad
\mu_{00}(0) = 0 \, .
\ee
We also match the enclosed mass and spherically-averaged density for the inner and outer halos at $r=r_m$.
This yields boundary conditions
\be \label{eq:boundary_conditions_mu_00}
\mu_{00}(r_m) = \frac{\sqrt{4\pi} G}{\sigma_0^2} M_m \, ,
\ee
\be
\label{eq:boundary_conditions_rho}
\int \frac{d\Omega}{4\pi} \, \rho_0 \, \exp\left( -\sum_{\ell m} \phi_{\ell m}(r_m) Z_{\ell m}(\Omega) - \phi_b(r_m,\Omega)\right) = 
\rho_m\,,
\ee
where the enclosed mass and spherically-averaged density at $r_m$ are
\be
M_m = M_{\rm cdm}(r_m) \, , \quad
\rho_m = \int \frac{d\Omega}{4\pi} \, \rho_{\rm cdm}(r_m, \Omega) \, .
\ee
Lastly, we check that our boundary conditions are consistent with physically-sensible solutions to Poisson's equation.
Note that form of Eq.~\eqref{eq:phi_lm} is modified in the $\ell=m=0$ case since the potential at the origin is subtracted out in Eq.~\eqref{eq:dimless_pot}.
The solutions are
\bea
\phi_{00}(r) &=& - \frac{4\pi G}{\sigma_0^2} \int_0^r dx \, \left( \frac{x^2}{r} - x \right) \rho_{00}(x) 
\\
\mu_{00}(r) &=& \frac{4\pi G}{\sigma_0^2}  \int_0^r dx \, x^2 \rho_{00}(x) \, .
\eea
At the origin, these are consistent with Eq.~\eqref{eq:boundary_condition_00}.
We may also consider a linear combination along the lines of Eq.~\eqref{eq:boundarycondition_rm}, arriving at a different constraint
\be \label{eq:not_a_bc}
\mu_{00}(r_m) + r_m \phi_{00}(r_m) = \frac{4\pi G}{\sigma_0^2} r_m \int_0^{r_m} dx \, x \, \rho_{00}(x)  \, .
\ee
In fact, it is unnecessary to impose this condition as it is redundant with Poisson's equation.
From the equations defined above, one can show that
\be
\frac{\partial }{\partial r} \left( \frac{\mu_{00}(r)}{r} + \phi_{00}(r) \right) = \frac{4\pi G}{\sigma_0^2} r \rho_{00}(r) \, .
\ee
Integrating both sides from $r=0$ to $r_m$ yields Eq.~\eqref{eq:not_a_bc}, provided $\mu_{00}(0) = 0$.

In summary, Eqs.~\eqref{eq:boundarycondition_0}, \eqref{eq:boundarycondition_rm}, \eqref{eq:boundary_condition_00}, \eqref{eq:boundary_conditions_mu_00}, \eqref{eq:boundary_conditions_rho} are the full set of
\be
(n_{\ell m} - 1) + (n_{\ell m} - 1) + 2 + 1 + 1  = 2 n_{\ell m} + 2
\ee
boundary conditions to determine our system.
Recall we started with $n_{\ell m}$ coupled second-order equations, plus two additional parameters to be determined.
We also note that due to the boundary conditions on the higher harmonics, the nonspherical isothermal Jeans model cannot be solved as an initial value problem and outside-in matching becomes a natural choice. 
We fix the outer halo $\rho_{\rm cdm}(\mathbf r)$ and solve Eqs.~\eqref{eq:dphi_dr} and \eqref{eq:dmu_dr} as a boundary value problem by imposing constraints at $r=0$ and $r_m$ using the relaxation method, as we now  describe.

\subsection{Relaxation method}

To implement the relaxation method introduced in Appendix \ref{app:relaxation}, to solve the nonspherical isothermal Jeans model,   we take a grid of $N$ points $r_i$,  from $r_1 = 0$ to $r_N = r_m$.
We denote our unknown functions on the grid as
\be
\phi_{\ell m}^i = \phi_{\ell m}(r_i) \, , \quad \mu_{\ell m}^i = \mu_{\ell m}(r_i) \, .
\ee
To solve for them, we follow a two step procedure.
First, we solve for $\phi_{00}, \mu_{00}$ in the nonspherical Jeans model truncated at $\ell_{\rm max} = 0$.\footnote{In general, the nonspherical Jeans model truncated at $\ell_{\rm max} = 0$ is {\it not} equivalent to the spherical Jeans model (as we have defined them).
For a general baryon potential $\Phi_b(\mathbf r)$, the former depends on $\int d\Omega e^{-\phi_b(\mathbf r)}$ in Eq.~\eqref{eq:F}, while the latter depends on the angular average of $\Phi_b$ itself.}
Our initial guess takes the same parameters $r_0^2, \sigma_0^2$, as well as the same spherically-averaged density and enclosed mass, to the spherical Jeans model without baryons.
Specifically, we set
\be
\phi_{00}^i = - h(r_i/r_0) - \log\left( \int \frac{d\Omega}{4\pi} \, e^{-\phi_b(r_i,\Omega)} \right) \, , 
\quad
\mu_{00}^i = - j(r_i/r_0) \, ,
\ee
where the functions $h(x)$, $j(x)$ are solutions to Eq.~\eqref{eq:jeans_no_baryons_2}. Then we  perform the relaxation algorithm until the solution converges.

Second, we use the preceding solution as an initial guess for the nonspherical Jeans model with $\ell_{\rm max} > 0$, taking $\phi_{00}^i, \mu_{00}^i, r_0^2, \sigma_0^2$ as solved for above. 
The remaining modes with $\ell > 0$ are initially set to $\phi_{\ell m}^i = \mu_{\ell m}^i = 0$.
Once again, we perform the relaxation algorithm until the solution converges.

In the remainder of this section, we tabulate the matrix blocks and vectors entering Eqs.~\eqref{eq:big_matrix_2} and \eqref{eq:big_E} for our algorithm.
Following Eq.~\eqref{eq:derivative1}, we have the $2n_{\ell m} \times 2n_{\ell m}$ blocks
\be
\mathbb{A}_i = \frac{1}{2} \left( r_{i+1} - r_i \right) \left( \begin{array}{c|c} 
\mathbb{0}_{n_{\ell m}\times n_{\ell m}}
&
\begin{array}{ccc}
\frac{\partial \mathcal{G}_{00}}{\partial\mu_{00}} & 0 & 0 \\[5pt] 
0 & \frac{\partial \mathcal{G}_{10}}{\partial\mu_{10}} & 0  \\
0 & 0 & \ddots 
\end{array} \\[5pt]
\hline
\\[-10pt]
\begin{array}{ccc}
\frac{\partial \mathcal{F}_{00}}{\partial \phi_{00}} & \frac{\partial \mathcal{F}_{00}}{\partial \phi_{10}} & \dots \\[5pt] 
\frac{\partial \mathcal{F}_{10}}{\partial \phi_{00}} & \frac{\partial \mathcal{F}_{10}}{\partial \phi_{10}} & \dots \\[5pt]
\vdots & \vdots & \ddots 
\end{array}
&
\mathbb{0}_{n_{\ell m}\times n_{\ell m}}
\end{array}
\right)_{\substack{\mathbf{y} = \bar{\mathbf{y}}_{i} \\ r = \bar{r}_{i} }}\,,
\ee
where 
\bea
\left. \frac{\partial \mathcal{F}_{\ell m}}{\partial \phi_{\ell' m'}} 
\right|_{\substack{\mathbf{y} = \bar{\mathbf{y}}_{i} \\ r = \bar{r}_{i} }}
&=& 
- \ell(\ell + 1) \delta_{\ell \ell'} \delta_{m m'} +  \frac{\bar{r}_i^2 }{ r_0^2}  \int d\Omega \,  Z_{\ell m}(\Omega) Z_{\ell' m'}(\Omega) \, e^{- \phi_b(\bar{r}_i,\Omega) - \bar\phi_{\rm dm}^{i}(\Omega) }\,,
\\
\left. \frac{\partial \mathcal{G}_{\ell m}}{\partial \mu_{\ell' m'}} \right|_{\substack{\mathbf{y} = \bar{\mathbf{y}}_{i} \\ r = \bar{r}_{i} }}
&=& 
- \frac{1}{\bar{r}_i^2} \delta_{\ell \ell'} \delta_{m m'} 
 \, .
\eea
Here it is helpful to denote mean values between grid points as
\be
\bar{\phi}_{\ell m}^i = \frac{\phi_{\ell m}^{i+1} + \phi_{\ell m}^{i}}{2} \, , \quad \bar{\mu}_{\ell m}^i = \frac{\mu_{\ell m}^{i+1} + \mu_{\ell m}^i}{2} \, , \quad
\bar{r}_i = \frac{r_{i+1} + r_i}{2} \, ,
\ee
and similarly
\be
\bar{\phi}_{\rm dm}^i(\Omega) = \sum_{\ell,m} Z_{\ell m}(\theta,\varphi) \, \bar{\phi}_{\ell m}^i \, .
\ee
From Eq.~\eqref{eq:derivative2}, we have $2n_{\ell m} \times 2$ blocks
\be
\mathbb{B}_i = (r_{i+1}-r_i)\left( \begin{array}{c} 
\mathbb{0}_{n_{\ell m} \times 2} 
\\
\hline
\\[-10pt]
\begin{array}{cc}
\frac{\partial \mathcal{F}_{00}}{\partial r_0^2} &
\frac{\partial \mathcal{F}_{00}}{\partial \sigma_0^2} \\[5pt]
\frac{\partial \mathcal{F}_{10}}{\partial r_0^2} &
\frac{\partial \mathcal{F}_{10}}{\partial \sigma_0^2} \\[5pt]
\vdots & \vdots 
\end{array}
\end{array}
\right)_{\substack{\mathbf{y} = \bar{\mathbf{y}}_{i} \\ r = \bar{r}_{i} }}\,,
\ee
where
\bea
\left.
\frac{\partial \mathcal{F}_{\ell m}}{\partial r_0^2} 
\right|_{\substack{\mathbf{y} = \bar{\mathbf{y}}_{i} \\ r = \bar{r}_{i} }}
&=& 
\frac{ \bar{r}_{i}^2 }{r_0^4}  \int d\Omega \,  Z_{\ell m}(\Omega) \, e^{ - \bar{\phi}_b(\bar{r}_i,\Omega) - \bar{\phi}_{\rm dm}^{i}(\Omega) } \,,\\
\left.
\frac{\partial \mathcal{F}_{\ell m}}{\partial \sigma_0^2}
\right|_{\substack{\mathbf{y} = \bar{\mathbf{y}}_{i} \\ r = \bar{r}_{i} }}&=& -
\frac{ \bar{r}_i^2 }{r_0^2 \sigma_0^2} \int d\Omega \,  Z_{\ell m}(\Omega) \, \bar{\phi}_b^{i} \, \, e^{ - \bar{\phi}_b(\bar{r}_i,\Omega) - \bar{\phi}_{\rm dm}^{i}(\Omega) } \, .
\eea
The upper half of $\mathbb{B}_i$ vanishes since $\frac{\partial \mathcal{G}_{\ell m}}{\partial r_0^2} = \frac{\partial \mathcal{G}_{\ell m}}{\partial \sigma_0^2} = 0$.
From Eq.~\eqref{eq:ODE_disc_1}, we have the $2n_{\ell m}$-component vectors
\be
\mathbf{E}_i = \left( \begin{array}{c}
\phi_{00}^{i+1} - \phi_{00}^i + (r_{i+1}-r_i) \mathcal{G}_{00}(\bar{\mathbf{y}}_i,\bar{r}_i,\mathbf p) \\[2pt]
\phi_{10}^{i+1} - \phi_{10}^i + (r_{i+1}-r_i) \mathcal{G}_{10}(\bar{\mathbf{y}}_i,\bar{r}_i,\mathbf p) \\
\vdots \\[5pt]
\hline
\\[-10pt]
\mu_{00}^{i+1} - \mu_{00}^i + (r_{i+1}-r_i)\mathcal{F}_{00}(\bar{\mathbf{y}}_i,\bar{r}_i,\mathbf p) \\[2pt]
\mu_{10}^{i+1} - \mu_{10}^i + (r_{i+1}-r_i) \mathcal{F}_{10}(\bar{\mathbf{y}}_i,\bar{r}_i,\mathbf p) \\
\vdots
\end{array}
\right) \, .
\ee

Next, we write the boundary conditions in the form of $\mathbf{H}(\mathbf y_1, \mathbf y_N, \mathbf p) = 0$ as demanded by Eq.~\eqref{eq:BC_form}, where
\be
\mathbf{H}(\mathbf y_1, \mathbf y_N, \mathbf p) = \left( \begin{array}{cl} 

\begin{array}{c}
\phi_{00}^1  \\ \phi_{10}^1 \\ \vdots  
\end{array}
&
\Bigg\} \; {n_{\ell m} \; {\rm rows}}
\\[5pt]
\hline
\\[-10pt]
\begin{array}{c}
\int \frac{d\Omega}{4\pi} \rho_{\rm dm}^N(\Omega) - \rho_m
\end{array}
&
\big\} \; {1 \; {\rm row}}
\\[5pt]
\hline
\\[-10pt]
\begin{array}{c}
\vdots\\
\mu_{\ell m}^N + r_m (\ell + 1) \phi_{\ell m}^N + \frac{4\pi G}{\sigma_0^2} r_m^{\ell + 1} J_{\ell m}
\\
\vdots \\
\end{array} 
&
\Bigg\} \; 
\begin{array}{c}
(n_{\ell m} -1) \; \textrm{rows}\\
\textrm{$\ell > 0$ only}
\end{array}
\\[5pt]
\hline
\\[-10pt]
\begin{array}{c}
\mu_{00}^1 \\
\mu_{00}^N - \frac{\sqrt{4\pi} G}{\sigma_0^2} M_m 
\end{array} 
&
\Big\} \; 2 \; {\rm rows}
\end{array}
\right)\,,
\ee
is a $(2n_{\ell m} + 2)$-component vector.
We have written the shorthand
\be
\rho_{\rm dm}^N(\Omega) = \rho_0 \exp\left( - \sum_{\ell m} \phi_{\ell m}^N Z_{\ell m}(\Omega) - \phi_b(r_N,\Omega) \right)
\ee
for the dark matter density at $r_N = r_m$.
Following Eq.~\eqref{eq:disc_BCs_blocks}, we have the $(2n_{\ell m} +2)\times 2 n_{\ell m}$ matrix blocks
\be
\mathbb{C}_1 =
\left( \begin{array}{c|c} \mathbb{1}_{n_{\ell m}} &  \mathbb{O}_{n_{\ell m} \times n_{\ell m}} \\[5pt]
\hline
\mathbb{O}_{n_{\ell m} \times n_{\ell m}} & \mathbb{O}_{n_{\ell m} \times n_{\ell m}} \\[5pt]
\hline
\mathbb{O}_{2 \times n_{\ell m}}  &
\begin{array}{cccc} 
1 & 0 & \dots & 0 \\
0 & 0 & \dots & 0
\end{array} 
\end{array} \right) \, , \quad
\mathbb{C}_N =
\left( \begin{array}{c|c} \mathbb{O}_{n_{\ell m} \times n_{\ell m}} &  \mathbb{O}_{n_{\ell m} \times n_{\ell m}} \\[5pt]
\hline
\\[-10pt]
\mathbb{L}_{n_{\ell m} \times n_{\ell m}} & \mathbb{K}_{n_{\ell m} \times n_{\ell m}} \\[5pt]
\hline
\\[-10pt]
\mathbb{O}_{2 \times n_{\ell m}}  &
\begin{array}{cccc} 
0 & \dots & 0 & 0\\
1 & 0 & \dots & 0 
\end{array} 
\end{array} \right) \, ,
\ee
where we have defined the $n_{\ell m} \times n_{\ell m}$ sub-blocks
\be
\mathbb{L}_{n_{\ell m} \times n_{\ell m}} 
=
\left(
\begin{array}{c|ccc}
-\frac{\rho_{00}^N}{4\pi} & \dots & -\frac{\rho_{\ell m}^N}{4\pi} & \dots \\[5pt]
\hline
\\[-10pt]
0 &  &  &  \\
\vdots & & \underbrace{r_m {\rm diag}(\ell+1)}_{\ell > 0 \; \textrm{only}} &  \\
0 & & &  \\
\end{array}
\right) 
\, , \quad
\mathbb{K}_{n_{\ell m} \times n_{\ell m}} 
=
\left(
\begin{array}{c|ccc}
0 & 0 & \dots & 0 \\[5pt]
\hline
\\[-10pt]
0 &  &  &  \\
\vdots & & \mathbb{1}_{(n_{\ell m}-1)} &  \\
0 & & &  \\
\end{array}
\right) \, ,
\ee
where $\rho_{\ell m}^N = \int d\Omega \, Z_{\ell m}(\Omega) \rho_{\rm dm}^N(\Omega)$.
Lastly, we have the $(2n_{\ell m} +2)\times 2$ matrix block
\be
\mathbb{D} = \left( \begin{array}{c} \mathbb{0}_{n_{\ell m} \times 2} \\[5pt]
\hline
\\[-10pt]
\begin{array}{c|c} 
-\int \frac{d\Omega}{4\pi} \, \rho_{\rm dm}^N(\Omega)/ r_0^2 & \int \frac{d\Omega}{4\pi} \, \rho_{\rm dm}^N(\Omega) \frac{1 + \phi_b(r_N,\Omega)}{\sigma_0^2} \\[5pt]
\hline
\\[-10pt]
0 & \vdots \\
\vdots & - \frac{4\pi G}{\sigma_0^4} r_m^{\ell+1} J_{\ell m}  \quad (\ell > 0 \; \textrm{only}) \\
0 & \vdots \\[5pt]
\hline
\\[-10pt]
0 & 0 \\
0 & \frac{\sqrt{4\pi} G}{\sigma_0^4} M_{\rm CDM}(r_m) 
\end{array}
\end{array}
\right) \, .
\ee

\subsection{Analytic solution for isothermal halo shapes}
\label{app:q_iso_semianalytic}

Here we derive an analytic solution for the shape profile for the nonspherical isothermal Jeans model.
The goal is to obtain the axial ratio $q$ for the inner halo in terms the solution to the {\it spherical} Jeans model and the shapes of the outer halo and baryon distribution.
Finally, we compare our analytic results to the exact numerical solution given above.

Our starting point is Eqs.~(\ref{eq:IRR}-\ref{eq:q_shell}).
First, we evaluate $M_{zz} / M_{RR}$.
We work to linear order in powers of $\log q$ and treat all deviations from sphericity as $\mathcal{O}(\log q)$. 
(Recall $M_{zz}$ and $M_{RR}$ implicitly depend on $q$ since they are integrals over a spheroidal shell with axis ratio $q$ and effective spheroidal radius $\rsph$.)
At this order, Eq.~\eqref{eq:r_theta} is
\be
r \approx \rsph \left( 1 + \frac{1}{3} (3\cos^2\theta - 1) \log q \right) \, .
\ee
The dark matter density is expanded as
\bea
\rho_{\rm dm}(\mathbf r) &=& \sum_{\ell m} \rho_{\ell m}(r) Z_{\ell m}(\theta,\varphi) \\
&\approx& \rho_{00}(\rsph) Z_{00} \left( 1 +  \frac{1}{3}  \frac{d \log \rho_{00}(\rsph)}{d \log \rsph} (3\cos^2\theta - 1) \log q \right) + \sum_{\ell > 0, m}  \rho_{\ell m}(\rsph) Z_{\ell m}(\theta,\varphi) \, .\notag
\eea
Plugging these expressions into Eqs.~\eqref{eq:IRR} and \eqref{eq:Izz}, the angular integral is performed straightforwardly to yield
\bea \label{eq:Mzz_linear}
M_{zz} &=& \frac{4\rsph^4 \Delta \rsph \rho_{00}(\rsph)}{3 \sqrt{4\pi}} \left[ 1 + \frac{4}{15} \left(5 + \frac{d \log {\rho}_{00}(\rsph)}{d \log \rsph} \right) \log q + \frac{2}{\sqrt{5}} \frac{\rho_{20}(\rsph)}{\rho_{00}(\rsph)} 
\right] \\
\label{eq:MRR_linear}
M_{RR} &=& \frac{4\rsph^4 \Delta \rsph \rho_{00}(\rsph)}{3 \sqrt{4\pi}} \left[ 1 - \frac{2}{15} \left(5 + \frac{d \log {\rho}_{00}(\rsph)}{d \log \rsph} \right) \log q - \frac{1}{\sqrt{5}} \frac{\rho_{20}(\rsph)}{\rho_{00}(\rsph)} 
\right] \, .
\eea
Finally, we take the ratio of Eqs.~\eqref{eq:Mzz_linear} and \eqref{eq:MRR_linear} and, setting it equal to $q^2 \approx 1 + 2 \log q$, solve for $\log q$ at linear order to obtain
\be \label{eq:q_analytic}
\log q(\rsph) = - \frac{3 \sqrt{5}}{2} \rho_{20}(\rsph) \left(\frac{d\rho_{00}(\rsph)}{d\log \rsph} \right)^{-1}\, .
\ee

Next, we must calculate the density moments $\rho_{00}(r)$ and $\rho_{20}(r)$.
The former is approximated at zeroth-order as $\rho_{00}(r) \approx \sqrt{4\pi} \bar{\rho}_{\rm dm}(r)$, where $\bar{\rho}_{\rm dm}$ is the spherical solution~\eqref{eq:matching}.
For the latter, the isothermal solution is
\be \label{eq:rho_20}
\rho_{20}(r) = \int d\Omega \, Z_{20}(\theta) \, \rho_0 e^{- \phi_{\rm dm}(\mathbf r) - \phi_{b}(\mathbf r) } \, .
\ee
Note Eq.~\eqref{eq:rho_20} is the usual solid angle integral over a sphere (not a spheroid).
We expand $\phi_{\rm dm}(\mathbf r)$ as in  Eq.~\eqref{eq:phi_dm_sph_exp} to write
\be \label{eq:rho_20_2}
\rho_{20}(r) \approx - \bar\rho_{\rm dm}(r) \left( \phi_{20}(r) - \int d\Omega \, Z_{20}(\theta) e^{-(\phi_b(\mathbf r) - \bar\phi_b(r))} \right) \, ,
\ee
where $\bar\phi_b(r) = \bar\Phi_b(r)/\sigma_0^2$ is the spherically-averaged (dimensionless) baryon potential.
The second term represents the effect on the inner halo shape from baryons.
For the first term, from Eq.~\eqref{eq:phi_lm}, we have
\be \label{eq:phi_20}
\phi_{20}(r) = - \frac{4\pi G}{5\sigma_0^2} \left( \frac{1}{r^3} \int_0^r dx \, x^4 \rho_{20}(x) + r^2 \int_r^\infty dx \, x^{-1} \rho_{20}(x) \right) \, .
\ee
The integration region $r > r_m$ represents the effect on the inner halo from the outer halo shape, while the region $r < r_m$ is the nonlinear back-reaction of the inner halo shape on itself.

The isothermal shape $q_{\rm iso}$ is obtain from Eq.~\eqref{eq:q_analytic} and plugging in the nonspherical isothermal solutions given in Eqs.~(\ref{eq:rho_20}-\ref{eq:phi_20}).
Combining everything, we delineate three contributions to the isothermal shape as
\be \label{eq:q_iso_three_terms}
\log q_{\rm iso} = (1 - f_{\rm dm}) \log q_b + f_{\rm dm} \log q_{\rm outer} + \Delta \log q_{\rm iso} 
\ee
where $f_{\rm dm}(r) = M_{\rm dm}(r)/M_{\rm tot}(r)$ is the enclosed dark matter fraction. 
First, the effective baryon shape arises from the second term in Eq.~\eqref{eq:rho_20_2} and is
\be \label{eq:log_qb_exact}
\log q_b(r) = \frac{15}{4} \left( \frac{G M_b(r)}{r \sigma_0^2} \right)^{-1} 
\int_0^\pi d\theta \, \sin\theta \, P_2(\cos \theta) \, \exp\left(- \frac{\Phi_b(\mathbf r) - \bar\Phi_b(r)}{\sigma_0^2}\right)  \, .
\ee
We further work to first order in $\Phi_b - \bar\Phi_b$, which yields
\be \label{eq:log_qb}
\log q_b(r) = - \frac{15}{4} \left( \frac{G M_b(r)}{r} \right)^{-1} 
\int_0^\pi d\theta \, \sin\theta \, P_2(\cos \theta) \, \Phi_b(\mathbf r)   \, ,
\ee
although in principle Eq.~\eqref{eq:log_qb_exact} may provide more accurate results for highly nonspherical baryon potentials.\footnote{We emphasize that $q_b$ represents the shape of the baryon potential, not the shape of the baryon density. For nonspherical distributions, the former is typically much rounder than the latter.}

Next, we delineate the inner and outer halo shape contributions by writing the second integral in Eq.~\eqref{eq:phi_20} as $\int_{r}^\infty dx = \int_{r}^{r_m} dx + \int_{r_m}^\infty dx$ for $r < r_m$.
The second part describes the effect of a nonspherical outer halo on the inner isothermal shape, given by
\bea
\log q_{\rm outer}(r) &=& - \frac{4\pi r^3}{5 M_{\rm dm}(r)} \int_{r_m}^\infty dx \, \bar\rho_{\rm dm}^\prime(x) \log q_{\rm cdm}(x) \notag \\
&=& - \frac{4\pi r^3}{5 M_{\rm dm}(r)} \left( \bar\rho_{\rm dm}(r_m) \log q_{\rm cdm}(r_m)  - \alpha \int_{r_m}^\infty \frac{dx}{x} \, \bar\rho_{\rm dm}(x) \right) \label{eq:log_q_outer}
\eea
where prime denotes a derivative and the second line follows using Eq.~\eqref{eq:q_cdm} and integration by parts.
Eq.~\eqref{eq:log_q_outer} is computed directly from the spherically-averaged solution and the assumed profile for the CDM outer halo.

Lastly, the back-reaction term from the inner halo is
\be \label{eq:q_iso_backreaction}
\Delta \log q_{\rm iso}(r) =
- \frac{4\pi r^3}{5 M_{\rm tot}(r)} \left( \frac{1}{r^2} \int_0^r dx\,  x^5 \bar\rho_{\rm dm}^\prime(x) \log q_{\rm iso}(x)
+ r^3 \int_r^{r_m} \frac{dx}{x} \bar\rho_{\rm dm}^\prime(x) \log q_{\rm iso}(x) \right) \, .
\ee
This term cannot be computed directly since it depends on $q_{\rm iso}$ itself.
If we approximate $q_{\rm iso}$ as constant, it can be pulled outside the integral and solved for algebraically in Eq.~\eqref{eq:q_iso_three_terms}.
However, instead of making this approximation, we proceed with an iterative solution.
We compute $q_{\rm iso}$ from Eq.~\eqref{eq:q_iso_three_terms} neglecting back-reaction, which we use to evaluate Eq.~\eqref{eq:q_iso_backreaction} numerically and update $q_{\rm iso}$.
The process is repeated until reaching convergence after a few iterations.

Next, we show some examples for $q_{\rm iso}$ and compare to our numerical solutions from the relaxation method given above.
For the outer halo, we consider NFW dark matter halos with virial mass $\Mvir = 2 \times 10^{12} \, \Msun$, concentration $c=10$, and either $q_0 = 0.6$ (oblate halo), $q_0 = 1$ (spherical halo), or $q_0=1.5$ (prolate halo), with $\alpha=0$, as per Eq.~\eqref{eq:q_cdm}.
For baryons, we consider disk-like Miyamoto-Nagai (MN) potentials~\cite{Miyamoto:1975zz}
\be
\label{eq:MN}
\Phi_{\rm MN}(r,\theta) = - \frac{G M_{d}}{\sqrt{r^2 \sin^2\theta + \left(a_d + \sqrt{r^2 \cos^2\theta + b_d^2}\right)^2}} \, ,
\ee
where $M_d = 5 \times 10^{10} \, \Msun$ is the (fixed) disk mass, and $a_b$, $b_d$ are the disk length scale and thickness, respectively.
We consider different parameter pairs $(a_d, b_d)$ and we label the corresponding potential as ${\rm MN}_{a_d,b_d}$, in units of $\kpc$.
Eq.~\eqref{eq:MN} spans oblate baryon distributions from the extreme case of a thin (Kuzmin) disk, ${\rm MN}_{a_d,0}$, to a spherically-symmetric Plummer sphere, ${\rm MN}_{0,b_d}$.
For prolate baryon distributions, following Ref.~\cite{2019MNRAS.486.3915A}, we consider spindle-like Miyamoto-Nagai (sMN) potentials as examples of self-consistent baryon distributions, given by
\be
\label{eq:sMN}
\Phi_{\rm sMN}(r,\theta) = - \frac{G M_{d}}{\sqrt{ \left(c_d + \sqrt{r^2 \sin^2\theta + d_d^2}\right)^2 + r^2 \cos^2\theta}} \, ,
\ee
where $c_d, d_d$ are distance parameters. We label the potentials by ${\rm sMN}_{c_d,d_d}$, in units of $\kpc$.

\begin{figure}[t]
\centering
\includegraphics[width=\textwidth]{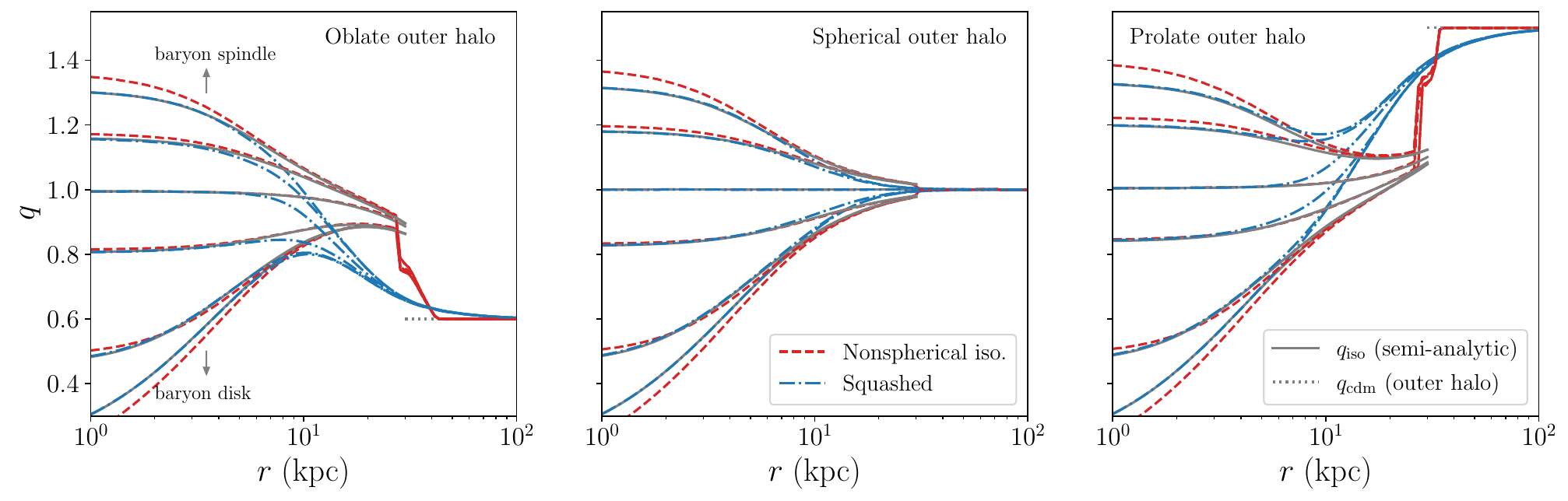}
\caption{\it Halo shape profiles for SIDM in nonspherical isothermal (red dashed) and squashed (blue dot-dashed) Jeans models are compared to our semi-analytic calculation of $q_{\rm iso}$ (solid gray) for different outer halos and baryon distributions.
Panels correspond to different outer halo shapes: oblate, $\qcdm = 0.6$ (left); spherical, $\qcdm=1$ (center); and prolate, $\qcdm = 1.5$.
In each panel, each grouping of curves corresponds to a different baryon potential, from disk-like (bottom) to spindle-like (top). 
See main text for further details.}
\label{fig:q_iso_plot}
\end{figure}

In Fig.~\ref{fig:q_iso_plot}, we compare our semi-analytic calculation for $q_{\rm iso}$ (gray solid lines) to our numerical solution from the nonspherical isothermal Jeans model, computed with the relaxation method with $\ell \le 10$ (red dashed lines).
We also include halo shapes inferred from the squashed Jeans model (blue dot-dashed lines) for comparison.
We have fixed the halo mass $\Mvir = 2 \times 10^{12} \, \Msun$, concentration $c = 10$, and matching radius $r_m = 30 \, \kpc$, which corresponds to $\sigma/m \approx 1 \cmg$.
Each panels represents a different outer halo shape, shown by the dotted horizontal lines: oblate, $\qcdm = 0.6$ (left); spherical, $\qcdm = 1$ (center); and prolate $\qcdm = 1.5$ (right).
In each panel, we consider six different baryon potentials
\be
\underbrace{ {\rm MN}_{5,0}, \; 
{\rm MN}_{5,1}, \;  
{\rm MN}_{5,5}}_\text{$\leftarrow$ disk-like}, \;  
\underbrace{{\rm MN}_{0,5}}_\text{spherical} , \;  
\underbrace{{\rm sMN}_{5,5}, \;  
{\rm sMN}_{5,3}}_\text{spindle-like $\rightarrow$} \, ,
\ee
which span from the extremely oblate thin disk to a prolate spindle-like potential, with a common disk mass $M_d = 5 \times 10^{10} \; \Msun$.

The key takeaways from Fig.~\ref{fig:q_iso_plot} are as follows:
\begin{itemize}
\item Within the inner halo, the halo shapes computed numerically (red dashed) and semi-analytically (gray solid) are in good agreement (except for the former's sharp discontinuity near the matching radius).
The largest differences occur for the most extreme shapes, which is to be expected since our expansion in $\log q$ is breaking down.
\item Differences in the baryon potential provide the main driver on the isothermal shape. However, we also see an impact from different outer halo shapes around $r\sim 25 \; \kpc$. Our $q_{\rm iso}$ calculation successfully accounts for this effect, which indicates that we have accurately modeled the shape due to both baryons and the outer halo.
\item The isothermal Jeans model shape (red dashed) does not smoothly match onto the outer halo shape (dotted lines) at the matching radius, as discussed previously. On the other hand, the squashed Jeans model (blue dot-dashed) provides a smooth transition between the isothermal inner halo and CDM outer halo as a function of the number of scatters.
\end{itemize}

\section{Squashed Jeans model elements}

\label{app:squashed}

\subsection{Collisional erasure of halo anisotropy}

In the squashed Jeans model, the halo shape $q(\mathcal{N} )$ is taken to be a function of the mean number of scatters per particle, $\mathcal{N} $.
Here we calculate $q(\mathcal{N} )$ analytically within a simplified toy model.
We take the boundary conditions
\be \label{eq:q_boundary_conditions}
q(\mathcal{N} ) = \left\{ \begin{array}{cc} 
\qcdm 
&
{\rm for \;} \mathcal{N}  \ll 1 \\
q_{\rm iso} 
&
{\rm for \; } \mathcal{N}  \gg 1
\end{array}
\right.
\ee
where $\qcdm$ is the halo shape in the absence of collisions, i.e., in the outer halo where collisions are negligible, and $q_{\rm iso}$ is the halo shape for an isothermal halo in hydrostatic equilibrium, computed from the nonspherical isothermal Jeans model.

Our calculation is in a similar spirit to that presented in Ref.~\cite{Agrawal:2016quu}.
In that work, the authors estimated the evolution of halo anisotropy due to self-interactions by computing the rate for hot and cold populations of SIDM to thermalize within a spatially homogeneous and isotropic toy model.
We improve on their argument by considering a genuinely anisotropic toy model and computing how the anistropy evolves due to collisions.
Ref.~\cite{Agrawal:2016quu} also assumed Rutherford-type self-interactions, whereas we assume contact-type interactions.

For our toy model, we consider a spatially homogeneous SIDM population with an anisotropic velocity distribution.
As an ansatz, we take an anisotropic Maxwell-Boltzmann distribution
\begin{equation} 
f(\mathbf{v}) = \frac{n}{(2\pi)^{3/2}\nu_x\nu_y\nu_z}\exp{\left(-\frac{v_x^2}{2\nu_x^2}-\frac{v_y^2}{2\nu_y^2}-\frac{v_z^2}{2\nu_z^2}\right)} \, ,
\end{equation}
where $n$ is number density and $(\nu_x,\nu_y,\nu_z)$ are 1D velocity dispersions along each Cartesian direction. 
We assume axial symmetry, $\nu_x = \nu_y$, and define the axial ratio 
\be
q = \frac{\nu_z}{\nu_x} = \frac{\nu_z}{\nu_y}
\ee
and mean velocity dispersion 
\be
\nu_0^2 = \frac{1}{3} \left( \nu_x^2 + \nu_y^2 + \nu_z^2 \right) \, .
\ee
Inverting these relations gives
\be \label{eq:invert}
\nu_x = \nu_y = \frac{\sqrt{3} \nu_0}{\sqrt{2 + q^2}} \, , \quad \nu_z = \frac{\sqrt{3} q \nu_0}{\sqrt{2 + q^2}} \, .
\ee
Since this setup is spatially homogeneous, we take the initial axis ratio in the absence of collisions to be $\qcdm = q_0$, i.e., a constant.

The effect of self-interactions is governed by the collisional Boltzmann equation
\be \label{eq:collisional_BE}
\frac{d f_1}{dt} = C = -\int d^3\mathbf{v_2}  \; \int d\Omega ' \; |\mathbf{v_1}-\mathbf{v_2}| \; \frac{d\sigma}{d\Omega '}\left(f_1 f_2-f_3 f_4  \right) \, ,
\ee
where subscripts denote the velocity argument, $f_i = f(\mathbf v_i)$.
The collision term $C$ arises from elastic two-body scattering, where $\mathbf{v}_{1,2}$ ($\mathbf{v}_{3,4}$) are the incoming (outgoing) particle velocities in the galaxy frame.
The differential cross section is $d\sigma(|\mathbf{v}_1 - \mathbf{v}_2|)/d\Omega^\prime$, where primes denote angles in the center-of-mass frame.
For a contact-type interaction, we have $d\sigma/d\Omega^\prime = \sigma/(4\pi)$.

The goal is to express the Boltzmann equation~\eqref{eq:collisional_BE} as as a differential equation for $q(t)$.
It suffices to take $\mathbf{v}_1 = \boldsymbol{0}$.
The left side of Eq.~\eqref{eq:collisional_BE} becomes
\be \label{eq:df1_dt}
\frac{df_1}{dt} = \frac{2 n \sqrt{2 + q^2}(q^2 - 1)}{(6\pi)^{3/2} \nu_0^3 q^2} \frac{dq}{dt} \, ,
\ee
where because of our assumption of spatial homogeneity, $n$ and $\nu_0$ must be time-independent by mass and energy conservation.
Next, to evaluate the collision term, we parametrize the remaining velocities as
\bea
\mathbf{v}_2 &=& v_2 (\sin\theta,0,\cos\theta) 
\notag \\
\mathbf{v}_3 &=& \frac{v_2}{2} (\sin\theta + \sin\theta^\prime \cos\phi^\prime,\sin\theta^\prime \sin\phi^\prime,\cos\theta + \cos\theta^\prime) 
\\
\mathbf{v}_4 &=& \frac{v_2}{2} (\sin\theta - \sin\theta^\prime \cos\phi^\prime,-\sin\theta^\prime \sin\phi^\prime,\cos\theta - \cos\theta^\prime) \, . \notag
\eea
Plugging into the collision term, the azimuthal integrals are trivial and we have
\bea
C = - \frac{n^2 \sigma}{2 (2\pi)^2 \nu_x^4 \nu_z^2} \int dv_2 \, v_2^3 \, \int d\cos\theta \int d \cos\theta^\prime \, \left[ \exp\left( - \frac{v_2^2 \sin^2\theta}{2\nu_x^2} - \frac{v_2^2 \cos^2\theta}{2\nu_z^2} \right) \right. \notag \\
\left. - \exp\left( - \frac{v_2^2 (\sin^2\theta + \sin^2\theta^\prime)}{4\nu_x^2} - \frac{v_2^2 (\cos^2\theta + \cos^2\theta^\prime)}{4\nu_z^2} \right) \right] \, .
\eea
These integrals can be evaluated analytically. They have a divergence at $\theta = \pi/2$, which translate to a divergence for $q\to0$. We do the integrals assuming $q\ne0$ and then analytically continue the result.   We obtain 

\be\label{eq:colfull}
C= \frac{n^2 \left(q^2+2\right) \sigma }{6 \pi ^2 \nu _0^2 q} \left(q \left(\frac{4 \tanh ^{-1}\left(\sqrt{\frac{q^2-1}{q^2+1}}\right)}{\sqrt{q^4-1}}-1\right)-\frac{\tanh ^{-1}\left(\frac{\sqrt{q^2-1}}{q}\right)}{\sqrt{q^2-1}}\right)\,.
\ee
Notice the $1/q$ divergence cancels with a factor of $1/q$ in the  term on the right hand side of Eq. ~\eqref{eq:df1_dt}. 
In principle we  can now use this result to evaluate the time evolution of the ellipticity of the halo. Unfortunately, the $q$-integral is difficult to do analytically.   
Assuming the system is not too far from isotropy, we can expand $C$ for $q$ close to $1$. Let us explore two approaches for this. First, if we expand in powers of $(q-1)$, to get
\be \label{eq:series_1}
C = - \frac{ 8 n^2 \sigma}{15 \pi^2 \nu_0^2} \left( (q-1)^2 - \frac{19}{21} (q-1)^3 + \frac{85}{84} (q-1)^4 - \frac{710}{693} (q-1)^5 + \frac{9001}{9009} (q-1)^6 + ... \right) \, .
\ee
This series  has $\mathcal{O}(1)$ coefficients, which implies a truncation will quickly break down for sizable anisotropy.
Fortunately, the situation is improved by taking an alternative expansion in powers of $\log q$ (around $\log q = 0)$.
The first several nonzero terms are 
\be \label{eq:series_2}
C = - \frac{ 8 n^2 \sigma}{15 \pi^2 \nu_0^2} \left( \log^2 q + \frac{2}{21} \log^3 q + \frac{5}{21} \log^4 q + \frac{82}{693} \log^5 q + \frac{1702}{45045} \log^6 q + ... \right) \, .
\ee
Because of the smaller series coefficients,  the leading term in Eq.~\eqref{eq:series_2}  gives a better approximation to $C$  in   \eqref{eq:colfull} compared to that of    Eq.~\eqref{eq:series_1}.

\begin{figure}[t]
\centering
\includegraphics[width=0.49\textwidth]{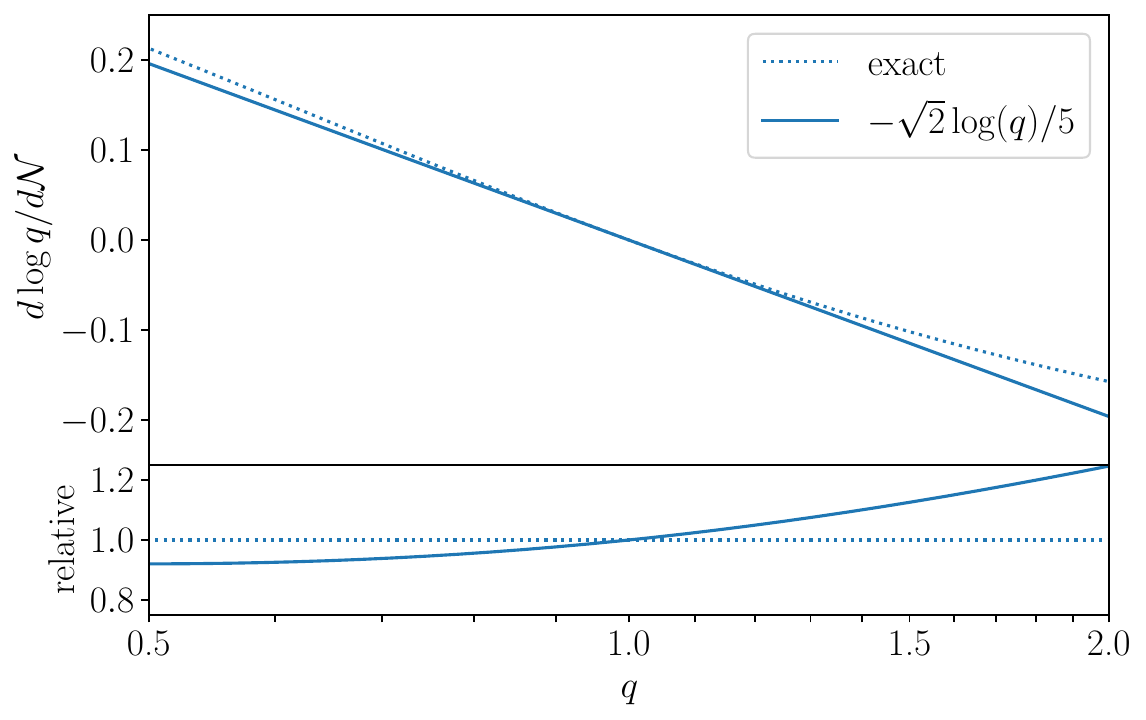}
\includegraphics[width=0.49\textwidth]{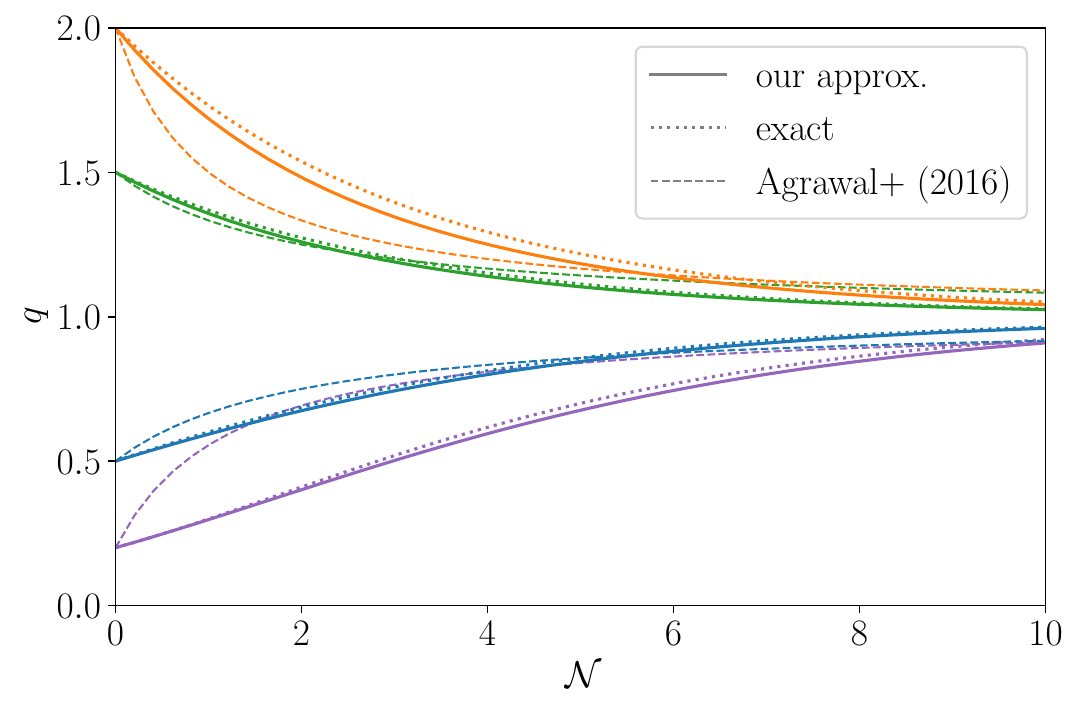}
\caption{\it Left: Top panel shows the anisotropy relaxation rate $d \log q/d\mathcal{N} $ for our simplified model with our small $\log q$ approximation (solid) and exact result (dotted). 
The bottom panel shows the ratio relative to the exact result. 
Right: Curves compare our results with those of Ref.~\cite{Agrawal:2016quu}, showing how different initial anisotropies $q_0 = 0.3, 0.5, 1.5, 2.0$ (bottom to top) are washed out with increasing number of collisions $\mathcal{N}$.
Solid lines denote our approximate result in Eq.~\eqref{eq:final_result} and dotted curves are corresponding exact results, while dashed lines are computed from Eq.~\eqref{eq:Agrawal_result} following Ref.~\cite{Agrawal:2016quu}.}
\label{fig:toy_model_plot}
\end{figure}

Next, we combine Eq.~\eqref{eq:df1_dt} and the leading term in Eq.~\eqref{eq:series_2} to give
\be \label{eq:approx}
\frac{d\log q}{d \mathcal{N}} \approx - \frac{\sqrt{2}}{5} \log q \, ,
\ee
where $d\mathcal{N} = R_{\rm scat} dt$ is the differential number of collisions and $R_{\rm scat} = 4 n \sigma \nu_0/ \sqrt{\pi}$ is the Boltzmann-averaged scattering rate.\footnote{Note $v_{\rm rel} = 4\nu_0 /\sqrt{\pi}$ is the mean relative velocity between two particles with Maxwell-Boltzmann distributions.}
The right side follows by further expanding the term in parentheses to leading order in $\log q$.
Fig.~\ref{fig:toy_model_plot} (left) compares our final approximate   result for $d\log q/d \mathcal{N}$ (solid line) and the exact  result obtained without doing a series expansion (dotted line) of  Eq. \eqref{eq:colfull}.
The lower panel shows the ratio of the two results.

Integrating  Eq.~\eqref{eq:approx} we get
\be \label{eq:final_result}
\log q(\mathcal{N}) = \log q_0 \, e^{-\sqrt{2} \mathcal{N}/5}
\ee
where $q_0$ is the initial anistropy at $\mathcal{N}=0$.
For comparison, Ref.~\cite{Agrawal:2016quu} followed a similar argument (albeit with different assumptions discussed above) to obtain the time evolution of halo ellipticity
\be
\varepsilon(t) = \frac{\varepsilon_0}{ \varepsilon_0 t / \tau+ 1}
\ee
where $\varepsilon_0$ is the initial ellipticity and $\tau = R_{\rm scat}^{-1}$ is the scattering time.
To compare to Eq.~\eqref{eq:final_result}, we write $\mathcal{N} = t/\tau$ and $\varepsilon = 1 - {\rm min}(q,1/q)$ to obtain
\be
\label{eq:Agrawal_result}
q(\mathcal{N}) = 1 - \frac{1 - q_0}{\mathcal{N}|q_0 - 1| + 1} \, .
\ee
In Figure~\ref{fig:toy_model_plot} (right), we compare these various results for $q(\mathcal{N})$ for different initial anisotropies $q_0$.
Our exact and approximate results (dotted and solid curves, respectively) are in good agreement, differing by less than 4\% for initial anisotropies shown, which justifies our use of the latter instead of the former in the context of other systematic uncertainties in our Jeans modeling.
We also show the results of Eq.~\eqref{eq:Agrawal_result} based on Ref.~\cite{Agrawal:2016quu} (dashed curves).
At face value, our results show that more extreme halo shapes (say, $\varepsilon_0 > 0.5$) relax more slowly with the onset of collisions compared to Ref.~\cite{Agrawal:2016quu}.
However, differences in the assumed scattering model (Rutherford vs contact interactions) may also play a role.

Lastly, the results presented so far demonstrate that halo shapes relax to isotropy in the limit $\mathcal{N} \gg 1$.
However, if there is an anisotropic baryon potential, the halo may relax to $q_{\rm iso}$ that may differ from unity.
We generalize Eq.~\eqref{eq:final_result} by the ansatz 
\be \label{eq:final_final_result}
\log q(\mathcal{N}
) = \log q_{\rm iso} 
+ \log (\qcdm/q_{\rm iso}) \, e^{-\sqrt{2} \mathcal{N}/5} \, ,
\ee
satisfying the boundary conditions in Eq.~\eqref{eq:q_boundary_conditions}.
Further theoretical support for Eq.~\eqref{eq:final_final_result} would likely require controlled numerical simulations, which is beyond the scope of the present work.

\subsection{Phase space anisotropy}

The squashed Jeans model assumes that the collisional relaxation rate for dark matter's spatial anisotropy follows that for its velocity anisotropy, given in the preceding section, according to the ansatz $q = \nu_z/\nu_x$.
Here we provide some theoretical justification.
Let us suppose that we have an anisotropic Maxwell-Boltzmann distribution function, 
\be \label{eq:nonspherical_f}
f(\mathbf r,\mathbf v) = \frac{n(\mathbf r)}{(2\pi)^{3/2} \nu_x^2 \nu_z} \exp\left( - \frac{v_x^2 + v_y^2}{2\nu_x^2} -\frac{v_z^2}{2\nu_z^2}\right) \, ,
\ee
where we have an anisotropic velocity distribution for $\nu_x \ne \nu_z$.
We also take a nonspherical spatial distribution, assuming that the number density $n(\mathbf r) = n(r_{\rm eff})$ is a function of the effective spherioidal radius
\be
r_{\rm eff} = \sqrt{q^{2/3} (x^2 + y^2) + q^{-4/3} z^2 }
\ee
where $q$ is the spatial shape of the halo.

Next, we suppose that our ansatz in Eq.~\eqref{eq:nonspherical_f} is a stationary state of the collisionless Boltzmann equation. 
This requires
\be \label{eq:collisionless_BE}
\mathbf{v} \cdot \frac{\partial f}{\partial \mathbf r} + \mathbf{g} \cdot \frac{\partial f}{\partial \mathbf v} = 0
\ee
where $\mathbf{g} = - \nabla \Phi$ is the gravitational acceleration.
Eq.~\eqref{eq:collisionless_BE} requires
\be \label{eq:conditions}
\left(\frac{v_x q^{2/3} x}{r_{\rm eff}} + \frac{v_y q^{2/3} y}{r_{\rm eff}} + \frac{v_z q^{-4/3} z}{r_{\rm eff}} \right) n^\prime
= \left(\frac{g_x v_x}{\nu_x^2} + \frac{g_y v_y}{\nu_x^2} + \frac{g_z v_z}{\nu_z^2} \right) n\, ,
\ee
where $n^\prime = \partial n/\partial r_{\rm eff}$.
Demanding that Eq.~\eqref{eq:conditions} is satisfied for any velocity $\mathbf{v}$, we get three separate equations
\be
\frac{g_x}{\nu_x^2} = \frac{n^\prime x q^{2/3}}{n r_{\rm eff}} \, , \quad
\frac{g_y}{\nu_x^2} = \frac{n^\prime y q^{2/3}}{n r_{\rm eff}} \, , \quad
\frac{g_z}{\nu_z^2} = \frac{n^\prime z q^{-4/3}}{n r_{\rm eff}} \, , 
\ee
and we can solve for the ratio
\be \label{eq:app_ratio}
\frac{\nu_z^2}{\nu_x^2} = \left(\frac{x g_z}{z g_x} \right) q^2 \, .
\ee
We may calculate $\mathbf{g}$ in the multipole expansion and the leading term in $q$ will be the monopole term. 
Since this term is radial, the term in parentheses in Eq.~\eqref{eq:app_ratio} will be unity.
Hence, we have $\nu_z/\nu_x \approx q$.

%%%%%%%%%%%%%%%%%%%%%%%% Bibliography %%%%%
%\clearpage
\bibliography{mega_dm_bib}
\bibliographystyle{helsevier}
%%%%%%%%%%%%%%%%%%%%%%%%%%%%%%%%%%%%%%%%
\end{document}